\documentstyle[11pt]{article}

\newcommand{\mrm}[1]{\rm{#1}}

\newcommand{\ttt}[1]{\tt{#1}}
 
\newcommand{\alphas}{\alpha_{\rm{s}}}
\newcommand{\alphaem}{\alpha_{\rm{em}}}

\renewcommand{\a}{\rm{a}}
\renewcommand{\b}{\rm{b}}
\renewcommand{\c}{\rm{c}}
\renewcommand{\d}{\rm{d}}
\newcommand{\e}{\rm{e}}
\newcommand{\f}{\rm{f}}
\newcommand{\g}{\rm{g}}
\newcommand{\hrm}{\rm{h}}

\newcommand{\n}{\rm{n}}
\newcommand{\p}{\rm{p}}

\newcommand{\s}{\rm{s}}
\renewcommand{\t}{\rm{t}}
\renewcommand{\u}{\rm{u}}
\newcommand{\A}{\rm{A}}
\newcommand{\B}{\rm{B}}
\newcommand{\D}{\rm{D}}

\renewcommand{\H}{\rm{H}}

\newcommand{\K}{\rm{K}}

\newcommand{\W}{\rm{W}}
\newcommand{\Z}{\rm{Z}}

\newcommand{\Jpsi}{\rm{J}/\psi}

{\end{list}}
\newcounter{enumct}

{\begin{list}{}{\setlength{\topsep}{0mm} \setlength{\itemsep}{0mm}
\setlength{\parskip}{0mm} \setlength{\parsep}{0mm}
\setlength{\leftmargin}{20mm} \setlength{\rightmargin}{0mm}
\setlength{\labelwidth}{18mm} \setlength{\labelsep}{2mm}}}%
{\end{list}}
{\begin{list}{}{\setlength{\topsep}{0mm} \setlength{\itemsep}{0mm}
\setlength{\parskip}{0mm} \setlength{\parsep}{0mm}
\setlength{\leftmargin}{10mm} \setlength{\rightmargin}{0mm}
\setlength{\labelwidth}{18mm} \setlength{\labelsep}{2mm}}}%
{\end{list}}

\newlength{\captivewidth}
\newcommand{\captive}[1]{\rule{5mm}{0mm}%
\begin{minipage}{\captivewidth}%
\caption[small]{#1}\end{minipage}}

\newlength{\tablinsep}
\newlength{\halfpagewid}
\def\lline{--------------------------------------------------------------------------------------------------------------------------------------------}
\def\sline{------------------------------------------------------------------------------------------------------------------------------------}
 
\input{epsf}

\begin{document}

\sloppy
\setcounter{page}{0}
\thispagestyle{empty}


\begin{figure}
\vspace{-11.0cm}
\epsfxsize=1638pt
\centerline{ \epsfbox{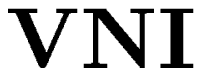} }
\vspace{-62.0cm}
\end{figure}

\begin{center}
{\LARGE {\bf  -- Version  4.1 -- $\;\;\;\;$ }}
\end{center}
\bigskip
\bigskip
\bigskip

\begin{center}
{\LARGE
{\bf Simulation of high-energy particle collisions in QCD:}}
\end{center}
\begin{center}
{\Large
Space-time evolution of $e^+e^-$ $\ldots$ $A+B$ collisions with
\\ ${}$ \\ -- parton-cascades -- \\ 
-- parton-hadron conversion --  \\  -- final-state hadron cascades--}
\end{center}
\bigskip
\bigskip
\bigskip
\bigskip
\bigskip

\begin{center}

{\Large
{\bf  Klaus Geiger$^{a}$, Ron Longacre$^{a}$, Dinesh K. Srivastava$^{b}$}
}
\medskip

$^{a}${\it Physics Department,
Brookhaven National Laboratory, Upton, N.Y. 11973, U.S.A.}

$^{b}${\it Variable Energy Cyclotron Centre, 1/AF Bidhan Nagar, 
       Calcutta 700 064, India}    

       email: klaus@bnl.gov, longacre@bnl.gov, dks@vecdec.veccal.ernet.in  
\end{center}
\smallskip

\begin{center}
       http://rhic.phys.columbia.edu/rhic/vni \\
       phone:  (516) 344-3791    \\
       fax:    (516) 344-2918    \\
\end{center}
\vspace{0.5cm}

\begin{center}
{\large {\bf Abstract}}
\end{center}
\bigskip

\noindent
$VNI$ is a general-purpose Monte-Carlo event-generator, which includes the
simulation of lepton-lepton, lepton-hadron, lepton-nucleus, 
hadron-hadron, hadron-nucleus, and nucleus-nucleus collisions.
It uses the real-time evolution of parton cascades in conjunction with
a self-consistent hadronization scheme, as well as  
the development of hadron cascades after hadronization.
The causal evolution from a specific initial state (determined by
the colliding beam particles) is followed by the time-development
of the phase-space densities of partons, pre-hadronic parton clusters,
and final-state hadrons, in position-space, momentum-space and color-space.
The parton-evolution is described in terms of a space-time generalization 
of the familiar momentum-space description of multiple (semi)hard interactions 
in QCD, involving $2 \rightarrow 2$ parton collisions, $2\rightarrow 1$ parton 
fusion processes, and $1\rightarrow 2$ radiation processes.  The formation 
of color-singlet pre-hadronic clusters and their decays into hadrons, 
on the other hand, is treated by using a spatial criterion motivated by 
confinement and a non-perturbative model for hadronization. Finally, the 
cascading of produced pre-hadronic clusters and of hadrons is includes a 
multitude of $2\rightarrow n$ processes, and is modeled in paralell to the 
parton cascade description.  This paper gives a brief review of the physics 
underlying $VNI$, as well as a detailed description of the program itself.
The latter program description emphasizes easy-to-use pragmatism and explains 
how to use the program (including simple examples), annotates input and 
control parameters, and discusses output data provided by it.
\bigskip
\bigskip

\noindent
{\it Program obtainable from:} 
http://rhic.phys.columbia.edu/rhic/vni

\noindent
{\it Originally  published in:} 
Computer Physics Communications {\bf 104}, 70 (1997)

\newpage

\tableofcontents

\newpage

\section{PROGRAM SUMMARY}
\label{sec:section1}
\bigskip

{\small

\noindent
{\bf Title of program:} VNI
\bigskip

\noindent
{\bf Computer fo which the program has been designed and
others on which it has been tested:}
IBM RS-6000, Sun Sparc, Hewlett Packard UX A-9000
\bigskip

\noindent
{\bf Operating systems:}
IBM-AIX, Sun-OS, and any other UNIX operating systems, as well as LINUX.
\bigskip

\noindent
{\bf Programming language used:}
Fortran 77
\bigskip

\noindent
{\bf Memory required to execute with typical data:} 2000 kwords
\bigskip

\noindent
{\bf No. of bits in a word:} 64
\bigskip

\noindent
{\bf No. of lines in distributed program:} 
25760 lines of main program, plus 49 and 244 lines for two example programs.
\bigskip

\noindent
{\bf Keywords:} 
Monte Carlo simulation, event generator, 
QCD kinetic theory, parton cascades, parton coalescence,
hadronic final states.
\bigskip

\noindent
{\bf Nature of physical problem:}

\noindent
In high-energy particle collisions certain phase-space regions
can be populated by a large number of quanta, such that
statistical correlations among them
(e.g., in space, momentum, or color) become of essential importance.
Examples are deep-inelastic lepton-hadron scattering and 
hadron-hadron collisions in the region of very small Bjorken-$x$, or,
collisions involving heavy nuclei in the central rapidity region.
In these cases the produced particles evolve in a 
complicated non-equilibrium environment created by the presence
of neighboring ones. The `deterministic' quantum evolution
of particle states due to self-interactions (depending only on the
particle itself), receives a new 
`statistical' kinetic contribution due to mutual interactions
(depending crucially on the local density).
The theoretical basis for addressing the solution for 
the dynamics of such particle systems is a quantum-kinetic
formulation of the QCD equations of motion,
an approximation that combines field-theoretical aspects
associated with the renormalization group (including
well-known resummation techniques) with aspects of
transport theory associated with non-equilibrium multi-particle
dynamics (including important quantum effects beyond the classical level).

\bigskip

\noindent
{\bf Method of solution:}

\noindent
The solution of the underlying quantum-kinetic equations
of motion for non-equilibrium multi-particle QCD 
by Monte-Carlo simulation of collisions allowing for a variety of 
combinations of beam and target particles.
To simulate the real-time evolution of the collision system in
position space and momentum space on the basis of the equations
of motions, the procedure is four-fold:
i) the construction of the initial state including the
decomposition of the beam and target particles
into their partonic substructure, (ii) the evolution
of parton cascades including multiple scatterings, emission- and
fusion-processes, (iii)  the self-generated
conversion of partons into hadrons using a phenomenological
model for parton-coalescence into pre-hadronic clusters and
subsequent decay into final-state particles,
and (iv) the final-state interactions among produced
clusters and hadrons including scattering, absorption, resonance
production.
\bigskip

\noindent
{\bf Restriction on the complexity of the problem:}

\noindent
For very high collision energy ($\sqrt{s} \gg 10$ TeV in
hadronic collisions,  and
$\sqrt{s} \gg 5$ TeV/nucleon in nuclear collisions)
numerical inaccuracies due to repeated Lorentz boosts, etc.,
may accumulate to cause problems, although
the program uses double precision throughout.
\bigskip

\noindent
{\bf Typical running time:}

\noindent
The CPU time for a typical simulation is strongly dependent on the type of
beam and target, the magnitude of collision energy, as well
as on the time interval $\Delta t$ chosen to follow an event in its real-time
evolution.
Examples  are (for $\Delta t = 35$ $fm$):
a) $e^++e^-$ at $\sqrt{s} = 100$ GeV:   10000 events/hour;
b) $p+\bar{p}$ at $\sqrt{s} = 200$ GeV:   5000 events/hour;
c) $p+ Au$ at $\sqrt{s} = 200$ GeV/nucleon:   100 events/hour;
d) $Au+Au$ at $\sqrt{s} = 200$ GeV/nucleon:   1 event/hour;
All of the above quotes are approximate, and 
refer to a typical
133 Mhz or 166 Mhz processor on a modern
Power-Workstation or Power-PC.
}
\bigskip

\newpage

\section{LONG WRITE-UP}
\label{sec:section2}
\bigskip

\subsection{Introduction}

VNI 
\footnote{
       The three letters VNI do not mean anything profound. VNI is
       pronounced "Vinnie", short for "Vincent Le CuCurullo Con GiGinello", 
       the little guy who likes to hang out with his pals, the quarks 
       ({\it cucu}rullos) and gluons ({\it gigi}nellos). 
       That is QCD in Wonderland, and that is the whole, true story.
}
is the Monte Carlo implementation 
of a relativistic quantum-kinetic approach \cite{ms39,ms42} to
the dynamics of high-energy particle collisions, inspired by the 
QCD parton picture of hadronic interactions \cite{kogut73,dok80,glr83}.
It is a product of several years of development in both
the improving physics understanding of high-energy multiparticle 
dynamics in QCD, 
as well as the technical implementation in the form of a computer simulation
program. The most relevant references
for the following are Refs. \cite{ms0,ms3,msrep,ms37,ms40,ms41}, 
where details of the
main issues, discussed below, can be found. 
The puropse of VNI is to provide  a comprehensive
description of particle collisions involving beams of
leptons, hadrons, or nuclei,
in terms of the space-time evolution of parton cascades and
parton-hadron conversion.
The program VNI is concepted as a useful {\it tool} (and nothing more) to study
the causal development of the collision dynamics in real time from 
a specified initial state of beam and target particles, all the way to the
final yield of hadrons and other observable particles.
The collision dynamics is traced in detail on the microscopic level 
of quark and gluon interactions in the framework of perturbative QCD,
supplemented by a phenomenological treatment of the non-perturbative dynamics,
including soft parton-collisions and parton-hadronization.  
The generic structure of the simulation concept is illustrated in Fig. 1.
\medskip

\begin{figure}
\epsfxsize=450pt
\centerline{ \epsfbox{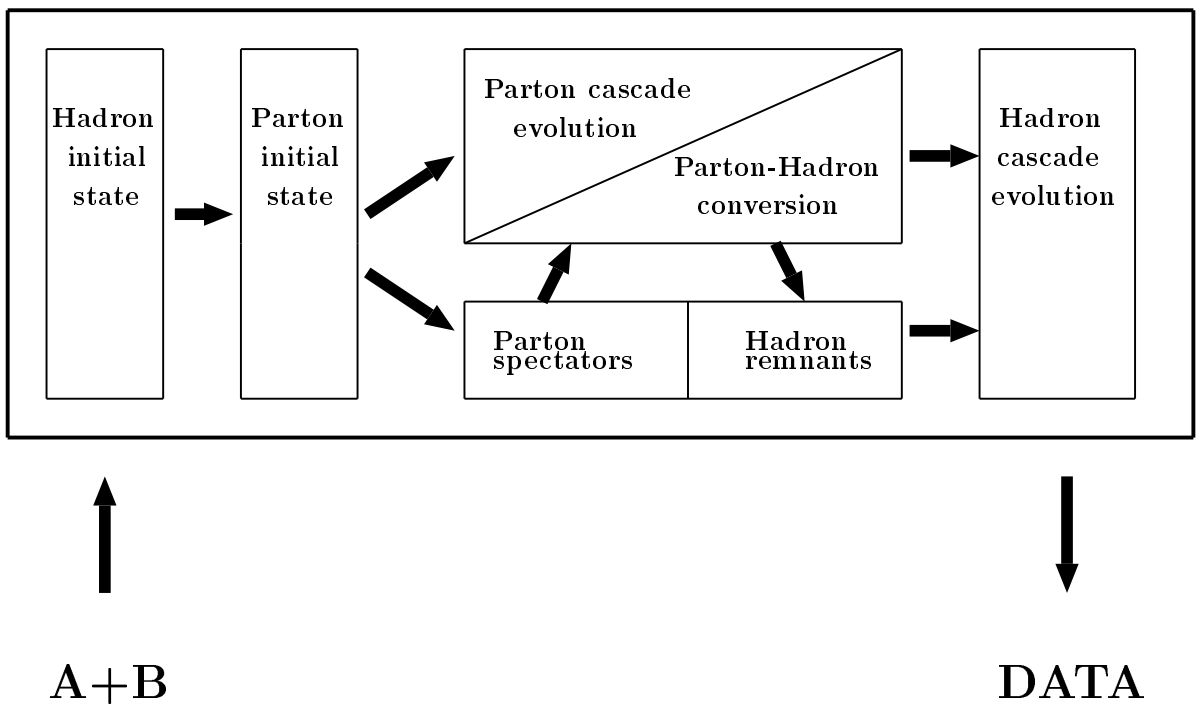} }
\caption{
Generic flow-chart of the simulation concept of VNI:
It starts with
the initial beam target particles $A$ and $B$, decomposing them
(except for leptonic $A$ and/or $B$) 
in their hadronic constituents with partonic substructure, then proceeds through
the parallely evolving stages of parton-cascade, parton-hadron conversion,
fragmentation of beam/target remnants, final-state hadron-cascade,
 and finishes up with
the final particle yield that reflects observables in a detector.
\label{fig:fig1}
}
\end{figure}

The main strength of VNI lies in addressing
the physics of high-density QCD, which  becomes an increasingly popular
object of research, both from the experimental, phenomenological
interest, and from the theoretical, fundamental point of view.
Presently, and in the near future, the collider facilities HERA
($ep$, $eA$?), Tevatron ($p\bar{p}$, $pA$), RHIC and LHC ($p\bar{p}$,
$AA$) are able to probe new regimes of dense quark-gluon matter 
at very small Bjorken-$x$ or/and at large $A$, 
with rather different dynamical properties.
The common feature of high-density
QCD matter that can be produced in these experiments, is an expected
novel exhibition
of the interplay between the high-momentum (short-distance) perturbative
regime and the low-momentum (long-wavelength) non-perturbative physics.
For example, with HERA and Tevatron experiments,
one hopes to gain insight into problems concerning
the saturation of the strong rise
of the proton structure functions at small Bjorken-$x$,
possibly due to color-screening effects that
are associated with the overlappping of a large number of small-$x$
partons.
Another example is the anticipated formation of a quark-gluon plasma in
RHIC and LHC heavy ion collisions, 
where multiple parton rescattering and cascading generates a 
high-density environment, in which the collective 
motion of the quanta imust be taken into consideration.
In this context the most advantageous and novel feature
of VNI
is the space-time cascade picture that provides a potentially powerful
tool to study high-density particle systems in QCD, where 
accounting for the dynamically
evolving space-time structure of the collisions is most important.
The {\it necessity of including space-time variables} in addition to
energy-momentum variables for high-density systems may
be common knowledge in the field of heavy-ion physics, where
the space-time aspect  of many-body transport theory is essential
ingredient to describe nucleus-nucleus collisions, but it is
new to many high-energy particle physicists which only recently
began to acknowleged the necessity to include time and space on top
of the commonly used, pure momentum space description of, e.g.,  lepton and
hadron collisions.
\bigskip

\noindent
{\it
Where does VNI fit into the diverse family of modern
event generators for physics simulation of particle collisions
(where `particle' stands for any beam/target particle from an electron 
to a heavy nucleus)?
}
\smallskip

\noindent
The Monte Carlo models for high-energy particle collisions that are on the 
market, may be crudely divided into two classes:
\begin{description}
\item{(i)}
The first class embodies event generators that restrict to
"clean" reactions  involving lepton or proton beams only,
and which aim at a high-precision description of experimental
tests of QCD's first principles.
Popular examples  are PYTHIA \cite{jetset},
HERWIG \cite {herwig}, ARIADNE \cite{ariadne}, LEPTO \cite{lepto},
and ISAJET \cite{isajet}.
The common feature of these event generators is a combination of 
well understood perturbative QCD parton-shower description and
a non-perturbative hadronization prescription to convert
the partonic final state into hadrons.
\item{(ii)}
The second class are event generators that aim to describe
"dirty" reactions involving nuclei, much less based on
first-principle knowledge, but instead rely on phenomenological
models to mimic the unknown details of the underlying physics.
Here the widely used concept is to visualize a nuclear
collision in terms of nucleon-nucleon collisions on the
basis of a constituent valence-quark picture plus
string-excitation and -fragmentation.
Examples for these models are
FRITIOF \cite{fritiof},  DPM \cite{dpm}, 
VENUS \cite{venus}, 
RQMD \cite{rqmd}, and HIJET \cite{hijet}.
Distinct from these is  HIJING \cite{hijing}, 
which also incorporates a perturbative QCD approach to multiple minijet
production, however it does not incorporate a space-time description.
\end{description}
With the exception of HERWIG, all of the
above Monte Carlo generators utilize some form of 
{\it string fragmentation phenomenology} to model 
the non-perturbative hadronization
and final-state particle production.
Most commonly used is the Lund string model \cite{string}.
HERWIG on the other hand is built on a very different 
{\it parton-cluster formation/fragmentation approach}, 
which forms the basis of the hadronization scheme developed in VNI.
\medskip

\noindent
{\it
What are the shortcomings of the above-mentioned Monte Carlo models
with respect to the particle dynamics in finite-density regions
created by high-energy collisions?
}
\smallskip

The high-energy particle physics generators of the first class
lack the inclusion space-time variables in the dynamics
description in both the
perturbative QCD parton evolution and the non-perturbative
process of parton-hadron transition.
These models therefore cannot account  for statistical
interactions due to the presence of a finite density of particles
close-by in space, such as rescattering, absorption of recombination
processes.
Hence, although these models to large extent use QCD's fundamental
quark-gluon  degrees of freedom, important aspects of
parton dynamics and interactions at finite density are left out,
because the particles are assumed to propagate unscathed in free space.

The event generators of the second class, on the other hand, 
mostly do utilize a space-time description, however on the level
of hadronic degrees of freedom (strings, baryons, mesons and resonances)
rather than partonic degrees of freedom.
For ultra-relativistic nucleus-nucleus collisions
the parton approach appears to be more realistic than the 
hadronic or string picture, as it
has been realized that short-range parton interactions play 
a  major role for heavy ion collisions at collider energies of
$\sqrt{s}  \, \lower3pt\hbox{$\buildrel >\over\sim$}\,100$ GeV,
at least during the early and most dissipative stage of the first few $fm$. 
Here copiously produced quark-gluon mini-jets cannot be considered as 
isolated rare events, but must be embedded in complicated multiple 
cascade-type processes.
Thus, the short range character of these  
interactions implies that perturbative QCD can and must be used, and 
that the picture of comparably large distance excitations
of strings or hadronic resonances does not apply in this kinematic regime.
\medskip

In view of this discussion,
the program VNI can be viewed in between the above two classes of
Monte Carlo models:
It provides a kinetic space-time description of parton evolution
by utilizing well-developed techniques for perturbative QCD simulations
at zero density or free space,
as the event generators of the first class. On the other hand,
it applies this concept also to the physics of
finite-density particle systems, e.g., in
collisions involving nuclei, as the event generators of the second class.
  Comparing VNI with the above-mentioned Monte Carlo models, 
the essential differences and partly new aspects are the following:
\begin{description}
\item{a)}
     the aspects of the space-time evolution of the particle distributions
     in addition to the evolution in momentum space \cite{msrep}, 
     and the concepts of quantum kinetic theory and statistical physics
     \cite{ms39,ms42}.
\item{b)}
     the self-consistent interplay of coherent (angular ordered) 
     perturbative 
     parton evolution according to the DGLAP \cite{DGLAP} equations, with the 
     fully dynamical cluster-hadronization according to the 
     phenomenological model of Ref. \cite{ms37}.
\item{c)}
     the microscopic tracing of color degrees of freedom and the effect of
     color-correlations by using explicit color-labels for each parton, 
     which allows to investigate final state interactions in the process 
     of hadron formation \cite{ms40}.
\item{d)}
     the diverse advantages of a stochastic simulation technique with
     which the various particle interaction processes are determined by 
     the dynamics itself, through the local density of particles as they 
     evolve causally in time.
\item{e)}
     the statistical many-particle description for general non-equilibrium
     systems, which allows to study thermodynamic behaviour of the bulk 
     matter \cite{ms2}, such as the evolution of macroscopic energy density, 
     pressure, etc., or the dynamical development of the sytem to 
     thermal/chemical equilibrium in heavy ion reactions.
\end{description}

In summary, the improvement to be expected from VNI
for the physics simulation of high-energy particle collisions 
lies clearly in the `dirty' high-density parton regime, 
where the space-time aspects are most important, and which
currently and in the future is
of central interest
in  experiments  at, e.g.,  HERA, RHIC and LHC.
On the other hand, VNI may also be perceived as a valuable alternative
to high-energy event generators for the
study of `clean', zero-density collisions as $e^+e^-$ annihilation
or $p\bar{p}$ collisions,
where the space-time aspects can provide useful additional insight in
the collision dynamics for experiments at, e.g., LEP or the Tevatron.
Recent applications to $e^+e^-$ and $ep$ collisions may be found in Refs. 
\cite{ms37,ms40,ms41}, and to heavy-ion collisions in Refs. \cite{dkskkg1,dkskkg2,longacre}.
\bigskip
\bigskip

\subsection{General concept}

The central element in the physics description implemented in VNI
is the use of
QCD transport theory \cite{msrep} and quantum field kinetics \cite{ms39}
to follow the evolution of a generally mixed multi-particle system
of partons and hadrons
in 7-dimensional phase-space $d^3 r d^3 k dE$.
Included are 
both the parton-cascade development
\cite{dok80,glr83,jetcalc,bassetto}
which embodies the renormalization-group improved evolution of 
multiple parton collisions
including inelastic (radiative) processes,
and the phenomenological
parton-hadron conversion model of Refs.  \cite{ms37,ms40,ms41},
in which the hadronization mechanism is described in terms of 
dynamical parton-cluster formation as
a local, statistical process that depends on the spatial separation and
color
of nearest-neighbor partons, followed by the decay of clusters into
hadrons.

In contrast to the commonly-used momentum-space description,
the microscopic history of the dynamically-evolving particle system
is traced in space-time {\it and} momentum space, so that
the correlations of partons in space,  time, and color can be taken
into account for both the  cascade evolution
and the hadronization mechanism.
It is to be emphasized,
that the interplay
between perturbative and non-perturbative regimes is controlled locally
by the space-time evolution of the mixed parton-hadron system itself
(i.e., the time-dependent local parton density),
rather than by an arbitrary global division
between parton and hadron degrees of freedom 
(i.e., a parametric energy/momentum cut-off).
In particular the parallel evolution of the mixed system
of partons, pre-hadronic clusters, and hadrons, with the relative proportions
determined by the dynamics itself, is a novel feature that
is only possible by keeping track of both space-time and 
energy-momentum variables.

Probably the greatest strength of this approach lies in its
application to the collision dynamics of
complicated multi-particle systems, as for example in  collisions involving
nuclei ($eA$, $pA$ and $AB$), 
for which a causal time evolution in position space
and momentum space is essential: Here statistical,
non-deterministic particle interactions are most important,
which  can only be accounted for by
following the time-evolution of the particle densities in space {\it and}
energy-momentum.
This approach allows to 
study the time evolution of an initially prepared beam/target collision system
in complete phase-space from the instant 
of collisional  contact, through the QCD-evolution of
parton distributions, up to the formation of final hadronic states.
It provides a self-consistent scheme to solve the underlying
equations of motion for the particle densities as determined by the 
microscopic dynamics.
\smallskip

The model as a whole consists of
four major building-blocks, which are illustrated schematically
in Fig. 2:
\begin{description}
\item[1.]
The {\it initial state} associated with the incoming
nuclei involves their decomposition into nucleons and of the nucleons into
partons on the basis of the experimentally measured  
nucleon structure functions and elastic form-factors.
This procedure translates the initial nucleus-nucleus system into
two colliding clouds of {\it virtual} partons according
to the well-established parton decpomposition of
the nuclear wavefunctions at high energy \cite{glr83}.
\item[2.]
The {\it parton cascade development}
starts from the initial interpenetrating parton clouds,
and involving  the space-time
development with mutual- and self-interactions of the 
system of quarks and gluons.
Included are multiple elastic and inelastic interaction processes, described 
as sequences of elementary $2 \rightarrow 2$ scatterings, $1\rightarrow 2$
emissions and $2 \rightarrow 1$ fusions.
Moreover,  correlations are accounted for between primary virtual
partons, emerging as unscathed remainders from the initial state, and
secondary real partons, materialized or  produced 
through the partonic interactions.
\item[3.]
The {\it hadronization dynamics} of the evolving system
in terms of parton-coalescence to color-neutral clusters
is decribed as a local, statistical process that depends on the spatial separation
and color of nearest-neighbor partons.
Each pre-hadronic parton-cluster fragments through isotropic two-body decay
into  primary hadrons, according to the density of
states, followed by the decay of the latter into final 
stable hadrons.
\item[4.]
The {\it `afterburner' hadron cascade} describes the evolution
of produced pre-hadronic clusters
and hadrons, emerging both from the hadronization
of cascading partons, as well as from primary remnant partons which
represent the fraction of unscathed initial state nucleons.
Pre-hadronic clusters can mutually rescatter, or scatter off close-by
hadrons, before they decay into stable final-state hadrons.
Similarly, already formed hadrons may deflected by
elestic collisions with other hadrons and clusters, or may be excited by inelastic
collisions and resonance formation/decay.
\end{description}

Such a pragmatical division, which assumes complex interference
between the different physics regimes to be negligible, is possible if
the respective dynamical scales are such that
the short-range (semi)hard parton interactions (scattering, radiation,
fusion) of perturbative nature,
and the non-perturbative
mechanism of hadron formation (parton-coalescence and cluster-decay),
occur on well-separated  space-time scales (or momentum scales, by
virtue of the uncertainty principle).
Loosely speaking, the typical momentum scale associated with
parton collisions, radiative emissions, or parton fusion, has to be 
larger than
the inverse `confinement length scale' $\sim1\,fm$ which
separates perturbative and non-perturbative domains.
Further discussion of this condition of validity can be found in, e.g.,
\cite{msrep,dokbook}.

\begin{figure}
\epsfxsize=500pt
\centerline{ \epsfbox{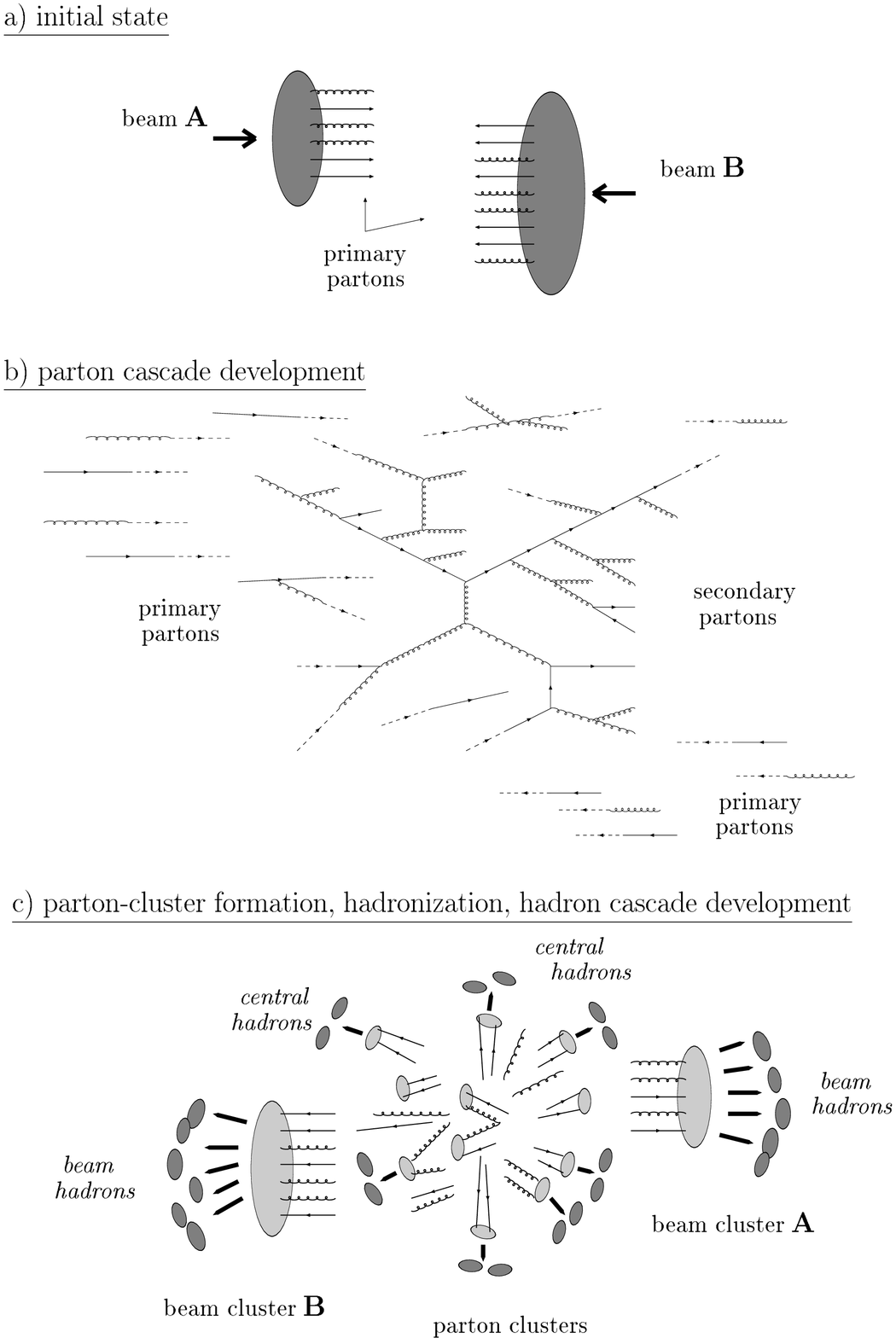} }
\caption{
The four components of the model:
a) the initial state constructed in terms of the parton distribution of
the incoming nuclei;
b) the time-evolution of parton cascades in 7-dimensional phase-space 
c) the formation of color neutral clusters from secondary partons
emerging from cascading, as well of the remnant primary partons from
the initial state, followed by
the fragmentation of the clusters into final hadrons.
d) the final-state hadron-cascade evolution of produced clusters and hadrons.
\label{fig:fig2}
}
\end{figure}
\bigskip

\subsection{Equations of motion from quantum kinetics for multi-particle
dynamics}

A firm theoretical basis for
the above space-time cascade description of a multiparticle system
in high-energy collisions can be derived systematically from
{\it quantum-kinetic theory} on the basis of
QCD's first principles in a stepwise approximation scheme 
(see  e.g., Refs. \cite{ms39,ms42} and references therein).
This framework allows  to
cast the time evolution of the mixed system of
individual partons, composite parton-clusters, and physical hadrons
in terms of a closed set of
integro-differential equations for
the phase-space densities of the different particle excitations.
The definition of these phase-space densities,
denoted by
$F_\alpha$, where $\alpha\equiv p, c, h$
labels the species of partons, pre-hadronic clusters, or hadrons,
respectively, is:
\begin{equation}
F_\alpha(r,k)\;\,\equiv\; \, F_\alpha (t, \vec r; E, \vec k)
\;\,=\;\,
\frac{dN_\alpha}{d^3r d^3k dE}
\;,
\label{F}
\end{equation}
where 
$r\equiv r^\mu = (t,\vec{r})$, $k\equiv k^\mu = (E,\vec{k})$,
and $k^2 = E^2 -\vec{k}^{\,2}$ can be off-shell 
(space-like $k^2 < m^2$, time-like $k^2 > m^2$) or on-shell ($k^2 = m^2$).
The densities (\ref{F}) measure the number of particles
of type $\alpha$ at time $t$ with position in $\vec r + d\vec{r}$,
momentum in $\vec k + d\vec{k}$,
and energy in $E + dE$ (or equivalently invariant mass in $k^2 + dk^2$).
The $F_\alpha$ are the quantum analogues of the
classical phase-space distributions, including both off-shell and on-shell
particles, and hence
contain the essential microscopic
information required for a statistical description
of the time evolution of a many-particle system in
complete 7-dimensional phase-space $d^3rd^3kdE$, 
thereby providing the basis for calculating
macroscopic observables.

The phase-space densities (\ref{F}) are determined by the
self-consistent solutions of
a set of {\it transport equations} (in space-time) coupled with
renormalization-group type {\it evolution equations} (in momentum space).
Referring  to Refs. \cite{ms37,ms39} for details,
these equations can be generically expressed as
convolutions of the densities $F_\alpha$ of particle species $\alpha$,
interacting with  specific cross sections $\hat{I}_j$ for the processes $j$.
The resulting coupled equations for the
space-time development of
the densities of partons $F_{p}$, clusters $F_c$ and
hadrons $F_h$ is a self-consistent set in which the change
of the densities $F_\alpha$ is governed by the balance of
the various possible interaction processes among the particles.
Fig. 3 represents these equations pictorially.
For the densities of {\it partons}, the {\it transport equation} 
(governing the space-time change with $r^\mu$) and the {\it evolution equation} 
(controlling the change with momentum scale $k^\mu$), read, respectively,
\begin{eqnarray}
k_\mu \frac{\partial}{\partial r^\mu}\; F_p(r,k)
&=&
F_{p''} F_{p'''}\circ 
\left[\frac{}{}
\hat{I}(p''p'''\rightarrow p p') \;+\;\hat{I}(p''p'''\rightarrow p)
\right]
\;-\; 
F_{p} F_{p'}\circ 
\left[\frac{}{}
\hat{I}(pp'\rightarrow p'' p''')\;+\; \hat{I}(pp'\rightarrow p'')
\right]
\nonumber \\
& &
-\;\;
F_{p} F_{p'}\circ \left[\frac{}{} 
\hat{I}(p'p''\rightarrow p)\;-\; \hat{I}(pp'\rightarrow p'')\right]
\;-\;
F_p\,F_{p'}\circ \hat{I}(p p'\rightarrow c)
\label{e1}
\\
k^2  \frac{\partial}{\partial k^2}\; F_p(r,k)
&=&
F_{p'}\circ \hat{I}(p'\rightarrow p p'')\;-\;
F_{p}\circ \hat{I}(p\rightarrow p' p'')
\label{e2}
\;.
\end{eqnarray}

\begin{figure}
\vspace{-2.5cm}
\epsfxsize=650pt
\centerline{ \epsfbox{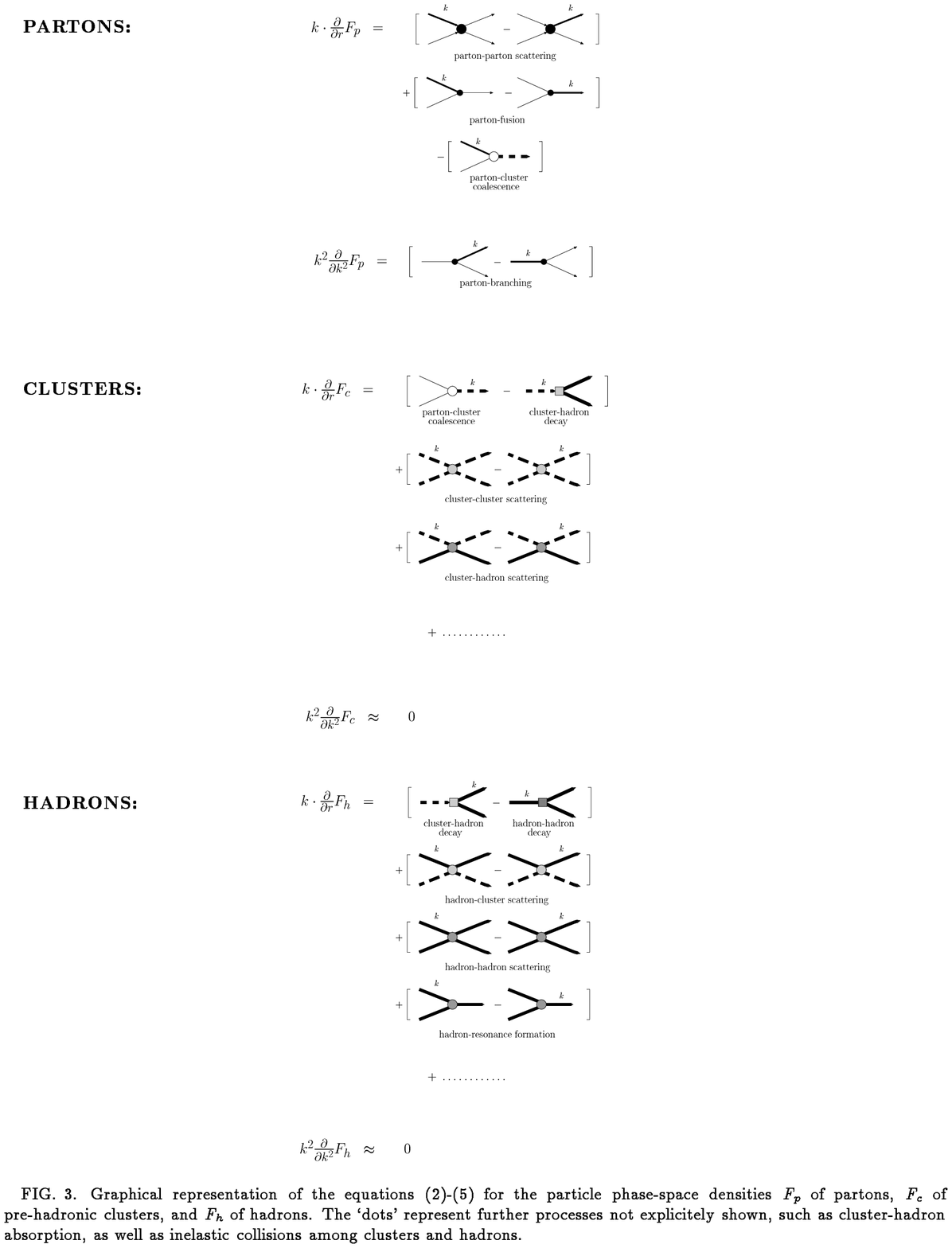} }
\vspace{-2.0cm}
\end{figure}

For the densities of {\it pre-hadronic clusters}  and
{\it hadrons}, the evolution equations  are homogeneous
to good approximation,
so that one is left with non-trivial transport equations only,
\footnote{
It is worth noting that eq. (\ref{e2})  embodies the momentum space
($k^2$) evolution of partons through
the renormalization of the phase-space densities $F_p$, determined
by their change $k^2 \partial F_p(r,k)/\partial k^2$
with respect to a variation of the mass (virtuality) scale $k^2$
in the usual QCD evolution framework \cite{dok80,jetcalc,bassetto}.
On the other hand,
for pre-hadronic clusters and hadrons, renormalization effects
are comparatively small, so that their
mass fluctuations $\Delta k^2/k^2$ can be ignored to first
approximation,
implying $k^2 \partial F_c(r,k) /\partial k^2
= k^2 \partial F_h(r,k) / \partial k^2  =0 $.
}
\footnote{
It is worth noting that eq. (\ref{e2})  embodies the momentum space
($k^2$) evolution of partons through
the renormalization of the phase-space densities $F_p$, determined
by their change $k^2 \partial F_p(r,k)/\partial k^2$
with respect to a variation of the mass (virtuality) scale $k^2$
in the usual QCD evolution framework \cite{dok80,jetcalc,bassetto}.
approximation,
implying $k^2 \partial F_c(r,k) /\partial k^2
= k^2 \partial F_h(r,k) / \partial k^2  =0 $.
}.
Hence, for the evolution of the cluster densities $F_c$, one has
\begin{eqnarray}
k_\mu \frac{\partial}{\partial r^\mu}\; F_c(r,k)
&=&
F_p\,F_{p'}\circ \hat{I}(p p'\rightarrow c)
\;-\;
F_c\circ \hat{I}(c\rightarrow h)
\;+\;
F_{c''} F_{c'''}\circ \hat{I}(c''c'''\rightarrow c c') \;-\; 
F_{c} F_{c'}\circ \hat{I}(cc'\rightarrow c'' c''')
\label{e3}
\\
& &
\;+\; 
F_{c'} F_{h'}\circ \hat{I}(c'h'\rightarrow c h) \;-\; 
F_{c} F_{h}\circ \hat{I}(ch\rightarrow c' h')
\;\;+\;\;\ldots\ldots
\nonumber
\\
k^2  \frac{\partial}{\partial k^2}\; F_c(r,k) &=& 0  
\;,
\end{eqnarray}
and similarly, for the evolution of the hadron densities $F_h$, the equations read
(Fig. 3)
\begin{eqnarray}
k_\mu \frac{\partial}{\partial r^\mu}\; F_h(r,k)
&=&
F_c \circ\hat{I}(c\rightarrow h)
\;+\;
\left[\frac{}{}
F_{h'}\circ \hat{I}(h'\rightarrow h)
\;-\;
F_h\circ \hat{I}(h\rightarrow h')
\right]
\;+\;
F_{h''} F_{h'''}\circ \hat{I}(h''h'''\rightarrow h h')
\label{e4} \\
& &
 \;-\; 
F_{h} F_{h'}\circ \hat{I}(hh'\rightarrow h'' h''')
\;+\; 
F_{c'} F_{h'}\circ \hat{I}(c'h'\rightarrow c h) \;-\; 
F_{c} F_{h}\circ \hat{I}(ch\rightarrow c' h')
\;\;+\;\;\ldots\ldots
\nonumber
\\
k^2  \frac{\partial}{\partial k^2}\; F_h(r,k) &=& 0
\;.
\end{eqnarray}
In (\ref{e1})-(\ref{e4}),
each convolution $F \circ\hat{I}$ of
the density of particles $F$ entering a particular vertex
${\hat I}$ includes a sum over contributing
subprocesses, and a phase-space integration
weighted with the associated subprocess probability distribution
of the squared amplitude. Explicit expressions are given in
Refs.  \cite{msrep,ms37}.
\bigskip
\bigskip
In (\ref{e1})-(\ref{e4}),
each convolution $F \circ\hat{I}$ of
the density of particles $F$ entering a particular vertex
${\hat I}$ includes a sum over contributing
subprocesses, and a phase-space integration
weighted with the associated subprocess probability distribution
of the squared amplitude. Explicit expressions are given in
Refs.  \cite{msrep,ms37}.
\smallskip

As mentioned before, the
equations (\ref{e1})-(\ref{e4}) reflect a {\it probabilistic
interpretation} of the multi-particle evolution in 
space-time and momentum space
in terms of sequentially-ordered interaction processes $j$,
in which the rate of change of the particle distributions $F_\alpha$
($\alpha=p,c,h$)
in a phase-space element $d^3rd^4k$
is governed by the balance of gain (+) and loss ($-$) terms.
The left-hand side
describes free propagation of a
quantum of species $\alpha$, whereas
on the right-hand side the interaction kernels $\hat{I}$
are integral operators that incorporate the effects of
the particles' self-  and mutual interactions.
This probabilistic  character 
is essentially an effect of time dilation, because in any frame
where the particles move close to the speed of light, the associated
wave-packets are highly localized to short space-time extent, so that
comparatively
long-distance quantum interference effects are generally very small.
\smallskip

The terms on the right-hand side
of the transport-  and evolution-equations (\ref{e1})-(\ref{e4})
correspond to one of the following categories (c.f. Fig. 3):
\begin{itemize}
\item
parton scattering and parton fusion  through 2-body collisions,
\item
parton multiplication through radiative emission processes
on the perturbative level,
\item
colorless cluster formation through parton coalescence
depending on the local color and spatial configuration,
\item
hadron formation by decay of the cluster excitations into final-state hadrons.
\item
scattering of pre-hadronic clusters or already formed hadrons
 with other clusters or  hadrons.
Note: this includes also (as indicated by the `dots') 
cluster-cluster fusion, cluster-hadron absorption,
as well as inelastic collisions among clusters and hadrons, in which
energy-momentum is transferred into excitation.
\end{itemize}
\bigskip
\bigskip

The equations (\ref{e1})-(\ref{e4}) reflect a {\it probabilistic
interpretation} of the multi-particle evolution in 
space-time and momentum space
in terms of sequentially-ordered interaction processes $j$,
in which the rate of change of the particle distributions $F_\alpha$
($\alpha=p,c,h$)
in a phase-space element $d^3rd^4k$
is governed by the balance of gain (+) and loss ($-$) terms.
The left-hand side
describes free propagation of a
quantum of species $\alpha$, whereas
on the right-hand side the interaction kernels $\hat{I}$
are integral operators that incorporate the effects of
the particles' self-  and mutual interactions.
This probabilistic  character 
is essentially an effect of time dilation, because in any frame
where the particles move close to the speed of light, the associated
wave-packets are highly localized to short space-time extent, so that
comparatively
long-distance quantum interference effects are generally small for
these short-distance interactions (however it does not mean they are
unimportant for global particle dynamics such as collective behavior).
\bigskip

It is worth emphasizing that the model goes {\it beyond a purely
classical cascade}! It incorporates  a number of
quantum mechanical features in the particle dynamics and
production, for example: 
\begin{itemize}
\item
the inclusion of off-shell particles, which treats
space-like or time-like particles as energy fluctuations
with finite life-time given by the dedgee of virtuality
$\propto 1/\sqrt{|k^2|}$;
\item
the feature of coherence in parton scatterings which
relates momentum transfer and impact parameter of
2-body scatterings and restricts the possible momentum
exchange to be consistent with the uncertainty relation;
\item
the concept of (angular ordered)
coherent parton evolution including important gluon interference
effects is used, which is well known to be an important
non-classical property of soft (small-$x$) partons,
\item
particle scatterings and decays (both partonic and hadronic)
are treated consistent with the uncertainty principle,
by requiring a scattering/formation time before particles
appear as incoherent quanta in the system;
\item
particles are produced {\it not} at a sharply localized point
in space and momentum space (as it would be classically, 
but impossible quantum mechanically); rather their position
and momentum at the production vertex is smeared with
$\exp[ -r^2/\sigma^2 - \sigma^2 p^2]$ (with $\sigma = 0.5 fm$
as inferred from Bose-Einstein data from $e^+e^-$
collisions), a prescription which imposes again
the uncertainty principle and guarantees $\Delta r \Delta p \gg \hbar$.
\end{itemize}
Therefore the particle's densities $F_a(r,k)$ are not
just the classical phase-space densities
(although they are just the number of countable particles per unit phase-space),
but incorporate
all the above {\it quantum mechanical features} implicitely.
The space-time and energy integrated $F_a$ are in fact 
observable quantities because after all
what is seen in a detector are a particle distributions.

\bigskip

\subsection{Scheme of solution in global Lorentz-frame of reference}

In the above kinetic approach to the multi-particle
dynamics,
the probabilistic character of the transport- and evolution equations
(\ref{e1})-(\ref{e4})
allows one to solve for the phase-space densities $F_\alpha(r,k)$ by
simulating
the dynamical development as a Markovian process causally in time.
Because it is an `initial-value problem', one must specify
some physically appropriate initial
condition $F_\alpha(t_0,\vec{r},k)$ at starting time $t_0$, such that
all the dynamics prior to this point is effectively embodied in this
initial form of $F_\alpha$.
The set of equations (\ref{e1})-(\ref{e4}) can
then be solved
in terms of the evolution of the phase-space densities $F_\alpha$
for $t > t_0$ using Monte Carlo methods to
simulate the time development of the mixed system
of partons, clusters, and hadrons
in position and momentum space \cite{ms37,msrep}.

With the initial state specified as discussed below,
the phase-space distribution of particles at $t=t_0 \equiv 0$ can be
constructed and then evolved in small time steps 
$\Delta t =O(10^{-3}\;fm)$ forward throughout the parallely evolving
stages of parton cascade, parton-cluster formation, and cluster-hadron decays,
until stable final-state hadrons and other particles (photons, leptons, etc.) 
are left as freely-streaming particles.
The partons propagate along classical trajectories until they interact,
i.e., collide (scattering or fusion process),
decay (emission process) or recombine to pre-hadronic composite
states (cluster formation).
Similarly, the so-formed pre-hadronic parton-clusters 
travel along classical paths until they convert into
primary hadrons (cluster decay), followed by the hadronic decays
into stable final state particles.
The corresponding probabilities and time scales of interactions are
sampled stochastically from the relevant probability distributions
in the kernels $\hat{I}$ of eq. (\ref{e1})-(\ref{e4}).

It is important to realize, that 
the spatial density and the momentum distribution of the 
particles are intimately connected: The momentum 
distribution continously changes through the interactions and
determines how the quanta propagate in coordinate space.
In turn, the probability for subsequent interactions depends on the 
resulting local particle density. Consequently, the development
of the phase-space densities is a complex
interplay, which - at a given point of time - contains implicitely the
complete preceeding history of the system.
\medskip

It is clear that
the description of particle evolution is
Lorentz-frame dependent, and a suitable reference frame
(henceforth called {\it global frame})
must be chosen (not necessarily the laboratory frame of an experiment).
When computing Lorentz-invariant quantities, such as
cross sections or final-state hadron spectra, the particular choice is
irrelevant, whereas for non-invariant observables, such as energy
distibutions
or space-time-dependent quantities, one must at the end transform
from the arbitrarily-chosen frame of theoretical description to the
actual frame of measurement.

For calculational convenience, it is most suitable
to choose the {\it global center-of-mass ($cm$) frame of the colliding 
beam particles}, with the collision axis in the $z$-direction.
In this global $cm$-frame, the incoming particles $A$, $B$ 
(= lepton, hadron, or nucleus) have four-momenta,
\begin{eqnarray}
& &
\;\;\;\;\;\;\;\;\;\;\;
P^\mu_{A,B} \; = \;\left( E_{A,B}, 0,0,\pm P_{cm}\right)
\nonumber
\\
E_{A,B} &=& \frac{1}{2 \sqrt{s}}\;\left(s \,+\,M_A^2\,\pm\,M_B^2\right)
\;\;\;\;\;\;\;\;\;\;\;\;\;\;\;
s\;=\; E_{cm}^2 = \left(P_A^\mu+P_B^\mu\right)^2
\label{cm}
\\
P_{cm}&=& \frac{1}{2\sqrt{s}}\;
\sqrt{\left[s-(M_A+M_B)^2\right]\; \left[s-(M_A- M_B)^2\right]}
\; ,
\nonumber
\end{eqnarray}
where $M_{A,B}$ are the masses of $A$ and $B$,
and so  incoming particles then carry well-defined fractions of $P_{cm}$,
having only a non-vanishing longitudinal momentum along 
the $z$-axis.
Particularly in the case of nuclei $A$, the daughter nucleons $N_i=1,\ldots A$, 
have momenta, $\vec P_{N_i}= (0,0,\pm P_{cm}/A)$.
\smallskip

In the following, the global $cm$-frame of $A$ and $B$ is assumed to be the
reference frame, with the initial energy-momentum of the collision system
given by (\ref{cm}).
Furtermore the terms `hadron', respectively `nucleon', are used to
distinguish initial states $A+B$ with $A$ and/or $B$ being 
a single hadron, respectively a nucleus with $A$ ($B$) nucleons.

\bigskip

\subsection{Initial state}

If one or both beam particles $A$, $B$ are leptons,
they are considered as  point-like objects
which carry the full beam energy, meaning
that  any QED or QCD substructure of the leptons,
as well as initial-state photon radiation by the leptons, is neglected.
For {\it lepton-lepton annihilation},
it is assumed that the colliding leptons produce a 
time-like $\gamma$ or $Z^0$ boson of invariant mass
$Q^2\equiv +q^2$ at time $t = -Q^{-1}$, so that $t=0$ characterizes the point
when the $\gamma$ ($Z^0$)  decays into a quark-antiquark pair.
Similarly, for {\it lepton-hadron (nucleus) collisions}, the lepton is
emitting a space-like virtual $\gamma$
of invariant mass $Q^2\equiv -q^2$ at time $t = -Q^{-1}$, and hence $t=0$ 
is the point when the $\gamma$ hits the hadron (nucleus).

In the general case of {\it collisions involving hadrons and/or nuclei},
the incoming particles $A$ and $B$ are
decomposed into their parton substructure by
phenomenological construction of the
momentum and spatial distributions of their daughter partons
on the basis of the known hadron (nucleon) structure functions
and elastic hadron (nucleon) form-factors.
In the $cm$-frame, where the two incoming particles
$A, B$ (= hadron, nucleus), are moving close to the speed of light, 
the parton picture is applicable and the parton substructure 
of the hadrons or nucleons
can be resolved with reference to some {\it initial resolution scale} $Q_0$. 
This resolution scale 
generally varies with beam-energy and mass number $A$, $B$, in that it depends
on the typical momentum and spatial density of partons as well as on
their interaction probability.
To be specific, it may be identified with
the statistically estimated expectation value for the 
interaction scale $Q^2$ of all {\it primary} parton-parton collisions 
(that are those in which at least one initial state parton is involved)
\cite{miklos}:
\begin{equation}
Q_0^2 \;\,\equiv\;\,
Q_0^2 (x,P,A) \,=\, A^\alpha \;\left(\frac{1}{\langle p_\perp^2 \rangle_{prim}}\,+\,
\frac{\langle p_\perp^2 \rangle_{prim}\,R_0^2}{2 x P}
\right)^{-1}
\;,
\label{Q0}
\end{equation}
where
$\langle p_\perp^2 \rangle_{prim}$
is the average relative transverse momentum squared generated
by primary parton-parton collisions.
and $R_0 = 1 GeV^{-1}$.
The pre-factor $A^\alpha$ ($0\le\alpha\le 4/3$ a parameter) 
accounts for the nuclear dependence for $A,B > 1$,
and 
$x$ is the parton's momentum fraction of the mother hadron or nucleon
which carries momentum $P$ ($= P_{cm}$ for single hadron, or $=P_{cm}/A$
for a nucleon in a nucleus $A$).
The scale $Q_0$ defines the initial point in momentum space above which the 
system of partons is treated as an ensemble
of incoherent quanta. The  dynamics prior to this point
is contained in the initial parton-structure function of the mother hadron
(nucleon).
Clearly, the convention (\ref{Q0}) yields only an average value 
dominated by the most probable (semi)hard parton collisions with
relatively small momentum transfers of a few GeV. 
However, primary parton collisions with a momentum scale
$Q^2\gg Q_0^2$, which correspond to relatively rare fluctuations, 
are taken into account individually
by the $Q^2$-evolution of the hadron (nucleon) structure functions through 
space-like and time-like radiation processes, discussed later.

The actual form of the  phase-space distribution of partons,
eq. (\ref{F}) is initially to be specified at the reference point $Q_0$.
It is constructed as the following superposition of the 
parton distributions in the individual 
hadrons at $Q_0$ and at time $t=t_0\equiv 0$, the point of collision of $A$ and $B$:
\begin{equation}
\left. F_a (r,k)\right|_{r^0  = t_0}\; = \; \sum_{i=1}^{N_{h}} 
P_a^{N_i} ( k, \vec P; Q_0^2) \cdot R_a^{N_i} ( k, \vec r,\vec R)
\left.\frac{}{}\right|_{r^0=t_0} 
\;\;\;.
\label{Fa0}
\end{equation}
The right hand side is a sum over all $1 \le N_i\le N_{h}$ 
hadrons or nucleons contained in the collision system $A+B$.
Each term in the sum is a  parton phase-space density given by a
convolution of an initial momentum distribution $P_a^{N_i}$ and a spatial
distribution $R_a^{N_i}$. 
The subscript $a = g, q_j, \bar q_j$ ($j= 1,\ldots , n_f)$
labels the species of partons, and $N_i$ refers to the type of the $i$-th
hadron (for nuclei this is just proton or neutron). The four-vectors
$k\equiv k^\mu=(E,\vec k)$ and $r \equiv r^\mu = (t, \vec r)$
refer to the partons. The  3-momenta and 
initial positions of the hadrons or nucleons are
denoted $\vec P$, respectively $\vec R$. All vectors are understood
with respect to the global $cm$-frame, at time $t = t_0 =0$.
The partons' energies
$E \equiv E_a(\vec{k}^{\,2},q^2) = \sqrt{\vec{k}^{\,2} + m_a^2 + q^2}$
take into account their initial space-like virtualities $q^2<0$
which are distributed around $\langle \vert q^2 \vert \rangle = -Q_0^2/4$,
under the constraint that for each hadron in the initial state,
the total invariant mass of the daughter partons must equal the mother 
hadron mass.  This mimics the fact that 
the initial partons are confined inside their parent hadrons or nucleons 
and cannot be treated as free particles, i.e. they have not enough energy to
be on mass shell, but  are off mass shell by a 
space-like virtuality $q^2$ (eq. (\ref{invmass}) below).
\smallskip

As suggestively illustrated in Fig. 2 and discussed in more detail now,
the initial state $A+B$ involving hadrons and/or nuclei,
appears in the global $cm$-frame as two approaching `tidal waves' of
large-$x$ partons (mainly valence quarks), where each
`tidal wave' has an extended tail
of low$-x$ partons (mostly gluons and sea-quarks).
\medskip

\subsubsection{Initial momentum distribution}

\noindent
For each hadron or nucleon the number of partons, the distribution of 
the flavors, their momenta and associated initial space-like virtualities,
are obtained from the function $P_a^{N_i}$ in (\ref{Fa0}).
Denoting  $P\equiv P_{cm}/N_{h}$, with $N_h$ the total number
of hadrons (nucleons) in $A+B$, it is represented in  the form
\begin{equation}
P_a^{N_i} (k,\vec P; Q_0^2)\;=\; 
\left(\frac{x}{\tilde x}\right) \; F_a^{N_i} (x,Q_0^2) \;\, \rho_a^A(x)\;\,
g (\vec k_{\perp}) 
\;\,\delta\left(P_z \,-\,P\right) \; \delta^2\left(\vec P_\perp\right)
\label{PaN}
\end{equation}
with the momentum and energy fractions
$
x = k_z/P
$,
$
\tilde x= E/P= \sqrt{x^2 + (k_\perp ^2 + m_a^2 + q^2)/P^2)}
$
and the normalizations
\begin{eqnarray}
&&
\;\;\;\;\;\;\;\;\;\;
\sum_a \int_0^1 dx \; x F_a^{N_i} (x, Q_0^2) \;= \; 1
\;,\;\;\;\;\;\;\;\;\;\;\;\;\;\;\;\;
\int_0^\infty d^2 k_\perp \;g(\vec k_\perp)  \;=\; 1
\label{norm1}
\\ 
&&
\;\;\;\;\;\;\;\;\;\;
\sum_a \int \, \frac{dk^2\,d^3 k}{(2\pi)^3 2 E}
\; E \; P_a^{N_i} ( k^2,\vec k, \vec P; Q_0^2) \;=\;
n^{N_i} (Q_0^2,\vec P) 
\label{norm3}
\;\; .
\end{eqnarray}

\noindent
The physics behind the ansatz (\ref{PaN}) is the following:
\begin{description}
\item[(i)]
The functions $F_a^{N_i} (x,Q_0^2)$ are the scale-dependent measured
hadron (nucleon) structure functions with $x$ being the fraction
of the parent hadron's or nucleon's longitudinal momentum $P$ carried by the parton.
The transverse momentum distribution $g(k_\perp)$ is specified below.
The factor
$x / \tilde x$ in (\ref{norm1}) is included to form the
invariant momentum integral $\int d^3 k / [(2 \pi)^3 2 E]$ out of
the distribution $P_a^{N_i}$ \cite{feyfld}.
In (\ref{norm3}) the quantity $n^{N_i}$ is the total
number of partons in a given nucleon with momentum $P=|\vec P|$ at $Q_0^2$.
\item[(ii)]
The function $\rho_a^A(x)$ in (\ref{PaN}) takes into account
nuclear shadowing effects
affecting  mainly soft (small $x$) partons in a nucleus
\footnote{This feature is optional in the simulation, by default it is
switched off.}.
These shadowing effects are evident 
in the observations of the European Muon Collaboration \cite{emc} as
a depletion of the nuclear structure functions at small $x$ relative to
those of a free nucleon.
Several mechanisms have been proposed to explain this nuclear shadowing
effect on the basis of the parton model \cite{nucshad4,nucshad2,nucshad3}.
However, here instead the phenomenological approach of Wang and Gyulassy 
\cite{hijing2} is adopted
which is based on the following parametrization \cite{nucshad3}
for the $A$ dependence of the shadowing for both quarks and gluons:
\begin{eqnarray} 
\rho_a^A(x)  &\equiv&
\frac{ {\sl F}_a^{A, N_i} ( x, Q_0^2)}{A \, F_a^{N_i}(x, Q_0^2)}
\;\,=\;\,
1 \,+ \,1.19 \,\left(\frac{}{}\ln A\right)^{1/6} \, \left[
x^3 - 1.5 (x_0 + x_L) x^2 + 3 x_0 x_L x \right]
\\
& &
\;\;\;\;\;\;\;\;\;\;\;\;\;\;\;\;\;\;\;\;\;\;\;\;
\;\;\;\;\;\;\;
- \left[ \;
\beta_A (R_\perp) - 
\frac{1.08 \,( A^{1/3} -1)}{\ln(A + 1)} \, \sqrt{x}\, \right]
\; \exp(-x^2 / x_0^2)
\;\;\;.
\label{rhoA}
\nonumber
\end{eqnarray}
where ${\sl F}_a^{A, N}$ and $F_a^{N}$ represent the structure function of
a nucleus $A$, respectively those of a free nucleon, and
$x_0 = 0.1$, $x_L = 0.7$. The function
$
\beta_A (r) = 0.1 \,\left(\frac{}{}A^{1/3} -1\right) 
\,\frac{4}{3} \,\sqrt{1 - \frac{R_\perp^2}{R_A^2}}
$
takes into account the impact parameter dependence, with $R_\perp$ labeling
the transverse distance of a nucleon from 
its nucleus center and $R_A$ the radius of the nucleus.
\item[(iii)]
The distribution $g(k_\perp)$ specifies
the primordial transverse momenta of partons  according to
a normalized Gaussian 
\begin{equation}
g (\vec k_{\perp})\; = \; \frac{1}{2\pi k_0^2} \; 
\exp \left[-\frac{\vert \vec k_\perp\vert^2}{k_0^2}\right] 
\;,
\label{gpT}
\end{equation}
independent of the type of parton or
hadron (nucleon). 
It takes into account the uncertainty of momentum 
(Fermi motion) due to the fact that the initial
partons are  confined within the nucleons. 
This intrinsic 
$k_{\perp}$ can be observed in Drell-Yan experiments
where it is found that the distribution is
roughly independent of $s$ and $Q^2$ \cite{pTprim}.
As inferred from these analyses, the
width in (\ref{gpT}) is 
$k_0 = 0.42$ GeV, corresponding to a mean
$\langle k_\perp \rangle \approx 0.38$ GeV. 
\end{description}

The scheme (i)-(iii) to sample the flavor and momentum distribution is carried out 
independently for each nucleon subject to the requirement
\begin{equation}
\left( \sum_i E_i \right)^2 - \left( \sum_i k_{x_i}\right) - \left( 
\sum_i k_{y_i}\right)^2 - \left( \sum_i k_{z_i}\right)^2\; = \; 
M_h^2 
\label{invmass}
\end{equation}
where the summation runs over all partons belonging to the same nucleon as determined by
(\ref{norm1}) and (\ref{norm3}),
and $M_h$ is the mother hadron (nucleon) mass.  
With the partons' 4-momenta distributed as outlined above,
the constraint (\ref{invmass}) determines the distribution
in the variable
$q^2 = k^2 - m_a^2 = E^2 -\vec k^2 -m_a^2 < 0$, 
the partons' initial space-like virtualities.
The resulting distribution in $q^2$ 
is a strongly peaked  Gaussian  with a mean value of $\approx Q_0^2/4$.
\medskip

\subsubsection{Initial spatial distribution}

\noindent
The initial spatial distribution of the partons, $R_a^{N_i}$, 
appearing in eq. (\ref{Fa0}), 
depends on the magnitude of their momenta, the
positions of their parent hadrons (nucleons), 
as well as on the spatial substructure
of the latter. It is represented as
\begin{equation}
R_a^{N_i} (\vec p, \vec r, \vec R) \;=\; 
\delta^3\left(\vec R \,-\,\vec R_{AB}\right)\;\,
\left[\frac{}{} \, h_a^{N_i} (\vec r) \;
\,H_{N_i} (\vec R) \, \right]_{boosted} 
\;\;,
\label{RaN}
\end{equation}
where the momentum dependence is here  purely due to
boosting the distributions to the global $cm$-frame. 
The components of the ansatz (\ref{RaN}) have the following meaning:
\begin{description}
\item[(i)]
The incoming beam particles (a single hadron, or a nucleus) 
are assumed to have initial $cm$-positions 
$\vec R_{AB} = (\pm \Delta Z_{AB}/2, \vec B_{AB})$,
corresponding to a chosen impact parameter $\vec B_{AB}=(B_x,B_y)$ 
and a minimum longitudinal separation $\Delta Z_{AB}$.
In the case where $A$ and/or $B$ is not a single hadron, but 
a nucleus,  the individual nucleons are 
assigned positions around the centers of the parent nucleus in its rest-frame,
according to a Fermi distribution for nuclei with mass number $A \geq 12$ and
a Gaussian distribution for nuclei $A < 12$,
\begin{equation}
H_{N_i} (\vec R)\;=\; \left\{ \,
\begin{array}{lr}
\, \frac{1}{4 \pi}\, \left( 1 + \exp \left[ (R-c)/a \right] \right)^{-1}
& \, ( A \geq 12 ) \\
\, \frac{1}{4 \pi}\, \frac{2}{\sqrt{\pi} b} \; \exp \left[ - R^2 / b^2 
\right]
& \, ( A < 12 ) 
\end{array}
\right.
\;\;.
\end{equation}
The parameters are $c = r_0 \, A^{1/3}$, $r_0 = 1.14$ $fm$, $a = 0.545$ $fm$ and
$b = \sqrt{2/3}\, R_A^{ms}$, where $R_A^{ms}$ is the mean square radius
of the nucleus.
\item[(ii)]
Next, the partons are distributed around the centers of their
mother hadrons or nucleons, still in the rest frame of the $A$, respectively $B$,
with an exponential distribution
\begin{equation}
h_a^{N_i} (\vec r)\;=\; 
\, \frac{1}{4 \pi}\,\frac{\nu^3}{8 \pi} \, \exp \left[ - \nu r \right]
\;\;,
\end{equation}
where $\nu = 0.84$ GeV corresponds to the measured 
elastic formfactor of the mother hadron or  nucleon, with a mean square radius
of $R_h^{ms} \equiv \sqrt{12/\nu} = 0.81$ $fm$.
\item[(iii)]
Finally, as indicated by the subscript {\it boosted} in eq. (\ref{RaN}),
the positions of the hadrons or nucleons and their associated valence quarks
are boosted into the global $cm$-frame of the colliding beam particles $A$ and $B$.
The valence quarks then occupy the Lorentz contracted region
$(\Delta z)_v \approx 2 R_A \,M_h/P$, whereas
the sea quarks and gluons are smeared out in the longitudinal
direction by an amount $(\Delta z)_{g,s} \approx 1/k_z < 2 R_A$ ($R_A=r_0 A^{1/3}$)
{\it behind}  the valence quarks. This is an important
feature of the partons when boosting a hadron or nucleus to high rapidities 
\cite{bj76,hwa86,mueller89}.
As a consequence, the parton positions are correlated in longitudinal
direction with their momenta, as required by the uncertainty principle.
This leads to an enhancement of the densities of gluons and sea quarks
with $x< 1/(2R_h M_h)$ proportional to $A^{1/3}$, because such partons from
different nucleons overlap spatially when the nucleons are at the same impact parameter.
\end{description}
\bigskip

\subsection{Parton cascade development}

With the above construction of the initial state,
the incoming beam particles $A$ and $B$ are decomposed in   
their associated parton content at the initial
resolution scale $Q_0^2$ at time $t=t_0\equiv0$, 
and the subsequent dynamical development of the
system for $t > t_0$
can now be traced according to the kinetic equations
(\ref{e1})-(\ref{e4}).
In the kinetic approach, the space-time evolution of  parton densities
may be  described in terms of
parton cascades. Fig. 4 illustrates  a typical parton cascade sequence. 
It is important to realize that, in general,
there can be many such cascade sequences, being internetted
and simultanously evolving (typical for hadron-nucleus or nucleus-nucleus
collisions). 

\begin{figure}
\epsfxsize=450pt
\centerline{ \epsfbox{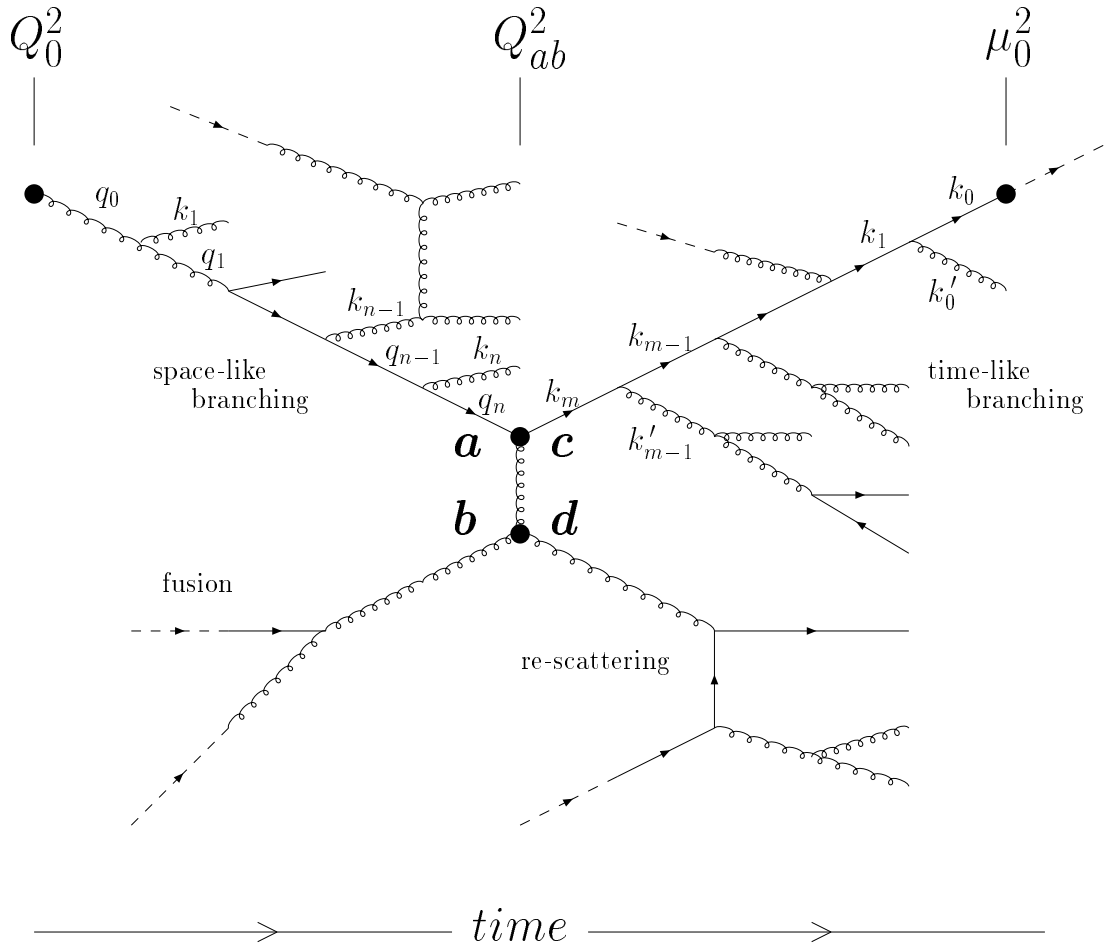} }
\caption{
Schematical illustration of 
a typical parton cascade development initiated by a 
collision of two partons $a$ and $b$.
The incoming primary parton $a$ evolves through
a space-like cascade from the 
initial resolution scale $Q_0^2$ up to
the scale $Q_{a b}^2$ at which it collides with parton $b$.
The outgoing partons $c$ and $d$ both initiate time-like
cascades which are described as a combination of 
multiple branchings (or emissions), fusions (or absorptions),
or secondary scatterings (or rescatterings).
A particular branch in a cascade tree terminates
(locally in space-time), when the partons in that branch 
recombine with neighboring ones to pre-hadronic clusters and 
their subsequent decay into hadrons.
\label{fig:fig4}
}
\end{figure}

Each cascade can be subdivided into elementary
$2 \rightarrow 2$ scatterings, $1 \rightarrow 2$ branchings (emissions), and
$2\rightarrow 1$ fusions (absorptions).
In Fig. 4, a primary parton $a$ that originates from one of the incident
nuclei, collides with another parton $b$ with some momentum transfer
characterized by the scale $Q_{a b}^2$ at the collision vertex.
The parton $a$ has evolved from the initial scale $Q_0^2$, at which it was
resolved in its parent hadron or nucleus, 
up to $Q_{a b}^2$ by successive space-like
branchings. From the scattering of $a$ and $b$ the partons $c$ and $d$ emerge,
both of which can initiate sequences of time-like branchings. These newly
produced partons can themselves branch,
rescatter, or undergo fusion with other partons.

In practice the dynamical interplay of $2 \rightarrow 2$, $2 \rightarrow 1$
and $1 \rightarrow 2$ processes can be simulated
as follows:
Since the time evolution of the parton system is described in
small discrete time steps $\Delta t = O(10^{-3} fm)$, 
the time scale of an interaction
process is compared with $\Delta t$ to decide whether the
interaction occurs within this time interval. 
A virtual parton
$e^\ast$ with momentum $k^\mu$, $k^2\ne 0$,
that is produced via $a + b \rightarrow e^\ast$, has a short
life-time $\tau_{e^\ast}\propto 1/\sqrt{|k^2|}$, 
if its invariant mass $|k^2|$ is large and it is likely
to decay within $\gamma \tau_{e^\ast} < \Delta t$
(the full formulae are derived  in \cite{ms1}).
Therefore, the $2 \rightarrow 2$ process
$a + b \rightarrow e^\ast \rightarrow c + d$ preferably occurs.
On the other hand, if $|k^2|$ is small, corresponding to a long
$\tau_{e^\ast}$, parton $e^\ast$ is likely not to decay within 
the time step $\Delta t$
and the fusion process $a + b \rightarrow e^\ast$ rather happens.
In this case, $e^\ast$ propagates as a quasi-stable particle until in
the following time step it has an increased decay probability \cite{ms1}.
This may then result in the
$1 \rightarrow 2$ decay $e^\ast \rightarrow c + d$. Alternatively,
$e^\ast$ might collide with another parton before its decay or 
may propagate freely until the following time step, and so on.
In this manner the elementary $2\rightarrow 2$, $2\rightarrow 1$,
and $1\rightarrow 2$ processes are treated on equal footing
and double counting is prevented:
either  a $2\rightarrow 2$, or a $2\rightarrow 1$,
and possibly a subsequent $1\rightarrow 2$ process can occur, 
but not both. The relative probability is  determined by
the uncertainty principle, i.e. by relating the momentum scale of 
the process to the time scale as explained.

For the collisional processes $2\rightarrow 2$ and $2 \rightarrow 1$,
the statistical occurence is determined by the 2-body cross-section
in Born apprpoximation, with higher-order inelastic
corrections effectively included in the parton evolution
before and after each collision.
This  parton-evolution is calculated
using the well-known jet calculus \cite{jetcalc,bassetto}
based
on the `modified leading logarithmic approximation' (MLLA)
to the QCD evolution of hard processes \cite{dokbook,dok88}.
Each individual parton-collision factorizes then in an elementary
2-body collision, in which the in- and out-going partons undergo
an ordered sequence
of elementary branchings $a\rightarrow b+c$, accounting for
higher-order radiative corrections to the lowest-order Born approximation.
These  branchingss can be described stochastically as a
Markov cascade in position and momentum space.
One distinguishes  initial-state, {\it space-like} branchings
of the two partons entering a collision vertex,
and final-state, {\it time-like} radiation off the 
collided partons after the collision.

The specific feature of the present approach is that, in addition to the
definite
virtuality and momentum,  each elementary vertex has a certain space and
time
position which is obtained by assuming that the partons in the cascade
propagate on
straight-line trajectories in between their interactions.
In the MLLA framework, the
basic properties of both space-like and time-like showers are
determined by the DGLAP equations \cite{DGLAP}, but with essential
differences in time ordering, kinematics and the treatment of
infrared singularities associated with soft gluon emission.
\medskip

\subsubsection{Elementary parton-parton collisions}

For the elementary  parton scatterings $a + b \rightarrow c + d$,
and fusion processes  $a + b \rightarrow c^\ast$, 
two distinct classes of processes 
are considered: 
\begin{description}
\item[(i)] truly perturbative QCD {\it hard}  parton collisions
with a sufficiently large momentum transfer $p_\perp^2$ or invariant mass
$\hat s$;
\item[(ii)] 
phenomenological treatment of {\it soft} parton collisions
\footnote{The inclusion of soft parton collisions is optional in the program,
and by default is switched off.} 
with low momentum
transfer $p_\perp^2$ or invariant mass $\hat s$.
\end{description}
The motivation for this differentiation of parton-parton collisions
is to regulate the singular behavior of the collision integrals
in (\ref{e1}) which results from the divergence of the 
associated Born-amplitudes squared
for small momentum transfers $Q^2$ (except for $\hat s$ channel
processes such as $q \bar q$ annihilation).
To render the parton-parton cross-sections finite, an invariant 
{\it hard-soft division scale} $p_0^2$ is introduced such that
collisions occurring at a momentum scale $Q^2 \geq p_0^2$
are treated as perturbative (semi)hard collisions,
whereas for those with $Q^2 < p_0^2$
a soft, non-perturbative  interaction is assumed to occur. That is, the total
parton-parton cross-section for a collision
between two partons $a$ and $b$ is represented as
\begin{equation}
\hat \sigma_{a b} ( \hat s ) \;=\;
\sum_{c, (d)} \, \left\{ \,\frac{}{}
\int_{0}^{p^2_0}  d Q^2
\left(
\frac{d \hat \sigma_{a b \rightarrow c (d)}^{soft}}{d Q^2}
\right)
\;+\;
\int_{p^2_0}^{\hat s}  d Q^2
\left(
\frac{d \hat \sigma_{a b \rightarrow c (d)}^{hard}}{d Q^2}
\right)
\, \right\}
\;\;\;,
\label{sigma}
\end{equation}
where
\begin{eqnarray}
Q^2 \,&\equiv& \,Q^2(\hat{s},\hat{t},\hat{u}) \;=\; \left\{
\begin{array}{cl}
p_\perp^2 \simeq  -\hat{t} (-\hat{u})
& \;\mbox{for} \;\hat{t} (\hat{u}) \; \mbox{channel}
\\
m^2 = \hat{s}
& \;\mbox{for} \;\hat{s} \;\mbox{channel}
\end{array}
\right.
\\
\hat s  = (p_a + p_b)^2
& &
,\;\;\;\;\;\;\;\;\;\;
\hat t = (p_a - p_c)^2
\;,\;\;\;\;\;\;\;\;\;\;
\hat u = (p_a - p_d)^2
\end{eqnarray}
That is, the scale $Q^2$ of the collision is set by
$p_\perp^2$ for scatterings
$a+b\rightarrow a+b$, or by $\hat s$ for annihilation
processes and $a+b\rightarrow c+d$ and fusion processes $a+b\rightarrow c$.
The sum over $c$, $(d)$ corresponds to summing over all possible reaction
channels (processes).
The specific value of $p_0$ is generally dependent on beam energy 
$E_{cm}=(P_A+P_B)^2$ of colliding hadrons (nuclei) $A+B$,
as well as on the  mass numbers $A$ and $B$. It is treated as a parameter
that is determined by existing experimental data. A suitable form is
\begin{equation}
p_0\;\,\equiv\;\,
p_0(E_{cm},A,B)\;=\; \frac{a}{4}\;\left(\frac{2 E_{cm}/\mbox{GeV}}
{A+B}\right)^b
\;\;\;\;\;\;\;\;
(a = 2.0 \;\mbox{GeV} , \;\;\;\;\;b= 0.27)
\;.
\end{equation}

The complementation of both hard and soft 
contributions renders the parton cross-section
finite and well defined for {\it all} $Q^2$.
It must be emphasized that the 
effect of soft parton collisions on the global dynamics is not essential
for collisions involving only leptons and/or hadrons, 
but plays an important role
in hadron-nucleus, or nucleus-nucleus collions.
The soft parton collisions
naturally involve comparably small momentum transfer, so that their contribution
to transverse energy production is small, but the effect on soft particle
production is significant.

In accord with the two-component structure of the cross-section (\ref{sigma}),
these parton collisions are distinguished depending on the momentum transfer:
\begin{description}

\item[(i)]
{\bf Hard collisions above} {\boldmath $p_{0}$}:
\smallskip

For the {\it perturbative (semi)hard collisions} above 
$p_{0}$, the momentum scale $Q^2$ is determined by the
corresponding differential cross-sections
\begin{equation}
\frac{d \hat \sigma_{ab \rightarrow cd}^{hard}}
{d  Q^2}
\,=\,
\frac{1}{16 \pi \hat s^2} \,
\left| \overline{M}_{ab \rightarrow cd} \right| ^2 
(\hat s, Q^2) 
\;\;\propto \frac{\pi \alpha_s^2(Q^2)}{Q^2}
\;,
\label{sigh}
\end{equation}
where $| \overline{M} | ^2$ 
is the process-dependent spin- and color-averaged squared matrix element
in Born approximation,  published in
the literature (see e.g. Refs. \cite{sig1} and for 
massive quarks Refs. \cite{sig2}).
\medskip

\item[(ii)]
{\bf Soft collisions below} {\boldmath $p_{0}$}:
\smallskip

In the case of a non-perturbative soft collision between two partons
it is assumed that a very low energy double gluon
exchange occurs. This provides a natural continuation
to the harder collisions above
$p_{0}$ where the dominant 
one gluon exchange processes
$g g \rightarrow g g$, $g q \rightarrow g q$ and
$q q \rightarrow q q$ have the same overall  structure \cite{gustaf82}.
A simple, and physically plausible, form for the soft cross-section
that continues the hard cross-section for $Q^2$ below $p_0$ down to $Q^2 = 0$,
may be modelled by introducing a screening mass-term $\mu^2$ in the
nominator of (\ref{sigh}),
\begin{equation}
\frac{d \hat \sigma_{ab \rightarrow cd}^{soft}}
{d  Q^2}
\;\propto\;
\frac{2\pi \alpha_s^2(p_0^2)}{Q^2 + \mu^2}
\label{sigs}
\end{equation}
where $\mu$ is a phenomenological parameter that 
governs the overall magnitude of the integrated  
$\sigma^{soft} \propto \ln[(p_0^2+\mu^2)/\mu^2]$.
The value of $\mu$ 
is not known precisely, but it can be estimated to be in the
range 0.3 - 1 GeV.
\end{description}

Notice that
both hard and soft scatterings are treated in this approach on completely equal footing.
That is, with the four momenta of the incoming partons known, the momentum
transfer and scattering angle are sampled from the respective
differential cross-sections $d\hat \sigma / d\hat t$
(\ref{sigh}) and (\ref{sigs}), assuming azimuthal symmetry of the
scattering geometry.
\medskip

\subsubsection{Space-like parton evolution}

\noindent
Space-like parton cascades are associated with
QCD radiative corrections due to brems-strahlung emitted 
by primary space-like partons, which are
contained in the initial state hadrons (nuclei),
and which encounter their very
first collision. Thereby such a primary, virtual parton is
struck out of the coherent hadron (nucleus) 
wavefunction to materialize to a real excitation.
A typical space-like cascade is illustrated in the left part of Fig. 4, 
where the parton $a_0$, originating from the initial hadron (nucleus) state, 
undergoes successive space-like branchings $a_0 \rightarrow a_1 a_1'$,
$a_1 \rightarrow a_2 a_2'$, ..., $a_{n-1} \rightarrow a_n a_n'$ to become
the parton $a \equiv a_n$ which then actually collides with another parton $b$.
The branching chain proceeds by increasing the virtualities 
$|q_i^2|$ of the partons $a_i$ in
the cascade, starting from $a_0$ with $|q_0^2|  \simeq Q_0^2/4$
up to $|q^2| \simeq  Q^2$ 
the space-like virtuality of the scattering parton $a$,
where $Q^2 \equiv Q_{ab}^2$ sets the scale of the collision
between partons $a$ and $b$, and hence for the evolution from $Q_0^2$ to $Q^2$.
The emitted partons $a_i'$ on the side branches with momenta $k_i$
on the other hand have, due to energy-momentum
conservation at the branching vertices, time-like virtualities and
each of them can initiate a time-like cascade, discussed afterwards.

The proper inclusion of both collinear, hard and coherent, soft branchings is
achieved by describing the space-like cascade in both space-time {\it and} 
momentum space by using
a so-called {\it angular-ordered} evolution variable 
$\tilde{q}_j^2$ (rather than the virtualities $|q_j^2|$)
\cite{webber88}
\begin{equation}
\tilde{q}_j^2 \;\equiv E_j^2\;\zeta_{j+1}
\;\;,\;\;\;\;\;\;\;\;\;\;\;
\zeta_{j+1} \;=\;\frac{q_0\cdot k_{j+1}}{\omega_0 \;E_{j+1}}
\;\simeq \;
1 - \cos \theta_{0,\;j+1}
\;\;\;\;\;\;\;\;\;(0 \le j \le n)
\;,
\label{tildep}
\end{equation}
where $q_j=(\omega_j,\vec{q}_j)$ and
$k_{j+1}=(E_{j+1},\vec k_{j+1})$ refer to
the $j^{th}$ branching $a_j \rightarrow a_{j+1} a_{j+1}'$ with
momentum assignment
$q_j\rightarrow q_{j+1} k_{j+1}$ (see Fig. 4).
The space-like cascade is then strictly ordered in the variable
$\tilde{q}_{j+1}^2 > \tilde{q}_j^2$, which is equivalent to the
ordering of emission angles,
$\omega_j \theta_{0, \;j+1^\prime} < \omega_{j+1} \theta_{0, \;j+2^\prime}$.

The space-like cascade terminates with parton $a_n \equiv a$ entering
the vertex of collision with parton $b$, that is, $Q_{ab}^2$ in Fig. 4.
The history of parton $a$ is however not known until after
it has collided with parton $b$, because it is this very collision
that causes the cascade evolution of parton $a$.
Therefore one must reconstruct the cascade {\it backwards} in
time starting from the time of the collisions at the vertex $Q_{ab}^2$ and
trace the history of the struck parton $a$ back to 
the initial state at time $t_0 =0$ at which it was 
originally resolved with $Q_0^2$ in its hadron (nucleon) mother.
The method used here is a space-time generalization of the
`backward evolution scheme'  \cite{backevol}.
To sketch the procedure, consider the space-like branching
$q_{n-1} \rightarrow q_n k_n$ which is closest to the 
collision vertex $Q_{ab}^2$  in Fig. 4.
The virtualities satisfy \cite{bassetto}
$\vert q_n^2 \vert > \vert q_{n-1}^2 \vert$, and $q_{n}^2, q_{n-1}^2 <
0$
(space-like) but $k_n^2 > 0$ (time-like).
The relative probability for a branching to occur  between
$\tilde{q}^2$
and $\tilde{q}^2 + d\tilde{q}^2$ is given by
\begin{eqnarray}
d {\cal P}_{n-1,\,n}^{(S)} ( x_{n-1}, x_{n}, \tilde{q}^2;\,\Delta t)
&=&
\frac{d \tilde{q}^2}{\tilde{q}^2}\, \frac{d z}{z}
\,
\frac{\alpha_s\left((1-z) \tilde{q}^2\right)}{2 \pi} \,
\,
\gamma_{n\mbox{-}1 \rightarrow n n^\prime} (z)
\;\;
\left(
\frac{F(r_{n-1}; x_{n-1}, \tilde{q}^2)}
{F(r_n; x_n, \tilde{q}^2)}
\right)
\;\,{\cal T}^{(S)}(\Delta t)
\;,
\label{PS}
\end{eqnarray}
where $x_j = (q_j)_z / P_z$  ($j=n, n-1$) are the
fractions of longitudinal momentum $P_z$ of the initial mother hadron (nucleon),
with $F(r_j;x_j, \tilde{q}^2)\equiv F(r_j,q_j)$
the corresponding parton distributions
defined by (\ref{F}),
and the variables
\begin{equation}
z \;=\;\frac{E_n}{E_{n-1}} \;\simeq \;\frac{x_n}{x_{n-1}}
\;\;\;,\;\;\;\;\;\;\;\;
1\;-\;z \;=\;\frac{E^\prime_n}{E_{n-1}} \;\simeq \;
\frac{x_{n-1}-x_{n}}{x_{n-1}}
\label{tildez}
\end{equation}
specify the fractional energy or longitudinal
momentum of parton $n$ and $n^\prime$, respectively, taken away from
$n-1$.
The function $\alpha_s/(2\pi)\, \gamma (z)$
is the usual DGLAP branching probability \cite{dok80,dokbook},
with $\gamma (z)$ giving the energy distribution in the variable $z$.
The last factor 
${\cal T}^{(S)}(\Delta t)$
in (\ref{PS}) determines the time interval in the global $cm$-frame,
$\Delta t = t_n - t_{n-1}$, that is associated with the branching
process
$a_{n-1}\rightarrow a_n a_n'$.
It accounts for the formation time of $a_n$ from $a_{n-1}$ on
the basis of the uncertainty principle:
$\Delta t = \Delta \omega/|q_n^2|$, 
$\Delta \omega \simeq  (x_n - x_{n-1})\,P_z$.
A very simple form is taken here, 
\begin{equation}
{\cal T}^{(S)}(\Delta t)
\;=\; \delta\left( \frac{x_n - x_{n-1}}{\vert q_n^2\vert}\, P_z
\;-\; \Delta t\right)
\label{delts}
\;.
\end{equation}
The backwards evolution of the space-like branching
$q_{n-1} \rightarrow q_n + k_n$ is expressed in terms of the
probability that parton $a_{n-1}$ did {\it not} branch between the
lower bound $\tilde{q}_0^2$, given by the initial resolution scale
$Q_0^2$, and
$\tilde{q}_n^2 \equiv \tilde{q}^2 \simeq Q_{ab}^2$.
In that case, parton $n$ can {\it not} originate from this branching,
but must have been produced otherwise or already been present in
the initial parton distributions.
This non-branching probability is given by the
{\it Sudakov form-factor for space-like branchings}:
\begin{equation}
S_n ( x_{n}, \tilde{q}^2, \tilde{q}_0^2;\,\Delta t)
\;=\;
\exp
\left\{
\,-\, \sum_{a}
\,
 \int_{\tilde{q}_0^2}^{\tilde{q}^2}
\,
\int_{z_- (\tilde{q}^{\prime})}^{z_+ (\tilde{q}^{\prime})}
\,
d {\cal P}_{n,\,n-1}^{(S)} ( x_n,  z, \tilde{q}^{\prime 2}; \,\Delta t)
\right\}
\;,
\label{slff2}
\end{equation}
where the sum runs over the possible species $a = g, q, \bar q$
of parton $a_{n-1}$.
The upper limit of the $\tilde{p}^2$-integration is set by
$\tilde{q}^2  \, \lower3pt\hbox{$\buildrel <\over\sim$}\,Q_{ab}^2$,
associated with the collision vertex with parton $b$.
The limits $z_\pm$ are determined by kinematics \cite{webber88}:
$z_-(\tilde{q})= Q_0/\tilde{q}$ and
$z_+(\tilde{q})= 1 - Q_0/\tilde{q}$.

The knowledge of the space-like formfactor 
$S_n ( x_{n}, \tilde{q}^2, \tilde{q}_0^2;\,\Delta t)$
is enough to trace
the evolution of the branching closest to the hard vertex backwards
from $q_n^2$ at  $t=t_n$, the time of collision in the global $cm$-frame,
$q_{n-1}^2$ at $t_{n-1}= t_n- x_n/|q_n^2|\,P_z$.
The next preceding branchings  $q_{n-2}\rightarrow q_{n-1}
k_{n-1}$, etc., are then reconstructed in exactly the same manner
with the replacements $t_n \rightarrow t_{n-1}$,
$x_n \rightarrow x_{n-1}$, $q_n^2 \rightarrow q_{n-1}^2$, and so forth,
until the initial point  $q_0^2$ at $t_0 = 0$ is reached.
\medskip

\subsubsection{Time-like parton evolution}

\noindent
Time-like parton cascades are initiated by secondary partons
that  emerge either from the side-branches of a 
preceding space-like or time-like cascade,
or directly from a scattering or fusion process.
Consider the time-like cascade
initiated by the parton $c$ in the right part of Fig. 4, 
with momentum $k = k_n$.
This parton has been produced in the collision
$a +b \rightarrow c + d$ with a time-like off-shellness 
$k_n^2 \simeq Q_{ab}^2$.

Again an {\it angular-ordered} (rather than virtuality-ordered) 
evolution in space-time {\it and} momentum-space
of the cascade is employed to incorporate interference effects of soft
gluons emitted along the time-like cascade tree of Fig. 4,
$c \equiv c_m \rightarrow c_{m-1} c_{m-1}'$, ..., 
$c_{1} \rightarrow c_0 c_0'$.
In contrast to (\ref{tildep}), the time-like version of the angular
evolution variable  is \cite{webber86}
\begin{equation}
\tilde{k}_j^2 \;\equiv E_j^2\;\xi_{j-1}
\;\;,\;\;\;\;\;\;\;\;\;\;\;
\xi_{j-1} \,=\,
\frac{k_{j-1} \cdot k^\prime_{j-1}}{E_{j-1} E^\prime_{j-1}} \,\simeq \,
1 - \cos \theta_{(j\mbox{-}1), (j\mbox{-}1)^\prime}
\;\;\;\;\;\;\;\;\;(m \ge j \ge 1)
\;.
\label{tildek}
\end{equation}
so that the  time-like cascade can be described by
a $\tilde{k}^2$-ordered (rather than $k^2$-ordered) evolution,
which corresponds to an angular ordering with decreasing emission angles
$\theta_{j, j^\prime} > \theta_{(j\mbox{-}1), (j\mbox{-}1)^\prime}$.

Proceeding analogously to the space-like case (c.f. (\ref{PS})),
the probability $d {\cal P}_{m,\,m-1}^{(T)}$ for the
first branching after the $\gamma q$ vertex,
$k_m \rightarrow k_{m-1} k_{m-1}^\prime$
with $k^2_{m-1}, k^{\prime \,2}_{m-1}$,
is given by the space-time extension \cite{ms39,msrep}
of the usual DGLAP probability distribution \cite{dok80,dokbook},
\begin{equation}
d {\cal P}_{m,\, m-1}^{(T)} (z,\tilde{k}^2;\,\Delta t) \,=\,
\frac{d \tilde{k}^2}{\tilde{k}^2}\, dz
\;
\frac{\alpha_s( \kappa^2 )}{2 \pi} \,
\gamma_{m \rightarrow (m\mbox{-}1),(m\mbox{-}1)^\prime} (z)
\;\,
{\cal T}^{(T)} (\Delta t)\,
\; ,
\label{PT}
\end{equation}
where
${\cal T}^{(T)} (\Delta t)$ is the probabilty that parton $m$ with
virtuality
$k_m^2$ and  corresponding proper lifetime $\tau_m \propto
1/\sqrt{k_m^2}$
 decays within a time interval $\Delta t$,
\begin{equation}
{\cal T}^{(T)} (\Delta t)
\;=\; 1\;  - \;\exp \left( - \frac{\Delta t}{t_m(k)}\right)
\label{deltt}
\;.
\end{equation}
The actual lifetime of the decaying parton $m$ in the global $cm$-frame
is then $t_m(k) = \gamma/\tau_m(k)$, where
$t_q(k) \approx 3 E/(2 \alpha_s k^2)$ for quarks and
$t_g(k) \approx  E/(2 \alpha_s k^2)$ for gluons \cite{ms3}.
As before,
$F_j$ denotes the local density of parton species $j=m,m-1$, and
$\alpha_s/(2\pi) \gamma(z)$ is the DGLAP branching kernel
with energy distribution $\gamma(z)$.
The probability (\ref{PT}) is formulated in terms of the  energy
fractions carried by the daughter partons,
\begin{equation}
z \;=\; \frac{E_{m-1}}{E_m}
\;\;\;,\;\;\;\;\;\;\;\;
1 - z \; = \; \frac{E_{m-1}^\prime}{E_m}
\;\;\; ,
\end{equation}
with the virtuality $k_m$ of  the parton $m$  related to $z$ and $\xi$
through
$k_m^2 = k_{m_1}^2 + k_{m-1}^{\prime \, 2} + 2 E_m^2 z ( 1 - z ) \xi$,
and the argument $\kappa^2$ in the running coupling $\alpha_s$ in
(\ref{PT})
is \cite{webber88}
$\kappa^2= 2 z^2 ( 1 - z )^2  E_m^2  \xi  \simeq k_\perp^2$.

The branching probability (\ref{PT}) determines the distribution
of emitted partons in both coordinate and momentum space, because
the knowledge of four-momentum and lifetime (or $\Delta t$ between
successive branchings)  give the spatial positions of the
partons, if they are assumed to propagate on straight paths between
the vertices.
The probability that parton $m$ does {\it not} branch between
$\tilde{k}^2$ and
a minimum value $\tilde{k}^2_0 \equiv \mu_0^2$
is given by the exponentiation of (\ref{PT}),
yielding the {\it Sudakov form-factor for time-like branchings}:
\begin{equation}
T_m (\tilde{k}^2,\tilde{k}_0^2; \, \Delta t)
\;=\;
\exp
\left\{
\,-\, \int_{\tilde{k}_0^2}^{\tilde{k^2}}
\, \sum_a
\, \int_{z_{-}(\tilde{k}^\prime)}^{z_{+}(\tilde{k}^\prime)}
\;\,
d {\cal P}_{m,\,m-1}^{(T)} (z, \tilde{k}^{\prime\,2};\,\Delta t)
\right\}
\;,
\label{tlff2}
\end{equation}
which is summed over the species $a = g , q, \bar q$
of parton $m-1$.
The integration limits $\tilde{k}_0^2$ and $z_\pm$
are determined by the requirement that the branching must terminate
when the partons enter the non-perturbative regime and begin
to hadronize.
This condition can be parametrized by
the confinement length scale $L_c = O(1 \,fm)$  with
$\tilde{k_0}^2  \, \lower3pt\hbox{$\buildrel >\over\sim$}\, L_c^{-2}
\equiv \mu_0^2$, and
$
z_{+} ( \tilde{k}_m ) =
1 -z_{-} ( \tilde{k}_m ) =
\mu_0/\sqrt{4 \tilde{k}_m^2}
$,
so that for
$z_{+} ( \tilde{k}_0^2)=
z_{-} ( \tilde{k}_0^2) = 1/2$
the phase space for the branching vanishes.

The time-like form factor 
$T_m (\tilde{k}^2,\tilde{k}_0^2; \, \Delta t)$
determines the four-momenta and
positions of the partons of a particular emission vertex
as sketched above for the first branching
from $k_m^2$ at  $t=t_m$, the time of production of parton
$c$ in the global $cm$-frame,
to $k_{m-1}^2$ at $t_{m-1}= t_m+ E_m/|k_m^2|$.
Subsequent branchings are described
completely analogously by replacing $t_m\rightarrow t_{m-1}$,
$k_m^2 \rightarrow k_{m-1}^2$, etc..
Hence $T(\tilde{k}^2,\tilde{k}_0^2; \, \Delta t)$
generates a time-like cascade
as sequential branchings starting from $t=0$ at the
hard vertex forward in time, until the partons eventually hadronize
as discussed below.
\bigskip

\subsection{Cluster formation and hadronization}

In view of lack of knowledge about the details of confinement dynamics 
and the non-perturbative hadronization mechanism, one  must
rely at present on  model-building.
In the present approach, the cluster-hadronization scheme of Ref. 
\cite{ms37,ms40,ms41} is employed.
This phenomenological scheme is inspired by the Marchesini-Webber model 
\cite{webber84}, however it works in space-time plus color-space.
On the other hand it is very different from
commonly used string-fragmentation models such as the Lund-model \cite{string}.

In VNI,
both the cluster-formation from the collection
of quarks and gluons at the end of the perturbative phase
and the subsequent cluster-decay into final hadrons
consist of two components:
\begin{description}
\item[1.]
The recombination of the {\it secondary time-like partons},
their conversion into colorless
{\it parton clusters} and the subsequent
decay into secondary hadrons.
\item[2.]
The recombination of the {\it primary space-like partons} that remained
spectators throughout the collision development into {\it beam clusters}
and the fragmentation of these clusters.
\end{description}

The important assumption here is that the process of hadron formation
depends only on the local space-, time-, and color-structure of the
parton system,
so that the hadronization mechanism can be modelled as the formation of
color-singlet clusters of partons as independent entities
(pre-hadrons),
which subsequently decay into hadrons.
This concept is reminiscent of the `pre-confinement' property
\cite{preconf} of parton evolution,
which is the tendency of the produced partons
to arrange themselves in color-singlet clusters with limited
extension in both position and momentum space, so that it is
suggestive to suppose that these clusters are the basic units out of
which hadrons form.
\medskip

\subsubsection{Cluster formation}

\noindent
{\bf (i)  Parton clusters:}
\smallskip

Parton clusters are formed from secondary partons, i.e. those
that have been produced by the hard interaction and the
parton shower development.
The coalescence of these secondary  partons to color-neutral clusters
has been discussed in detail in Refs. \cite{ms37,ms40}.
Throughout the dynamically-evolving
parton cascade development,  every parton and its nearest
spatial neighbour are considered as as potential candidates for
a 2-parton cluster, which, if {\it color neutral}, plays the role
of a `pre-confined' excitation in the process of hadronization.
Within each single time step, the probability for parton-cluster
conversion is determined for each nearest-neighbor pair by
the requirement that the total color charge of the two partons must
give a composite color-singlet state (if necessary by accompanying
gluon emission), and the condition that
their {\it relative spatial distance} $L$ exceeds the critical
{\it confinement length scale} $L_c$.
The scale $L$ is defined as
the Lorentz-invariant distance $L_{ij}$ between parton $i$
and its nearest neighbor $j$:
\begin{equation}
L\;\equiv\; L_{ij} \;= \;
\mbox{min} (\Delta_{i 1}, \ldots , \Delta_{i j}, \ldots , \Delta_{i n})
\;,
\label{L}
\end{equation}
where $\Delta_{ij} = \sqrt{ (r_{ij})_0^2 + 
(r_{ij})_x^2 + (r_{ij})_y^2 + (r_{ij})_z^2}$, 
with $r_{ij}^\mu = r_i^\mu - r_j^\mu$,
and the probability for the coalescence of the two partons $i$, $j$ to
form
a cluster is modelled by a distribution of the form
\begin{equation}
\Pi_{ij\rightarrow c}\;\propto \; \left(\frac{}{}
1\,-\, \exp\left(-\Delta F\;L_{ij}\right)
\right)
\;\,\simeq \;\,
1\;-\;\exp\left(\frac{L_0-L_{ij}}{L_c-L_{ij}} \right)
\;\;\;\;\;\mbox{if $L_0 \;<\;L_{ij}\;\le\;L_c$}
\;,
\label{Pi2}
\end{equation}
and $\Pi_{ij\rightarrow c}= 0\; (1)$ if $L_{ij} < L_0$ ($L_{ij} > L_c$).
Here $\Delta F$ is the local change in the free energy
of the system that is associated with the
conversion of the partons to clusters,
and the second expression on the right side is a parametrization
in terms of $L_0 = 0.6$ $fm$ and $L_c = 0.8$ $fm$
that define the transition regime.
As studied in Ref. \cite{ms40}, the aforementioned color constraint,
that only colorless 2-parton configurations may produce a cluster,
can be incorporated by allowing
coalescence for any pair of color charges, as determined by the
space-time separation $L_{ij}$ and the probability (\ref{Pi2}),
however, accompanied by the additional emission
of a gluon or quark that carries away any unbalanced net color
in the case that the two coalescing partons are not in a colorless
configuration.
\medskip

\noindent
{\bf (ii)  Beam clusters:}
\smallskip

If one or both of the colliding beam/target particles $A$ and $B$ were
a hadron or a nucleus, the one, respectively two beam clusters
are formed from the spectator partons that represent
the receding beam/target remnants of the original particles $A$ and $B$.
More precisely,
the remaining fraction of the longitudinal momentum and energy that has
not materialized or been redirected 
and harnessed during the course of the collision, is
carried by those primary partons of the initial-state hadrons or nuclei, which
remained spectators throughout. In the present approach
these partons maintain their originally assigned momenta
and their space-like virtualities. Representing the  beam remnants
of $A$ and/or $B$, they
may be pictured as the coherent relics of  the original
hadron (or nucleus) wavefunctions.
Therefore the primary virtual partons must be treated differently
than the secondary partons which are real excitations that contribute
incoherently to the hadron yield.
In the global $cm$-frame,  the primary partons are
grouped together to form a  massive beam cluster with its four-momentum
given by the sum of the parton momenta and its position given by
the 3-vector mean of the  partons' positions.
\bigskip

\subsubsection{Hadronization of clusters}

\noindent
{\bf (i) Parton clusters:}
\smallskip

For  the decay of each parton cluster into final-state hadrons,
the scheme presented in detail in Refs. \cite{ms37}
is employed:
If a cluster is too light to
decay into a pair of hadrons, it is taken to represent
the lightest single meson that corresponds to its
partonic constituents. Otherwise, the cluster
decays isotropically in its rest frame into
a pair of hadrons, either mesons or baryons, whose combined
quantum numbers correspond to its partonic constituents.
The corresponding  decay probability is chosen to be
\begin{equation}
\Pi_{c\rightarrow h}\;=\;
\;{\cal T}_c(E_c,m_c^2) \;
\;\,{\cal N}
\int_{m_h}^{m_c}
\frac{dm}{m^3}\;\exp\left(-\frac{m}{m_0}\right)
\;,
\label{pi3}
\end{equation}
where ${\cal N}$ is a normalization factor,
and the integrand is a Hagedorn spectrum \cite{hagedorn} that
parametrizes quite well
the density of accessible hadronic states below $m_c$ which are
listed in the particle data tables, and $m_0 = m_{\pi}$.
In analogy to (\ref{deltt}), ${\cal T}_c$ is a
life-time factor giving the probability
that a cluster of mass $m_c^2$ decays  within
a time interval $\Delta t$ in the global $cm$-frame,
\begin{equation}
{\cal T}_c(E_c,m_c^2)\;=\; 1\;-\;
\exp\left( - \frac{\Delta t}{t_c(E_c,m_c^2)}\right)
\;,
\label{lft2}
\end{equation}
with the Lorentz-boosted life time
$t_c= \gamma_c \tau_c\simeq E_c/m_c^2$.
In this scheme, a particular cluster
decay mode is obtained from (\ref{pi3})
by summing over all possible decay channels,
weighted with the appropriate spin, flavour, and
phase-space factors, and then choosing the actual decay mode
acording to the relative probabilities of the channels.
\medskip

\noindent
{\bf (ii) Beam clusters:}
\smallskip

The fragmentation of the beam clusters containing the spectator partons
mimics in the present model what is commonly termed the `soft underlying event',
namely, the emergence of those final-state hadrons that are
associated with the non-perturbative physics which underlies the
perturbatively-accessible dynamics of the hard interaction
with parton shower fragmentation.

In the spirit of the Marchesini-Webber model \cite{webber88},  
a version (suitably modified
for the present purposes) of the  soft hadron production model
of the UA5 collaboration \cite{UA5}, which is based on a parametrization
of the CERN $p\bar p$ collider data for minimum-bias hadronic
collisions.
The parameters involved in this model are set to give a good
agreement with those data.

Soft hadron production is known to be a universal mechanism
\cite{Mdiff} that is
common to all high-energy collisions that involve beam hadrons in the
initial state,
and that depends essentially on the total energy-momentum of
the fragmenting final-state beam remnant.
Accordingly, one may assume that the fragmentation of the final-state
beam cluster depends solely on its invariant mass $M$, and that it
produces a charged-
particle multiplicity with a binomial distribution \cite{UA5},
\begin{equation}
P(n) \;=\; \frac{\Gamma(n+k)}{n! \Gamma(k)} \;
\frac{ (\overline{n}/k)^n}{(1 + \overline{n}/k)^{n+k}}
\label{ndist}
\;,
\end{equation}
where the mean charged multiplicity
$\overline{n}\equiv\overline{n}(M^2)$
and the parameter $k\equiv k(M^2)$
depends on the invariant cluster mass
\footnote{
Notice that in this model $M$ fluctuates statistically, as a result
of fluctuations of the initial-state parton configuration in
the incoming hadrons (or nuclei), as well as
due to the fluctuating number of remnant partons during the space-time
evolution.
Hence the distribution (\ref{ndist}) and the mean multiplicity
(\ref{nk})
vary from event to event. This is in contrast to the original UA5 model,
in which the fixed beam energy $\sqrt{s}/2$ controls the energy
dependence of
soft hadron production.
}
according to the
following particle data parametrization \cite{UA5},
\begin{equation}
\overline{n}(M^2) \;=\;  10.68 \; (M^2)^{0.115} \;-\; 9.5
\;\;\;\;\;\;\;\;\;
k(M^2) \;=\;  0.029 \; \ln(M^2) \;-\; 0.064
\label{nk}
\;.
\end{equation}
Adopting the scheme of Marchesini and Webber \cite{webber88}, the
fragmentation of a beam cluster of mass $M$ proceeds then as
follows:
First, a particle multiplicity $n$ is chosen from (\ref{ndist}), and the
actual charged particle multiplicity is taken to be $n$ plus the
modulus of the beam cluster charge.
Next, the beam cluster is split into
sub-clusters $(q_1 \bar{q}_2), (q_2 \bar{q}_3), \ldots$ ($q_i = u, d$),
which are
subsequently hadronized in the beam cluster rest frame,
in the same way as the parton clusters described above.
To determine the sub-cluster momenta,  the following mass distribution
is assumed,
\begin{equation}
P(M) \;=\; c \; (M-1) \; \exp \left[ -a (M-1) \right]
\label{mdist}
\;,
\end{equation}
with $c$ a normalization constant and $a = 2$ GeV$^{-1}$, resulting
in an average value of $\langle M \rangle \approx 1.5$ GeV.
The transverse momenta are taken from the distibution
\begin{equation}
P(p_\perp) \;=\; c^\prime \; p_\perp \;
\exp \left[ -b \sqrt{p_\perp^2 +M^2}\right]
\label{pTdist}
\;,
\end{equation}
with normalization $c^\prime$ and slope parameter
$b = 3$ GeV$^{-1}$, and
the rapidities $y$ are drawn from a simple flat distribution
$P(y) \propto const.$ with an extent of
0.6 units  and Gaussian tails with 1 unit standard deviation at
the ends.
Finally, all  hadronization products of the sub-clusters are
boosted from the rest frame of the original beam cluster back into the
global $cm$-frame.
\bigskip
\bigskip

\subsection{Cluster/hadron cascade development}

The `afterburner' hadron cascade development is concepted in
in close analogy with the parton cascade evolution.
For a description of the physics ingredients and the technical
details, the interested reader is referred to Ref. \cite{hijet}.
There are two sources for producing primary hadrons,
namely, first, the  pre-hadronic parton-clusters 
(those  emerging from coalescence of 
materialized, interacted partons), 
and second, the beam/target fragmentation clusters
(those  resulting from reassembling of  the spectator partons
that have not interacted at all). 
Both these classes of clusters and hadrons participate in the
hadron cascading. The clusters or hadrons
propagate along classical paths until they either scatter off or absorb
other clusters or hadrons, or form resonant states
with subsequent decays. At each step the newly produced particles 
can reinteract themselves.
The collisions among clusters and hadrons are treated on equal footing,
and the corresponding interaction probabilities are calculated 
from the `additive quark model' \cite{aqm} within which one can
associate a specific cross-section for any pair of colliding hadrons 
according to their valence quark content.
Cluster-cluster and cluster-hadron collisions are straightforward to include
in this approach, since  each cluster that emerges from parton
coalescence has a definite quark content, depending on the flavor of the
mother partons.
Again, each newly produced hadron becomes a `real' particle only after
a characteristic formation time 
$\Delta t_{h} = \gamma / M_{h}$ depending on their invariant mass $M_h$ and
their energy through $\gamma= E/M_h$. Before that time has passed, a hadron
must be considered as a still virtual object that cannot interact
incoherently until it has formed according to the uncertainty principle.
\bigskip
\bigskip

\section{PROGRAM DESCRIPTION}
\label{sec:section3}
\bigskip

\subsection{The package VNI}

The  program package VNI
is a completely new written code, using only fragments of its precedessors 
VNI-1.0 (Ref. \cite{ms0}) and VNI-2.0 (Ref. \cite{ms3}). 
The current program is a 
first release of a long term project that aims at unifying the simulation 
of a wide class of high energy QCD processes, ranging from elementary 
particle physics processes, e.g., $e^+e^-$-fragmentation, deep inelastic 
$ep$-scattering, via hadronic collisions such as $pp$ ($p\bar{p}$), 
to reactions involving (heavy) nuclei, 
e.g. deep inelastic $eA$ scattering, $pA$, or $AA$ ($AB$) collisions.
The types of collision processes that are available 
and are discussed in more detail
in Section 3.4 below, include
collisions involving
$e^\pm,\mu^\pm$ for leptons, $p,n,\pi,\Lambda,\Sigma\ldots$ for hadrons,
and
$\D\e, \H\e, {\rm O}, \ldots,
\A\u, {\rm P}\b, {\rm U}$ for nuclei.
Some of these collision processes have of course been 
investigated before; here 
they are revisited within the space-time picture of 
parton cascades and cluster dynamics in order to provide an alternative 
method for Monte Carlo simulation and also to check the consistency with 
the wide literature on both experimental measurements and theoretical 
predictions. 

\subsection{Overall structure of the program}

  The program package VNI is written entirely in Fortran 77, and should 
run on any machine with such a compiler.  The program VNI adopts the 
common block structure, particle identification codes, etc.,
from the Monte Carlos of the Lund family, e.g.,
the PYTHIA program \cite{jetset}. 
In addition, various program elements 
of the latter are used in the sense of a library, 
however modified to incorporate 
the additional aspects of the space-time and color label information. 
VNI also employs certain (significantly modified) parts of the 
HERWIG Monte Carlo \cite{herwig}, used for the 
final decays of clusters into hadronic resonances and stable particles.
It is important to stress that the correspondence of 
VNI with PYTHIA and HERWIG is rather loose and should not 
lead to naive identification, in particular since VNI is taylored for a 
larger set of degrees of freedom that is necessary to describe the 
space-time dynamics and the information on color correlations. 
Nevertheless, by truncating the additional information of the particle 
record of VNI, a direct interface to the particle records of the Lund 
Monte Carlo PYTHIA, and to the program HERWIG, is actually 
straightforward. Instructions are given below in  Section 3.10.
\smallskip

  The package VNI consists of the following files:

\begin{description}
\item{(1)}  the documentation {\bf vni.ps} (this one)
\item{(2)}  the program VNI: {\bf vni.f}
\item{(3)}  three include files {\bf vni1.inc}, {\bf vni2.inc}, {\bf vni3.inc}
\end{description}

  Further useful, but supplementary software (see http://rhic.phys.columbia.edu/rhic/vni):
\begin{description}
\item{(4)}  various example programs for applications of VNI.
\item{(5)}  a simple, portable histogram package {\bf vnibook.f }.
\item{(5)}  a program to insert 2-particle probes into VNI, {\bf vniprob.f }.
\item{(5)}  a program to study Bose-Einstein physics and HBT, {\bf vnihbt.f }.
\item{(6)}  the PYTHIA program {\bf pythia.f}, to serve as interface with 
            VNI if desired
\end{description}

  The program {\it vni.f} is the Fortran source code that
contains all the subroutines, functions, and block data, as categorized
above. All default settings and particle data are automatically loaded 
by including the three include files {\it vni1.inc} - {\it vni3.inc}.
The example programs, as well as the histogram package
{\it vnibook.f}, the programs {\it vniprob.f}, {\it vnihbt.f},  
and {\it pythia.f}, are not integral part of the program, i.e. are 
distributed supplementary as useful guiding tools.
\medskip

  The program source code {\it vni.f} is organized in 9 parts, in 
which subroutines and functions that are related in their performance duties, 
are grouped together
(a summary list of all components with a brief description of their 
purpose is given in Appendix E):

\begin{description}
\item[part 1:]
            Main steering routines VNIXRIN (initialization), VNIXRUN 
           (event generation), and VNIXFIN (finalization), plus other
	    general-duty subroutines.
           VNIXRIN and VNIXFIN are only to be called once, while
	    VNIXRUN
           is to be called anew for each event, in which the 
           particle system is evolved in discrete time steps and 
           7-dimensional phase-space.
\item[part 2:]
           Initialization routines. They include, e.g. the process
           dependent event-by-event initialization of the chosen type 
           of collision process, and various other slave routines 
           associated with the initial set up of an event.
\item[part 3:]
           Package for parton distributions (structure functions). A
           portable and self-contained collection of routines with
           a number of structure function parametrizations, plus a
           possible interface to the CERN PDFLIB, plus various slave 
           routines. 
\item[part 4:]
           Evolution routines for the parton cascade / cluster-hadronization 
	   simulation, including both (process-dependent) 
           particle evolution routines and  (process-independent)
           universal routines for space-like and time-like shower evolution, for
           parton scatterings, parton recombination to clusters, and 
           cluster decays to hadrons and other final state particles.
\item[part 5:]
           Diverse utility routines and functions performing the 
           slave-work for the part of perturbative parton evolution.
\item[part 6:]
           Diverse utility routines and functions performing the 
           slave-work for the part of cluster fragmentation to hadrons.
\item[part 7:]
           Evolution routines for the `afterburner' hadron cascade as
	   a self-contained integrated package. This part was developed
	   by R. Longacre, the {\it Wizard of Poquott}.
\item[part 8:]
           Event study and analysis routines for accumulating, analysing
           and printing out statistics on observables.
\item[part 9:]
           Dummy routines and system dependent utilities.
\end{description}
\medskip

\subsection{Special features and machine dependence}

Important to remark are the following special features, being 
not strictly common in standard Fortran 77:

\begin{description}
\item[1.]
  The program  uses  of a {\it machine-dependent timing routine},
called $VNICLOC$, which measures the elapsed CPU time 
\footnote{
The usage of the time measurement is not 
necessary for the performance of the program as-a-whole, although it 
is a nice convenience.}.
Two alternatives are provided: 
a) the $TIMEX$ routine of the CERN library, in which case the 
   latter must be linked with the program, or,
b) the $MCLOCK$ routine which is specific to IBM-AIX compilers. 
If neither of these options are desired, or does not work in the users 
local environment, the lines involving 
calls for $TIMEX$ ($MCLOCK$) calls can easily commented out.
\footnote{
The subroutine $VNICLOC$ is listed at the very end in {\it vni.f}.
}.
\item[2.]
All common block variables and dimensions are defined globally within
the {\it include files  vni1.inc, vni2.inc, vni3.inc}.
This greatly simplifies for instance possible modifications 
(e.g., enlarging or decreasing) of global common block arrays.
Consequently, all subroutines contain INCLUDE statements for these include
files, and therefore
when compiling and linking the program, the files 
{\it vni1.inc}, {\it vni2.inc}, and {\it vni3.inc}  must be properly accessible.
\item[3.]
  {\it Double precision} is assumed throughout since version 4.0. 
For applications at extremely high 
energies, single precision for any real variable starts to become a 
problem. The main source of numerical precision-loss arises from 
multiple Lorentz boosts of particles throughout the simulation. 
Therefore the code has been converted now to full double precision.
\item[4.]
{\it Implicit variable-type assignment} is assumed, such that variables
or functions beginning with letters I - M are of {\it integer}$\ast$8
type, and variables or functions with first letters A - H and O - Z
are of type {\it real}$\ast$8 (double precision).
\item[5.]
{\it  SAVE statements} have been included in accordance with the Fortran
standard. Since most ordinary machines take SAVE for granted, this
part is not particularly well tried out, however. Users on machines
without automatic SAVE are therefore warned to be on the lookout for
any variables which may have been missed.
\end{description}
\medskip

\subsection{The main subroutines}

  There is a minimum of three routines that a user must know: 
\begin{description}
\item[1.] VNIXRIN for the overall initialization of a sample of 
          collision events, 
\item[2.] VNIXRUN for the subsequent actual generation of each event,
\item[3.] VNIXFIN for finishing up simulation.
\end{description}
These three routines, which are briefly descibed below, 
are to be called in that order by a user's program as is exemplified by the
example program in Section 3.10.
\smallskip


\begin{verbatim}
SUBROUTINE VNIXRIN(NEVT,TFIN,FRAME,BEAM,TARGET,WIN)
\end{verbatim}

\noindent
{\it Purpose}: to initialize the overall simulation procedure.
\medskip

\begin{description}
\item[{\boldmath $NEVT$} :] 
    integer specifying the number of collision events to be 
    simulated for a selected physics process.
\item[{\boldmath $TFIN$} :] 
    final time (in $fm$) in the global Lorentz frame up to which each
    collision event is followed.
\item[{\boldmath $FRAME$} :] 
    a character variable used to specify the global Lorentz frame  of the
    experiment. Uppercase and lowercase letters may be freely mixed.
\\
{\bf = 'CMS' :} colliding beam experiment in CM frame, with beam momentum
        in +$z$ direction and target momentum in $-z$ direction.
\\
{\bf = 'FIXT':} fixed target experiment, with beam particle momentum
        pointing in $+z$ direction.
\\
{\bf = 'USER' :} full freedom to specify frame by giving beam
momentum in \tt{PSYS(1,1)}, \tt{PSYS(1,2)} and \tt{PSYS(1,3)} and target
momentum in \tt{PSYS(2,1)}, \tt{PSYS(2,2)} and \tt{PSYS(2,3)} in
commonblocks \tt{VNIREC1} and \tt{VNIREC2}. 
Particles are assumed on the mass shell,
and energies are calculated accordingly.
\\
{\bf = 'FOUR' :} as \tt{'USER'}, except also energies should be
specified, in \tt{PSYS(1,4)} and \tt{PSYS(2,4)}, respectively. The
particles need not be on the mass shell; effective masses are 
calculated from energy and momentum. (But note that numerical
precision may suffer; if you know the masses the option \ttt{'FIVE'}
below is preferrable.)
\\
{\bf = 'FIVE' :} as \tt{'USER'}, except also energies and masses
should be specified, i.e the full momentum information in
\tt{PSYS(1,1) - PSYS(1,5)} and \tt{PSYS(2,1) - PSYS(2,5)} should be given for
beam and target, respectively. Particles need not be on the mass
shell. Space-like virtualities should be stored as $-\sqrt{-m^2}$.
Four-momentum and mass information must match (see Appendix C for details). 
\item[{\boldmath $BEAM$, }] {\boldmath $TARGET$} : 
     character variables to specify beam and target particles.
    Uppercase and lowercase letters may be freely mixed. An
    antiparticle may be denoted either by "$\tilde{}$" (tilde) or 
    "$\bar{}$" (bar) at the end of
    the name. It is also possible to leave out the charge for neutron
    and proton. 
\\
{\bf = 'e--'   :} electron $e^-$.
\\
{\bf = 'e+'   :} positron $e^+$.
\\
{\bf = 'mu--'  :} muon $\mu^-$.
\\
{\bf = 'mu+'  :} antimuon $\mu^+$.
\\
{\bf = 'gamma':} real photon $\gamma$ (not yet implemented).
\\
{\bf = 'pi+'  :} positive pion $\pi^+$.
\\
{\bf = 'pi--'  :} negative pion $\Pi^-$.
\\
{\bf = 'pi0'  :} neutral pion $\pi^0$.
\\
{\bf = 'n0'   :} neutron $n$.
\\
{\bf = 'n{\boldmath $\tilde{}$}0'  :} antineutron $\overline{n}$.
\\
{\bf = 'p+'   :} proton $p$.
\\
{\bf = 'p{\boldmath $\tilde{}$}--'  :} antiproton $\overline{p}$.
\\
{\bf = 'Lambda0':} $\Lambda$ baryon.
\\
{\bf = 'Sigma+' :} $\Sigma^+$ baryon.
\\
{\bf = 'Sigma--' :} $\Sigma^-$ baryon.
\\
{\bf = 'Sigma0' :} $\Sigma^0$ baryon.
\\
{\bf = 'Xi--'   :} $\Xi^-$ baryon.
\\
{\bf = 'Xi0'   :} $\Xi^0$ baryon.
\\
{\bf = 'Omega--':} $\Omega^-$ baryon.
\\
{\bf = 'De' :} $^2_{1}\D\e$ (Deuterium) nucleus.
\\
{\bf = 'He' :}  $^4_2\H\e$ (Helium) nucleus.
\\
{\bf = 'Ox' :}  $^{16}_8{\rm O}$ (Oxygen) nucleus.
\\
{\bf = 'Al' :}  $^{27}_{13}{\rm Al}$ (Aluminium) nucleus.
\\
{\bf = 'Su' :}  $^{32}_{16}{\rm S}$ (Sulfur) nucleus.
\\
{\bf = 'Ca' :}  $^{40}_{20}{\rm Ca}$ (Calcium) nucleus.
\\
{\bf = 'Cu' :}  $^{64}_{29}{\rm Cu}$ (Copper) nucleus.
\\
{\bf = 'Kr' :}  $^{84}_{36}{\rm Kr}$ (Krypton) nucleus.
\\
{\bf = 'Ag' :}  $^{108}_{47}\A\g$ (Silver) nucleus.
\\
{\bf = 'Ba' :}  $^{137}_{56}{\rm Ba}$ (Barium) nucleus.
\\
{\bf = 'Wt' :}  $^{184}_{74}{\rm W}$ (Tungsten) nucleus.
\\
{\bf = 'Pt' :}  $^{195}_{78}{\rm Pt}$ (Platinium) nucleus.
\\
{\bf = 'Au' :}  $^{197}_{79}\A\u$ (Gold) nucleus.
\\
{\bf = 'Pb' :}  $^{208}_{82}{\rm P}\b$ (Lead) nucleus.
\\
{\bf = 'Ur' :}  $^{238}_{92}{\rm U}$ (Uranium) nucleus.
\item[{\boldmath $WIN$} :] 
    related to energy of system (in GeV). Exact meaning depends on FRAME, 
    i.e. for
    FRAME='CMS' it is the total energy $\sqrt{s}$ of system, and for
    FRAME='FIXT' it is the  absolute 3-momentum $P = \sqrt{\vec{P}^2}$ of 
    beam particle.
\end{description}
\bigskip

\begin{verbatim}
SUBROUTINE VNIXRUN(IEVT,TRUN)
\end{verbatim}
\medskip

\noindent
{\it Purpose}: to generate one event of the type specified by the VNIXRIN. 
    This is the main routine, which administers the overall
    run of the event generation and calls a number of other routines 
    for specific tasks.

\begin{description}
\item[{\boldmath $IEVT$} :] 
    integer labeling the current event number.
\item[{\boldmath $TRUN$} :] 
    time (in $fm$) $TRUN\le TFIN$ in the global Lorentz frame up to which 
    the space time evolution
    of the current event $IEVT$ is carried out. If $TRUN < TFIN$, then
    the particle record contains the history up to this time,
    and repeated calls resume from $TRUN$ of the previous call 
    (c.f. Section 3.10).
\end{description}
\medskip

\begin{verbatim}
SUBROUTINE VNIXFIN()
\end{verbatim}

\noindent
{\it Purpose}:
    to finish up overall simulation procedure, to write data 
    analysis results to files, evaluate CPU time, close files, etc..
    This routine must be called after the last event has been
    finished (c.f. Section 3.10).

\bigskip
\bigskip

For a deeper understanding of the 
physics routines and their connecting structure,
I refer to Appendix A for a brief description.
Generally, for 
each of the included collision processes, there is an initialization 
routine that sets up the initial state, and an evolution 
routine  that carries out the time evolution of the particle 
distributions in phase-space, starting from the initial state. 
Within the evolution routines, a number of 
universal slave routines (i.e. independent of the process under consideration)
perform the perturbative parton evolution in terms of space-like branchings, 
time-like branchings, and parton collisions, as well as
the cluster hadronization. 
\bigskip

  Finally, the program provides some useful
pre-programmed event-analysis routines which collect, analyze and print out information
on the result of a simulation, and give the user immediate access
to calculated particle data, spectra, and other observables. 
A more detailed description of their duties, and how
to call them can be found in Appendix B.
\newpage

\begin{table}[ptb]
\captive{Available collision processes.
\protect\label{t:ipro} } \\
\vspace{1ex}
\begin{center}
\begin{tabular}{|rc|cl|}
\hline\hline
    IPRO  &&&             type of collision process
\\
\hline\hline
&&& \\
     1    &&&       $l^+l^- \rightarrow  \gamma/Z_0  \rightarrow 2-jets \rightarrow hadrons$
\\
     2    &&&       $l^+l^- \rightarrow Z_0 \rightarrow W^+W^- \rightarrow 4-jets \rightarrow hadrons \;\;\;\;\;\;$
\\
     3    &&&       $\gamma + h \rightarrow jets \rightarrow hadrons$      
\\
     4    &&&       $\gamma + A \rightarrow jets \rightarrow hadrons$     
\\
     5    &&&       $l + h  \rightarrow  l + jets + X  \rightarrow  hadrons$ 
\\
     6    &&&       $l + A  \rightarrow  l + jets + X  \rightarrow  hadrons$ 
\\
     7    &&&       $h + l  \rightarrow  l + jets + X  \rightarrow  hadrons$ 
\\
     8    &&&       $h + h'  \rightarrow  jets + X  \rightarrow  hadrons$ 
\\
     9    &&&       $h + A  \rightarrow  jets + X  \rightarrow  hadrons$ 
\\
    10    &&&       $A + l  \rightarrow  l + jets + X  \rightarrow  hadrons$
\\
    11    &&&       $A + h  \rightarrow  jets + X  \rightarrow  hadrons$ 
\\
    12    &&&       $A + A'  \rightarrow  jets + X  \rightarrow  hadrons$ 
\\ &&& 
\\ \hline\hline
\end{tabular}
\end{center}
\end{table}

\begin{table}[ptb]
\captive{Available beam and target particles.
\protect\label{t:bt} } \\
\vspace{1ex}
\begin{center}
\begin{tabular}{|c|c|c|}
\hline\hline
  $\;\;\;$ leptons $l$ $\;\;\;$ & $\;\;\;$ 
hadrons $h$ $\;\;\;$ & $\;\;\;$ nuclei $A$ $\;\;\;$
\\
\hline\hline
&& \\
$ e^\pm $ & $\pi^\pm, \pi^0$ &  $^2_{1}\D\e$
\\
$ \mu^\pm $ & $n, \overline{n}$ &  $^4_2\H\e$ 
\\
$  $ & $p, \overline{p}$ &  $^{16}_8{\rm O}$
\\
$  $ & $\Lambda^0$ &  $^{32}_{16}{\rm S}$
\\
$  $ & $\Sigma^\pm, \Sigma^0$ &  $^{40}_{20}{\rm Ca}$
\\
$  $ & $\Xi^-, \Xi^0$ &  $^{64}_{29}{\rm Cu}$ 
\\
$  $ & $\Omega^-$ &  $^{84}_{36}{\rm Kr}$ 
\\
$  $ & $  $  &  $^{108}_{47}{\rm Ag}$ 
\\
$  $ & $  $  &  $^{137}_{56}{\rm Ba}$ 
\\
$  $ & $  $  &  $^{184}_{74}{\rm Wt}$ 
\\
$  $ & $  $  &  $^{195}_{78}{\rm Pt}$ 
\\
$  $ & $  $  &  $^{197}_{79}{\rm Au}$ 
\\
$  $ & $  $  &  $^{208}_{82}{\rm Pb}$ 
\\
$  $ & $  $  &  $^{238}_{92}{\rm U}$ 
\\ && 
\\ \hline\hline
\end{tabular}
\end{center}
\end{table}

\subsection{The physics processes}

The program is structured to incorporate different classes of particle 
collision processes in a modular manner. The 
general classes that are part of the program are summarized in
Table 1.
All of the generic collision processes in Table 1
 allow a further specification 
of beam and/or target particle, .e.g., $l = e, \mu$ for leptons, 
$h = p, n, \pi, \ldots$ for hadrons, or $A =De, He, O, 
\ldots$  for nuclei. They are summarized in Table 2.
\medskip

As explained before, the
actual evolution of any beam/target-particle collision
is simulated
on microscopic level of the system of partons, clusters and hadrons, and 
their interactions. For the parton interaction processes, a selection 
of elementary hard/soft $2\rightarrow 2$ scatterings and $2\rightarrow 1$ 
fusions are included 
in VNI, and future extensions are planned. In addition 
there are the standard space-like and time-like $1\rightarrow 2$ branching 
processes, which, when combined with the elementary tree processes, 
provide the usual parton cascade method of including
higher order, real and virtual, corrections to the 
elementary tree processes. The parton-cluster
formation and hadronization scheme includes a number of 2-parton
recombination processes and the decay of the formed clusters into
final state hadron. 

\begin{table}[ptb]
\captive{Elementary partonic subprocesses.
\protect\label{t:isub} } \\
\vspace{1ex}
\begin{center}
\begin{tabular}{|lc|cc|}
\hline\hline
$ \;\;\;$ Class &    ISUB $\;\;\;\;\;$ &&             type of subprocess $\;\;\;\;$
\\
\hline\hline
&&& \\
  a) $2 \rightarrow> 2$ : 
&                  1  &&  $ q_i q_j \rightarrow  q_i q_j$
\\
&                  2  &&  $ q_i \bar{q}_i \rightarrow  q_k \bar{q}_k    $    
\\
&                  3  &&  $ q_i \bar{q}_i \rightarrow  g g         $   
\\
&                  4  &&  $ q_i \bar{q}_i \rightarrow  g \gamma     $  
\\
&                  5  &&  $ q_i \bar{q}_i \rightarrow  \gamma \gamma $ 
\\
&                  6  &&  $ q_i g \rightarrow  q_i g         $ 
\\
&                  7  &&  $ q_i g \rightarrow  q_i \gamma    $ 
\\
&                  8  &&  $ g g  \rightarrow  q_k \bar{q}_k      $ 
\\
&                  9  &&  $ g g  \rightarrow  g g          $ 
\\
&                 10  &&   soft  scattering 
\\
&&& \\
    b) $2 \rightarrow 1$ :
&                 11  &&  $ q_i \bar{q}_i \rightarrow  g^\ast$
\\
&                 12  &&  $ q_i g \rightarrow  q_i^\ast$
\\
&                 13  &&  $ g g  \rightarrow  g^\ast$
\\
&&& \\
    c) $1 \rightarrow 2$ (space-like) :
&                 21  &&  $ g^\ast \rightarrow  q_i q_i$
\\
&                 22  &&  $ q_i^\ast \rightarrow  q_i g$
\\
&                 23  &&  $ g^\ast \rightarrow  g g$
\\
&                 24  &&  $ q_i^\ast \rightarrow  q_i \gamma$
\\
&&& \\
    d) $1 \rightarrow 2$ (time-like) :
&                 31  &&  $ g^\ast \rightarrow  q_i \bar{q}_i$
\\
&                 32  &&  $ q_i^\ast \rightarrow  q_i g$
\\
&                 33  &&  $ g^\ast \rightarrow  g g$
\\
&                 34  &&  $ q_i^\ast \rightarrow  q_i \gamma$
\\ 
&&&
\\
\hline\hline
\end{tabular}
\end{center}
\end{table}
 
It is possible to select a combination of partonic subprocesses to 
simulate. For this purpose, all subprocesses are numbered according
to an $ISUB$ code. The list of allowed codes is given below.
In the following $g$ denotes a gluon, $q_i$ represents a quark of 
flavour $i=1,\ldots,n_f$, i.e. for $n_f=6$ this translates
 to $d, u, s, c, b, t$. 
A corresponding antiquark 
is denoted $\bar{q}_i$. The notation $\gamma$ is for a real photon, i.e. on shell.  
An asterix $\ast$ denotes an off-shell parton.
\bigskip

\subsection{The particle record}

  Each new event generated is in its entity stored in the commonblocks
VNIREC1 and VNIREC2, which thus forms the event record. Here each particle that
appears at some stage of the time evolution of the system, will
occupy one line in the matrices. The different components of this line
will tell which particle it is, from where it originates, its
present status (fragmented/decayed or not), its momentum, energy and
mass, and the space-time position of its production vertex.
The structure of the particle record VNIREC1/VNIREC2 follows closely the
one of the Lund Monte carlos \cite{jetset,ariadne,lepto},
employing the same overall particle classification of
particle identification, status code, etc.,
however with important differences
concerning color- and space-time
degrees of freedom..

  Note: When in the following reference is made to certain switches 
and parameters MSTV or PARV, these are described in Sec. 3.7 below. 

\begin{verbatim}
     PARAMETER (NV=100000)
     COMMON/VNIREC1/N,K(NV,5),L(NV,2)
     COMMON/VNIREC2/P(NV,5),R(NV,5),V(NV,5)
\end{verbatim}

\noindent
{\it Purpose}: 
Contains the event record, i.e. the complete list of all 
         partons and particles in the current event.
\medskip

\begin{description}
\item[{\boldmath $N$} :]  number of lines in the $K$, $L$, $P$, $R$, and $V$ 
    arrays occupied by the current
    event. $N$ is continuously updated as the definition of the original
    configuration and the treatment of parton cascading, cluster fragmentation
    and hadron production proceed.
    In the following, the individual parton/particle number, running
    between 1 and $N$, is called $I$.
    The maximum $N$ is limited by the dimension $NV$ of the arrays.
\item[{\boldmath $K(I,$}1) :] status code ($KS$), which gives the current status of the
    parton/particle stored in the line. The ground rule is that codes
    1 - 10 correspond to currently existing partons/particles, while
    larger codes contain partons/particles which no longer exist.
\\
{\bf = 0 :} empty line.
\\
{\bf = 1 :} an undecayed particle or an unfragmented parton.
\\
{\bf = 2 :} an unfragmented parton, which is followed by more partons in 
        the same color-singlet parton subsystem.
\\
{\bf = 3 :} an unfragmented parton with special color-flow information
        stored in $K(I,4)$ and $K(I,5)$, such that color connected partons 
        need not follow after each other in the event record.
\\
{\bf = 4 :} an unfragmented parton 1, associated with an externally
        injected 2-particle probe, i.e. $q\bar q$, $gg$, $q\gamma$
        (only in conjunction with the package VNIEVPR).
\\
{\bf = 5 :} an unfragmented parton 2, associated with an externally
        injected 2-particle probe, i.e. $q\bar q$, $gg$, $q\gamma$
        (only in conjunction with the package VNIEVPR).
\\
{\bf = 6 :} a final state hadron resulting from decay of clusters
       that have formed from materialized (interacted) partons.
\\
{\bf = 7 :} a final state hadron resulting from soft cluster decay of 
     beam/target remnants that have formed from left-over (non-interacted)
     initial state partons.
\\
{\bf = 8 :} a final state hadron resulting from bound-state formation 
       of $q\bar q$ pairs as externally inserted probes 
        (only in conjunction with the package VNIEVPR).
\\
{\bf = 9 :} a final state hadron resulting from jet fragmentation
       of $q$, $\bar q$, or $g$ as externally inserted probes 
        (only in conjunction with the package VNIEVPR).
\\
{\bf = 10 :} a prehadronic cluster that has been reconstructed
             from its original hadronic decay products if those
             have not passed through their formation time yet.
\\
{\bf = 11 :} a decayed particle or a fragmented parton, c.f. = 1.
\\
{\bf = 12 :} a fragmented jet, which is followed by more partons in the 
        same color-singlet parton subsystem, c.f. = 2.
\\
{\bf = 13 :} a parton which has been removed when special color-flow
        has been used to rearrange a parton subsystem, cf. = 3.
\\
{\bf = 14 :} as = 4, but decayed or fragmented.
\\
{\bf = 15 :} as = 5, but decayed or fragmented.
\\
{\bf = 16 :} an unstable and removed hadron resulting from decay of clusters
       that formed from materialized (interacted) partons, c.f. = 6.
\\
{\bf = 17 :} an unstable and removed hadron 
     resulting from soft cluster decay of 
     beam/target remnants formed from left-over (non-interacted)
     initial state partons, c.f. = 7.
\\
{\bf = 18 :} as = 8, but decayed or fragmented.
\\
{\bf = 19 :} as = 9, but decayed or fragmented.
\\
{\bf = 20 :} as = 10, but decayed or fragmented.
\\
{\bf {\boldmath $<$} 0 :}
     Particle entries with negative status codes (e.g. -1, -6, -7) parallel
     in their meaning those from above, but refer to particles
     which are only virtually present and become active only after
     a certain formation time, upon which their status code is set to
     the appropriate positive value.
\\
\item[{\boldmath $K(I,$}2) :] 
    Flavor code ($KF$) for partons, hadrons and electromagnetic 
    particles included in the current version of VNI.  The particle code 
    is summarized in Tables 3 -- 9. It is based on
    the flavor and spin classification of particles, following the 1988
    Particle Data Group numbering conventions. Nuclei are attributed
    a non-standard code that is internal to VNI, given by 
    $KF = 1000000 + 1000 \cdot N_{neutrons} + N_{protons}$. 
    A negative $KF$ code, where existing, 
    always corresponds to the antiparticle of the one listed in the Tables
    3 -- 9.
\item[{\boldmath $K(I,$}3) :] 
    line number of parent particle or jet, where known, else 0.
    Note that the assignment of a particle to a given jet of a jet
    system is unphysical, and what is given there is only related to
    the way the event was generated.
\item[{\boldmath $K(I,$}4) :] 
    Special color-flow information (for internal use only) of the 
    form $K(I,4) =$ MSTV(2)$\cdot  ICFR + ICTO$, where $ICFR$ and $ICTO$ give 
    the line numbers of the partons {\it from}
     which the color comes and {\it to} where it goes, respectively. 
\item[{\boldmath $K(I,$}5) :] 
    Special color-flow information (for internal use only) of 
    the form $K(I,5) =$ MSTV(2)$\cdot JCFR + JCTO$, where 
    $JCFR$ and $JCTO$ give 
    the line numbers of the partons {\it from} which the anticolor comes 
    and {\it to} where it goes, respectively. 
\\
\item[{\boldmath $L(I,$}1) :] 
    color label ($L=1,\ldots, NC$) of a parton, where $NC$ is number of
    colors specified by the parameter MSTV(5). Is = 0 for antiquarks 
    and all non-colored particles. 
\item[{\boldmath $L(I,$}2) :] 
    anticolor label ($L=1,\ldots, NC$) of a parton. Is = 0 for quarks 
    and all non-colored particles.
\\
\item[{\boldmath $P(I,$}1) :] 
    $p_x$, momentum in the $x$ direction in GeV.
\item[{\boldmath $P(I,$}2) :] 
     $p_y$, momentum in the $y$ direction in GeV.
\item[{\boldmath $P(I,$}3) :] 
     $p_z$, momentum in the $z$ direction in GeV.
\item[{\boldmath $P(I,$}4) :] 
     $E$, energy, in GeV.
\item[{\boldmath $P(I,$}5) :] 
     $m = p^2$, mass in GeV. For partons with space-like virtualities, 
    i.e. where $Q^2 = - m^2 > 0$, its value is 
    $P(I,5) = -\sqrt{|m^2|} = -Q$.
\\
\item[{\boldmath $R(I,$}1) :] 
    current $x$ position in frame of reference in 1/GeV.
\item[{\boldmath $R(I,$}2) :] 
    current $y$ position in frame of reference in 1/GeV.
\item[{\boldmath $R(I,$}3) :] 
    current $z$ position in frame of reference in 1/GeV.
\item[{\boldmath $R(I,$}4) :] 
    time (in 1/GeV) of active presence in the system, 
    i.e. $t_{pres}= t - t_{prod}$. 
    Is set equal to PARW(2) at time of production $t_{prod}$ of the
    particle and then increased
    by PARW(2) in each timestep until the particle fragments or decays.
    If $ \langle$ 0, giving the remaining formation time for virtually produced,
    but not yet real particles. 
\item[{\boldmath $R(I,$}5) :] 
    number and type of interactions of a particle, and is equal to
    MSTV(3)$^2\cdot NCO +$ MSTV(3)$\cdot NSB + NTB$, 
    where -- for partons -- $NCO$ is 
    the number of 2-body collisions, $NSB$ of space-like branchings, and 
    $NTB$ of time-like branchings, undergone up to current time. For 
    clusters, hadrons and unstable particles, $NSB$ is zero and $NTB$ 
    counts the number of 2-body decays.
\\
\item[{\boldmath $V(I,$}1) :] 
    $x$ position of production vertex, in 1/GeV.
\item[{\boldmath $V(I,$}2) :] 
    $y$ position of production vertex, in 1/GeV.
\item[{\boldmath $V(I,$}3) :] 
    $z$ position of production vertex, in 1/GeV.
\item[{\boldmath $V(I,$}4) :] 
    time $t$ of production, in 1/GeV.
\item[{\boldmath $V(I,$}5) :] 
    encodes origin of production as 
    $10\cdot$MSTV(4)$\cdot IMO + 10\cdot ITM + OR$, 
    where $IMO$ is the direct mother, $ITM$ the time of production, and $OR$ 
    the generation of particle $IP$ in a cascade tree, i.e. the genetical 
    origin with respect to the production vertex of its original 
    `Ur'-mother. (This keeps track of the production history even if the 
    decayed mother has been removed and allows to follow the genealogy
    of the system.)
\end{description}
\medskip

\noindent
{\bf  Special cases:} 
\smallskip

  For the simulation of collisions involving hadrons or nuclei in 
the initial state, additional initial state information on the 
status and specific variables of "primary partons" of the incoming 
hadron (nucleus) are stored as follows. Note however, that after
the first interaction of a "primary parton", this initial state
information is lost and the assignments for the components of 
the arrays $K,R,V$ are the same as above.
\begin{description}
\item[{\boldmath $K(I,$}4) :] 
    $KF$ flavour code of hadron (nucleon), which is mother of 
    the primary parton.
\item[{\boldmath $K(I,$}5) :] 
    $KF$ flavour code of nucleus from which a primary parton originates.
    (For hadrons $K(I,4)$ and $K(I,5)$ are equal). 
\item[{\boldmath $R(I,$}5) :] 
    Is equal to $-|x|$, where $x$ is the longitudinal momentum 
    fraction of primary parton that it carries off its mother hadron. 
\item[{\boldmath $V(I,$}4) :] 
    encodes location $IB$ in the particle record of a primary 
    sea-quark's brother, i.e. the sea-antiquark belonging to the same
    vacuum polarization loop,  as 
    MSTV(4)$^2\cdot IB + $MSTV(4)$\cdot ITM + OR$,
    analogous to  $V(I,5)$ above. For primary valence quarks and gluons 
    it is 0.
\end{description}
{\it  Additional remark}: The arrays $K(I,3)-K(I,5)$ and the arrays $P, R$, 
and $V$ may temporarily take special meaning other than the above, 
for some specific internal use.
\bigskip

In Section 3.10 below, a typical event listing of the
particle record is printed, which 
serves as an example for the organization of the 
particle record, exhibiting the information for 
{\it momentum space}:
It lists the particles 
contained in VNIREC1/VNIREC2 
at a certain point during the evolution of an event, 
where $KS$ and $KF$ give status and flavor codes, $C$ and $A$ the color and 
anticolor index, $P$ the energy-momentum-mass in GeV
for each particle that appeared at some point in the event history.
An analogous  listing for  the {\it position space}
particle record (not printed here) contains the spatial
coordinates and time of active presence for each particle, as well as 
the particle's rapidity and transverse momentum
with respect to the $z$-axis (jet-axis, or beam-axis).
Note: both listings 
can  be obtained at any point by calling the routine VNILIST,
described in Appendix B.

\begin{table}[thb]
\captive{Quark, lepton and gauge boson codes.
\protect\label{t:codeone} } \\
\vspace{1ex}
\begin{center}
\begin{tabular}{|c|c|c||c|c|c||c|c|c|@{\protect\rule{0mm}{\tablinsep}}}
\hline \hline
KF & Name & Printed & KF & Name & Printed & KF & Name & Printed \\
\hline\hline
    1 & $\d$ & \ttt{d}   &  11 & $\e^-$ &  \ttt{e-}  & 21 & $\g$ & \ttt{g} \\  
    2 & $\u$ & \ttt{u}   &  12 & $\nu_{\e}$ &  \ttt{nu\_e} & 22 & $\gamma$ & \ttt{gamma} \\
    3 & $\s$ & \ttt{s}   &  13 & $\mu^-$ &  \ttt{mu-}  & 23 & $\Z^0$ & \ttt{Z0} \\
    4 & $\c$ & \ttt{c}   &  14 & $\nu_{\mu}$ & \ttt{nu\_mu} & 24 & $\W^+$ & \ttt{W+} \\
    5 & $\b$ & \ttt{b}   &  15 & $\tau^-$ & \ttt{tau-}  & 25 & $\H^0$ & \ttt{H0} \\
    6 & $\t$ & \ttt{t}   &  16 & $\nu_{\tau}$ &  \ttt{nu\_tau}   & 26 & &  \\
    7 & & & 17 & & & 27 & & \\
    8 & & & 18 & & & 28 & & \\
    9 & & & 19 & & & 29 & &\\
   10 & & & 20 & & & 30 & &\\
\hline \hline
\end{tabular}
\end{center}
\end{table}
\begin{table}[bhp]
\captive{Special codes.
\protect\label{t:codetwo} }   \\
\vspace{1ex}
\begin{center}
\begin{tabular}{|c|c|c|@{\protect\rule{0mm}{\tablinsep}}}
\hline \hline
KF & Printed & Meaning \\
\hline \hline
   90 &  \ttt{CMsoft}  & Center-of-mass of beam/target remnant system before
	 soft fragmentation  \\
   91 &  \ttt{cluster}  & Cluster from coalescence of secondary partons  \\
   92 &  \ttt{Beam-REM}   & Cluster from remnant primary partons of beam particle\\
   93 &  \ttt{Targ-REM}   & Cluster from remnant primary partons of target particle\\
   94 &  \ttt{CMshower} & Four-momentum of time-like showering system  \\
   95 &  \ttt{SPHEaxis} & Event axis found with \ttt{VNISPHE}  \\
   96 &  \ttt{THRUaxis} & Event axis found with \ttt{VNITHRU}  \\
   97 &  \ttt{CLUSjet}  & Jet (cluster) found with \ttt{VNICLUS}  \\
   98 &  \ttt{CELLjet}  & Jet (cluster) found with \ttt{VNICELL}  \\
   99 &      &  \\
  100 &           &   \\
\hline \hline
\end{tabular}
\end{center}
\end{table}

\newpage
 
{\footnotesize
\begin{table}[thb]
\captive{Meson codes, part 1.
\protect\label{t:codethree} }  \\
\vspace{1ex}
\begin{center}
\begin{tabular}{|c|c|c||c|c|c|@{\protect\rule{0mm}{\tablinsep}}}
\hline \hline
KF & Name & Printed & KF & Name & Printed \\
\hline \hline
  211 & $\pi^+$ & \ttt{pi+}        & 213 & $\rho^+$ & \ttt{rho+}  \\
  311 & $\K^0$ & \ttt{K0}          & 313 & $\K^{*0}$ & \ttt{K*0}  \\
  321 & $\K^+$ & \ttt{K+}          & 323 & $\K^{*+}$ & \ttt{K*+}  \\
  411 & $\D^+$ & \ttt{D+}          & 413 & $\D^{*+}$ & \ttt{D*+}  \\
  421 & $\D^0$ & \ttt{D0}          & 423 & $\D^{*0}$ & \ttt{D*0}  \\
  431 & $\D_s^+$ & \ttt{D\_s+}     &
433 & $\D_{\s}^{*+}$ & \ttt{D*\_s+}  \\
  511 & $\B^0$ & \ttt{B0}          & 513 & $\B^{*0}$ & \ttt{B*0}  \\
  521 & $\B^+$ & \ttt{B+}          & 523 & $\B^{*+}$ & \ttt{B*+}  \\
  531 & $\B_s^0$ & \ttt{B\_s0}     &
533 & $\B_{\s}^{*0}$ & \ttt{B*\_s0}  \\
  541 & $\B_c^+$ & \ttt{B\_c+}     &
543 & $\B_{\c}^{*+}$ & \ttt{B*\_c+}  \\
  111 & $\pi^0$ & \ttt{pi0}        & 113 & $\rho^0$ & \ttt{rho0}  \\
  221 & $\eta$ & \ttt{eta}         & 223 & $\omega$ & \ttt{omega}  \\
  331 & $\eta'$ & \ttt{eta'}       & 333 & $\phi$ & \ttt{phi}  \\
  441 & $\eta_{\c}$ & \ttt{eta\_c} & 443 & $\Jpsi$ & \ttt{J/psi}  \\
  551 & $\eta_{\b}$ & \ttt{eta\_b} &
553 & $\Upsilon$ & \ttt{Upsilon}  \\
  661 & $\eta_{\t}$ & \ttt{eta\_t} & 663 & $\Theta$ & \ttt{Theta}  \\
  130 & $\K_{\mrm{L}}^0$ & \ttt{K\_L0} & & &  \\
  310 & $\K_{\mrm{S}}^0$ & \ttt{K\_S0} & & &  \\
\hline \hline
\end{tabular}
\end{center}
\end{table}
} 

\vspace{-0.75cm}

{\footnotesize
\begin{table}[thb]
\captive{Meson codes, part 2.
\protect\label{t:codefour} }  \\
\vspace{1ex}
\begin{center}
\begin{tabular}{|c|c|c||c|c|c|@{\protect\rule{0mm}{\tablinsep}}}
\hline \hline
KF & Name & Printed & KF & Name & Printed \\
\hline \hline
10213 & $\b_1$ & \ttt{b\_1+}              &
10211 & $\a_0^+$ & \ttt{a\_0+}  \\
10313 & $\K_1^0$ & \ttt{K\_10}            &
10311 & $\K_0^{*0}$ & \ttt{K*\_00}  \\
10323 & $\K_1^+$ & \ttt{K\_1+}            &
10321 & $\K_0^{*+}$ & \ttt{K*\_0+}  \\
10413 & $\D_1^+$ & \ttt{D\_1+}            &
10411 & $\D_0^{*+}$ & \ttt{D*\_0+}  \\
10423 & $\D_1^0$ & \ttt{D\_10}            &
10421 & $\D_0^{*0}$ & \ttt{D*\_00}  \\
10433 & $\D_{1 \s}^+$ & \ttt{D\_1s+}      &
10431 & $\D_{0 \s}^{*+}$ & \ttt{D*\_0s+}  \\
10113 & $\b_1^0$ & \ttt{b\_10}            &
10111 & $\a_0^0$ & \ttt{a\_00}  \\
10223 & $\hrm_1^0$ & \ttt{h\_10}            &
10221 & $\f_0^0$ & \ttt{f\_00}  \\
10333 & $\hrm'^0_1$ & \ttt{h'\_10}          &
10331 & $\f'^0_0$ & \ttt{f'\_00}  \\
10443 & $\hrm_{1 \c}^0$ & \ttt{h\_1c0}      &
10441 & $\chi_{0 \c}^0$ & \ttt{chi\_0c0}  \\  \hline
20213 & $\a_1^+$ & \ttt{a\_1+}            &
  215 & $\a_2^+$ & \ttt{a\_2+}  \\
20313 & $\K_1^{*0}$ & \ttt{K*\_10}        &
  315 & $\K_2^{*0}$ & \ttt{K*\_20}  \\
20323 & $\K_1^{*+}$ & \ttt{K*\_1+}        &
  325 & $\K_2^{*+}$ & \ttt{K*\_2+}  \\
20413 & $\D_1^{*+}$ & \ttt{D*\_1+}        &
  415 & $\D_2^{*+}$ & \ttt{D*\_2+}  \\
20423 & $\D_1^{*0}$ & \ttt{D*\_10}        &
  425 & $\D_2^{*0}$ & \ttt{D*\_20}  \\
20433 & $\D_{1 \s}^{*+}$ & \ttt{D*\_1s+}  &
  435 & $\D_{2 \s}^{*+}$ & \ttt{D*\_2s+}  \\
20113 & $\a_1^0$ & \ttt{a\_10}            &
  115 & $\a_2^0$ & \ttt{a\_20}  \\
20223 & $\f_1^0$ & \ttt{f\_10}            &
  225 & $\f_2^0$ & \ttt{f\_20}  \\
20333 & $\f'^0_1$ & \ttt{f'\_10}          &
  335 & $\f'^0_2$ & \ttt{f'\_20}  \\
20443 & $\chi_{1 \c}^0$ & \ttt{chi\_1c0}  &
  445 & $\chi_{2 \c}^0$ & \ttt{chi\_2c0}  \\  \hline
30443 & $\psi'$     & \ttt{psi'}     &   &   &  \\
30553 & $\Upsilon'$ & \ttt{Upsilon'} &   &   &  \\
\hline \hline
\end{tabular}
\end{center}
\end{table}
}
 
\begin{table}[thp]
\captive{Baryon codes.
\protect\label{t:codefive} }  \\
\vspace{1ex}
\begin{center}
\begin{tabular}{|c|c|c||c|c|c|@{\protect\rule{0mm}{\tablinsep}}}
\hline \hline
KF & Name & Printed & KF & Name & Printed \\
\hline \hline
      &  &                            &
 1114 & $\Delta^-$ & \ttt{Delta-} \\
 2112 & $\n$ & \ttt{n0}                     &
 2114 & $\Delta^0$ & \ttt{Delta0} \\
 2212 & $\p$ & \ttt{p+}                     &
 2214 & $\Delta^+$ & \ttt{Delta+} \\
      &  &                            &
 2224 & $\Delta^{++}$ & \ttt{Delta++} \\
 3112 & $\Sigma^-$ & \ttt{Sigma-}           &
 3114 & $\Sigma^{*-}$ & \ttt{Sigma*-} \\
 3122 & $\Lambda^0$ & \ttt{Lambda0}         & & &  \\
 3212 & $\Sigma^0$ & \ttt{Sigma0}           &
 3214 & $\Sigma^{*0}$ & \ttt{Sigma*0} \\
 3222 & $\Sigma^+$ & \ttt{Sigma+}           &
 3224 & $\Sigma^{*+}$ & \ttt{Sigma*+} \\
 3312 & $\Xi^-$ & \ttt{Xi-}                 &
 3314 & $\Xi^{*-}$ & \ttt{Xi*-} \\
 3322 & $\Xi^0$ & \ttt{Xi0}                 &
 3324 & $\Xi^{*0}$ & \ttt{Xi*0} \\
      &  &                            &
 3334 & $\Omega^-$ & \ttt{Omega-} \\
 4112 & $\Sigma_{\c}^0$ & \ttt{Sigma\_c0}   &
 4114 & $\Sigma_{\c}^{*0}$ & \ttt{Sigma*\_c0} \\
 4122 & $\Lambda_{\c}^+$ & \ttt{Lambda\_c+} & & &  \\
 4212 & $\Sigma_{\c}^+$ & \ttt{Sigma\_c+}   &
 4214 & $\Sigma_{\c}^{*+}$ & \ttt{Sigma*\_c+} \\
 4222 & $\Sigma_{\c}^{++}$ & \ttt{Sigma\_c++} &
 4224 & $\Sigma_{\c}^{*++}$ & \ttt{Sigma*\_c++} \\
 4132 & $\Xi_{\c}^0$ & \ttt{Xi\_c0}         & & &  \\
 4312 & $\Xi'^0_{\c}$ & \ttt{Xi'\_c0}       &
 4314 & $\Xi_{\c}^{*0}$ & \ttt{Xi*\_c0}  \\
 4232 & $\Xi_{\c}^+$ & \ttt{Xi\_c+}         & & &  \\
 4322 & $\Xi'^+_{\c}$ & \ttt{Xi'\_c+}        &
 4324 & $\Xi_{\c}^{*+}$ & \ttt{Xi*\_c+}  \\
 4332 & $\Omega_{\c}^0$ & \ttt{Omega\_c0}   &
 4334 & $\Omega_{\c}^{*0}$ & \ttt{Omega*\_c0}  \\
 5112 & $\Sigma_{\b}^-$ & \ttt{Sigma\_b-}     &
 5114 & $\Sigma_{\b}^{*-}$ & \ttt{Sigma*\_b-}  \\
 5122 & $\Lambda_{\b}^0$ & \ttt{Lambda\_b0} & & &  \\
 5212 & $\Sigma_{\b}^0$ &\ttt{Sigma\_b0}   &
 5214 & $\Sigma_{\b}^{*0}$ & \ttt{Sigma*\_b0}  \\
 5222 & $\Sigma_{\b}^+$ & \ttt{Sigma\_b+}   &
 5224 & $\Sigma_{\b}^{*+}$ &  \ttt{Sigma*\_b+}  \\
\hline \hline
\end{tabular}
\end{center}
\end{table}
\vspace{-0.5cm}
\begin{table}[bhp]
\captive{Nucleus codes.
\protect\label{t:codesix} }  \\
\vspace{1ex}
\begin{center}
\begin{tabular}{|c|c|c||c|c|c|@{\protect\rule{0mm}{\tablinsep}}}
\hline \hline
KF & Name & Printed & KF & Name & Printed \\
\hline \hline
& & & & &\\ 
1001001 & $^2_1\D\e$ & \ttt{de}  & 1061047 & $^{108}_{47}\A\g$ & \ttt{ag} \\
& & & & &\\ 
1002002 & $^4_2\H\e$ & \ttt{he} & 1110074 & $^{110}_{74}{\rm W}$ & \ttt{wt} \\
& & & & &\\ 
1008008 & $^{16}_8{\rm O}$ & \ttt{ox} & 1117078 & $^{195}_{78}{\rm Pt}$ & \ttt{pt} \\
& & & & &\\ 
1014013 & $^{27}_{13}{\rm Al}$ & \ttt{al}  & 118079 & $^{197}_{79}\A\u$ & \ttt{au} \\
& & & & &\\ 
1016016 & $^{32}_{16}{\rm S}$ & \ttt{su}   & 1126082 & $^{208}_{82}{\rm P}\b$ & \ttt{pb} \\
& & & & &\\ 
1020020 & $^{40}_{20}{\rm Ca}$ & \ttt{ca}   & 1146092 & $^{238}_{92}{\rm U}$ & \ttt{ur} \\
& & & & &\\ 
1035029 & $^{64}_{29}{\rm Cu}$ & \ttt{cu}   &         &                      &          \\
& & & & &\\ 
 \hline \hline
\end{tabular}
\end{center}
\end{table}

\newpage

\subsection{The general input and control parameters}
\label{sec:section5}
\bigskip

The VNIDA0 common block contains the status code and parameters
that regulate the performance of the program. All of them are
provided with sensible default values, so that a novice user can
neglect them, and only gradually explore the full range of
possibilities. 

\begin{verbatim}
COMMON/VNIDA0/MSTV(200),PARV(200),MSTW(200),PARW(200)
\end{verbatim}
 
\noindent
{\it Purpose:} to give access to status code and parameters that
regulate the performance of the program. If the default values,
denoted below by (D=\ldots), are not satisfactory, they must in
general be changed before the VNIXRIN call. Exceptions, i.e.
variables that can be changed for each new event, are denoted by
(C).
\bigskip
\bigskip
 
\begin{center}
{\bf MSTV(200), PARV(200): control switches and physics parameters}
\end{center}
\bigskip

\noindent
\lline

\noindent{\bf 1. General:}

\noindent
\lline

\begin{description}

\item[ MSTV(1)]  : (D=100000) number of lines available in the commonblock
    VNIREC1/VNIREC2. Should always be changed if the dimensions of the
    $K,P,R,V$ arrays are changed.
\item[ MSTV(2)]  : (D=25000) is used to trace internal color flow information 
    in the array $K$.
\item[ MSTV(3)]  : (D=10000) is used to store number and type of interactions
    of a particle in the array $R$ of VNIREC1/VNIREC2.
\item[ MSTV(4)]  : (D=1000) is used to trace origin of a particle's production
    in the array $V$ of VNIREC1/VNIREC2.
\item[ MSTV(5)]  : (D=3) number of colors $N_c$ included in the simulation.
\item[ MSTV(6)]  : (D=6) number of quark flavors $N_f$ included.
\item[ MSTV(7)] : (D=1) energy and momentum conservation options. Since 
    both the initial generation of initial hadronic or nuclear parton
    system and the final cluster-hadron decay scheme require occasional 
    reshuffling of 4-momentum, an energy and/or momentum imbalance can 
    occur. This can be corrected, if desired, by renormalizing particle 
    energies and momenta.\\
    {\bf = 0 :} no explicit conservation of any kind.\\
    {\bf = 1 :} energy conservation is enforced at the end of each event
        (i.e. once only per event):
        particles share energy imbalance compensation according
        to their energy (roughly equivalent to boosting event to CM
        frame).\\
    {\bf = 2 :} as =1, but now energy conservation is enforced at the end
        of each time step in each event (i.e. continously throughout an event). \\
    {\bf = 3 :} as =1, but in addition also 3-momentum conservation the end
        of each event
        (i.e. once only per event),
        however at the cost of some 
        particles shifted off-mass shell.\\
    {\bf = 4 :} as =3, but in addition also 3-momentum conservation the end
        of each  timestep in each event (i.e. continously throughout an event). 
\item[ MSTV(8)] : (D=0) type of particles to be included in energy 
       and momentum conservation options under MSTV(7).\\
    {\bf = 0 :} all living  particles are included.\\
    {\bf = 1 :} only partons.\\
    {\bf = 2 :} only hadrons.
\item[ MSTV(9)] :  - not used -
\item[ MSTV(10)] : (D=111111) initial seed for random number generation.
\item[ MSTV(11)] : (D=5000000) maximum number of collision events $A+B$, 
    after which program is forcibly terminated.
\item[ MSTV(12)] : (D=2500) maximum number of time steps per collision 
    event, after which a new collision event is generated. When
    increased, the dimension of the arrays in commonblock $VNITIM$ 
    need to be altered accordingly.
\item[ MSTV(13)] : (D=0) choice of time grid $TIME(I)$ and time increase
    $TINC(I)$. Generically, $t(i)=t(i-1)+[f(i)-f(i-1)]$, with 
    $i=0,\ldots,MSTV(12)$, and $t(0)=t_0=$PARV(12), $t_f=$PARV(13). \\
    {\bf = 0 :} power law increase $f(i)=a\cdot(i)^b$, where $a=$PARV(14) and 
    $b=$PARV(15). \\
    {\bf = 1 :} logarithmic increase $f(i)=a\cdot \ln(i)^b$, where $a=$PARV(14), 
    $b=$PARV(15).
\item[ MSTV(14)] : (D=0) Option to select or switch on/off certain physics
    processes $ISUB$ in the simulation, by specifying the array $MSUB$ 
    (c.f. Sec. 3.8 below), where $MSUB(ISUB) = 0\; (1)$ switches a 
    subprocess off (on).\\ 
    {\bf = 0 :} the relevant processes are  initialized  automatically.\\ 
    {\bf = 1 :} user selection of included processes is required.
\item[ MSTV(15)] : (D=0) Option to smear out classical (sharply localized) 
     positions and three-momenta of particles at their production vertex,
     according to Gaussian distribution $exp[-r^2/a^2 - p^2/b^2]$, 
     where the widths are 
     $a=$ PARV(16) in fm, and $b=$ PARV(17) in GeV.\\
    {\bf = 0 :} switched off, i.e. no smearing.\\
    {\bf = 1 :} smearing of particle {\it position} according to Gaussian 
     with width PARV(16).\\
    {\bf = 2 :} as = 1, plus smearing of particle {\it momentum} 
     according to Gaussian with width PARV(17).
\item[ MSTV(16)] : (D=0) Option to use PYTHIA for generating the initial state
     particle system rather than VNI. Presently this applies only to
     $l^+l^-$ collisions, i.e., IPRO =1 or 2 in Table 1.
     NOTE: The subroutine package {\it vnipythia.f}  must be linked in
     order to use this option. Also, in this case, one must comment out the 
     dummy routines PYINIT and PYEVNT at the very end of {\it vni.f}.\\
    {\bf = 0 :} switched off.\\
    {\bf = 1 :} switched on.
\medskip

\noindent
\sline

\item[ PARV(1)]{\bf - PARV(11)} : -- not used --
\item[ PARV(12)] : (D=0. fm) choice of initial point of time $t_i =$ PARV(12) 
    when evolution of collision event begins.
\item[ PARV(13)] : (D=0. fm) choice of final point of time $t_f =$ PARV(13).
    If = 0., then it is set automatically to $TIME$(MSTW(2)), where MSTW(2)
    is the selected number of time steps. Otherwise, if $> 0.$, it specifies
    $t_f$ such that the minimum  $\min(t_f , t_{max})$ is taken as the point when the
    evolution of a collision event stops, where $t_{max} = TIME$(MSTV(12)).
\item[ PARV(14)] : (D=0.05) prefactor $a$ of the time-increment function $f$, see
    MSTV(13).
\item[ PARV(15)] : (D=1.1) exponent $b$ of the time-increment function $f$, see
    MSTV(13). For longer collision time scales as  in hadron-nucleus
    or nucleus nucleus collisions, it is automatically scaled by
    PARV(15) to 1.5 $\times$ PARV(15) to extend the time range.
\item[ PARV(16)] : (D=0.1 fm)
     For MSTV(15) = 1, width parameter $a$=PARV(16) of 
     Gaussian $exp[-r^2/a^2]$ for smearing out classical (sharply localized) 
     positions of particles at their production vertex.
     (Reasonable values are $a = 0.1 - 1$ fm, 
     c.f. Zhang et. al, nuc-th/9704041).
\item[ PARV(17)] : (D=0.1 GeV)
     For MSTV(15) = 2, width parameter $b$=PARV(17) of 
     Gaussian $exp[-p^2/b^2]$ for smearing out classical (sharply localized) 
     momenta of particles at their production vertex.
     (Note, unertainty relation requires PARV(17) $\simeq$ 1/PARV(16) GeV/fm).

\noindent
\sline

\end{description}
\medskip

\noindent
\lline

\noindent {\bf 2. Initial state of collision system:}

\noindent
\lline

\begin{description}
\item[ MSTV(20)] : (D=0) choice of model for initial parton distributions.  \\
    {\bf = 0 :} standard structure functions are used.\\
    {\bf = 1 :} other parametrization, e.g. a la McLerran {\it et al.}
    (not implemented yet).
\item[ MSTV(21)] : (D=1) choice of nucleon structure-function 
      set (c.f. MSTV(22)). \\
    {\bf =  1 :} GRV94LO (Lowest order fit).\\
    {\bf =  2 :} GRV94HO (Higher order: MSbar fit).\\
    {\bf =  3 :} GRV94DI (Higher order: DIS fit).\\
    {\bf =  4 :}  - not used - \\
    {\bf =  5 :} CTEQ2M (best higher order MSbar fit).\\
    {\bf =  6 :} CTEQ2MS (singular at small $x$).\\
    {\bf =  7 :} CTEQ2MF (flat at small $x$).\\
    {\bf =  8 :} CTEQ2ML (large $\Lambda$).\\
    {\bf =  9 :} CTEQ2L (best lowest order fit).\\
    {\bf = 10 :} CTEQ2D (best higher DIS order fit).
\item[ MSTV(22)] : (D=1) choice of nucleon structure-function library.  \\
    {\bf = 1 :} the internal one with parton distributions chosen according 
    to MSTV(21).\\
    {\bf = 2 :} the PDFLIB one with MSTV(21)=$1000\cdot NGROUP+NSET$
    (requires PDFLIB to be linked). 
\item[ MSTV(23)] : (D=1) choice of pion structure-function set (c.f. MSTV(24)).\\
    {\bf = 1 :} Owens set 1.\\
    {\bf = 2 :} Owens set 2.\\
\item[ MSTV(24)] : (D=1) choice of pion structure-function library.  \\
    {\bf = 1 :} the internal one with parton distributions chosen according 
    to MSTV(23).\\
    {\bf = 2 :} the PDFLIB one with $MSTV(23)=1000\cdot NGROUP+NSET$
    (requires PDFLIB to be linked). 
\item[ MSTV(25)] : (D=2) choice of minimum $Q$-value $Q_{0}$ used 
in parton distributions of initial hadronic or nuclear state.\\
    {\bf = 0 :} fixed at $Q_{0}=$PARV(23).\\
    {\bf = 1 :} determined by the average momentum tranfer of primary parton 
    collisions $Q_0=\langle Q_{prim}\rangle$, 
    with initial value $Q_{0}=$PARV(23) before the first event.\\
    {\bf = 2 :} as =1, but in addition varying with total energy $\sqrt{s}$ of 
    the overall collision system, with 
    $\widetilde{Q}_{0} = \max[Q_0, \frac{a}{4} \cdot \left(E_h/GeV\right)^b]$, 
    where $Q_0=$PARV(23), $a=$PARV(24), $b=$PARV(25), and 
    $E_h=\sqrt{s}/\mbox{hadron}$ the average energy per hadron (nucleon).
\item[ MSTV(26)] : (D=1) option to choose $x$-dependent $Q_{in}$ 
in parton densities.  \\
    {\bf = 0 :} switched off, $Q_{0}$ according to selection of MSTV(25) 
    is used.\\
    {\bf = 1 :} varying with Bjorken-$x$ of struck initial state partons, as 
    well as with total energy $\sqrt{s}$ of collision system, with  
    $Q_{0}=Q(x)$ where $1/Q^2(x) = 1/Q_{0}^2+Q_{0}^2/(x E_{had})^2$, and 
    $E_h=\sqrt{s}/\mbox{hadron}$ the average energy per hadron (nucleon).
\item[ MSTV(30)] : (D=0) sub-selection of initial state for the following 
     collision processes $A+B$, where specification of beam/target particles 
     $A,B$ is not unambigous, because collision may proceed via 
     alternative channels (c.f. Table 1): 
\begin{itemize} 
\item[a)] {\bf $l^+l^- \rightarrow \,X\,\rightarrow \,hadrons$:}
\\   {\bf =  0 :} $e^+e^- \rightarrow \gamma/Z^0 \rightarrow q\bar{q}$
\\   {\bf =  1 :} $e^+e^- \rightarrow W^+W^- \rightarrow q\bar{q}' \,l \,\nu$
\\   {\bf =  2 :} $e^+e^- \rightarrow W^+W^- \rightarrow q\bar{q}' \,q'\bar{q}$
\end{itemize} 
\item[ MSTV(31)] : (D=0) initial separation along the collision ($z$-)axis of 
    beam and target particle in the overall CM frame.\\
    {\bf =  0 :} fixed at $\Delta z=$PARV(31).\\
    {\bf$>$ 0 :} shifted towards minimal separation, such that beam and 
    target particles almost touch with $\Delta z =$MSTV(31)$\cdot$PARV(31).
\item[ MSTV(32)] : (D=2) selection of impact parameter $b$ transverse to the 
    collision ($z$-) axis of beam and target particle.\\
    {\bf = 0 :} fixed at $b =$ Max[PARV(32),PARV(33)].\\
    {\bf = 1 :} randomly sampled in between $b_{min}\le b\le b_{max}$, with 
    $b_{min} = $PARV(32) and $b_{max} = $PARV(33).\\
    {\bf = 2 :} as 1, but with $b_{min} =$ PARV(32)$\cdot\sqrt{\sigma_{nd}/\pi}$, 
    and $b_{max} =$ PARV(33)$\cdot\sqrt{\sigma_{nd}/\pi}$, 
    where $\sigma_{nd}$ is the non-diffractive crossection for given 
    beam/target and beam energy.
\item[ MSTV(33)] : (D=2) Choice of nuclear matter distribution for collisions
    involving one or two nuclei, i.e. of nucleons' positions within nucleus.\\
    {\bf = 0 :} uniform distribution with `sharp edge'.\\
    {\bf = 1 :} Gaussian distribution for light nuclei $A\le 12$, 
    and `smeared edge' distribution for intermediate and heavy nuclei 
    $A > 12$ (c.f. PARV(27)).\\
    {\bf = 2 :} Gaussian distribution for $A\le 12$, and Fermi distribution for
    $A > 12$ (c.f. PARV(28)).
\item[ MSTV(34)] : (D=1) spatial distribution of initial partons within their
    parent hadrons of beam and target particles, in the hadron restframe
    within a sphere of the hadron radius.\\
    {\bf = 0 :} uniform distribution.\\
    {\bf = 1 :} exponential distribution with width given by PARV(34).\\
    {\bf = 2 :} Gaussian distribution with width given by PARV(35).
\item[ MSTV(35)] : (D=2) Lorentz-boosted spatial distribution of initial partons.\\
    {\bf = 0 :} spatial positions of all partons (valence, sea, glue) inside 
    parent hadron are Lorentz-boosted with the parent hadron's (nucleon's) 
    $\gamma$-factor.\\ 
    {\bf = 1 :} only valence quarks are boosted with the parent  hadron's 
    (nucleon's) 
    $\gamma$-factor; seaquarks and gluons are smeared out symmetrically(!) 
    in hadron (nucleon) direction of motion around the valence quark disc.  \\ 
    {\bf = 2 :} as = 1, but now seaquarks and gluons are 
    distributed all behind(!) the valence quark disc.
\item[ MSTV(36)] : (D=1) choice of primordial $k_\perp$-distribution of 
    initial partons.\\
    {\bf = 0 :} no primordial $k_\perp$.\\
    {\bf = 1 :} exponential $k_\perp$-distribution with width 
    specified by PARV(37).\\
    {\bf = 2 :} Gaussian $k_\perp$-distribution with width specified by PARV(38).
\item[ MSTV(37)] : (D=1) choice of masses of initial partons.\\
    {\bf = 0 :} on mass shell with quark current masses.\\
    {\bf = 1 :} off mass shell with space-like virtuality as determined by
    the excess of the partons' momentum over their energies.
\item[ MSTV(38)] : (D=0) switch for including effects of parton shadowing in
    nuclear parton structure functions by multiplication with effective
    shadowing factor $R_A(x)=f_A(x)/f_N(x)$ as a simple parametrization
    of EMC data, where $f_A$ and $f_N$ are the measured nuclear and nucleon 
    structure functions, respectively.\\
    {\bf = 0 :} parton shadowing switched off.\\
    {\bf = 1 :} switched on.
\item[ MSTV(39)] : (D=0) option for asymmetric (in mass) collision
    systems involving nuclei, $h+A$ with $h=\pi,p,n,\ldots$, 
    or $A+B$ with $A\ne B$,  to
    choose instead of global c.m. frame or fixed target frame,
    the frame of equal (hadron)nucleon-nucleon velocity, that is the frame
    in which hadron (nucleons) from beam side have equal but opposite
    directed momentum than those from target side.\\
    {\bf = 0 :} switched off.\\
    {\bf = 1 :} switched on.
\medskip

\noindent
\sline

\item[ PARV(21)] : (D=1. GeV$^{-1}$) the minimum resolvable partonic 
    Bjorken-$x$, denoted $x_c$,
    is set by the `confinement radius' $R_c = $PARV(21), which defines the 
    maximum possible longitudinal spread of a confined parton within 
    $\Delta z = 1/xP < 1/(x_c P) \equiv R_c$, where $P$ is the longitudinal 
    momentum of the mother hadron.
\item[ PARV(22)] : (D=0.95)  the minimum 
    summed $x$-value $\left(\sum_i x_i\right)_{min}$, 
    when `filling' a hadron with partons of energy(momentum) fractions
    $x_i$ of the hadron. The value depends, slightly on the value of $Q_0$ 
    (c.f. PARV(23)-PARV(25)), or if cuts on $x$ are made. It should be 
    adjusted such that when adding one more parton, the average 
    $\langle \sum_i x_i \rangle$ 
    should be = 1. If the sum is smaller than PARV(22), another parton 
    is added to the hadron, i.e. PARV(22) = $1 - \langle \sum_i x_i \rangle$. 
\item[ PARV(23)] : (D=1. GeV) minimum value $Q_0$ that is used in initial parton
    distributions (structure functions or alternative distributions).
    I.e., if $Q < Q_0$, then the value PARV(23) is used.
\item[ PARV(24)], {\bf PARV(25)} : (D=2.50,0.25) 
    possibility of using effective, energy-dependent
    $Q_{in}$ 
    value in structure functions if MSTV(25)=2, according to the function
    $\widetilde{Q}_{0} = \max[Q_0,\frac{a}{4}\cdot \left(E_h/GeV\right)^b]$, 
    where  $Q_0 = $PARV(23), $a = $PARV(24), 
    $b= $ PARV(25), and
    $E_h=\sqrt{s}/\mbox{hadron}$ the average energy per hadron (nucleon).
\item[ PARV(26)] : (D=1.18 fm) nuclear radius parameter $r_0$, with 
    $R(A) = r_0 A^1/3$. 
\item[ PARV(27)] : (D=3.0 fm) parameter $R_d$ in `smeared-edge' distribution of 
    nucleons' positions in nucleus with $A > 12$ (for MSTV(33) = 1).
\item[ PARV(28)] : (D=0.545 fm) parameter $R_a$ in Fermi distribution of 
    nucleons' positions in nucleus with $A > 12$ (for MSTV(33) = 2).
\item[ PARV(31)] : (D=0.25 fm) initial separation along the collision ($z$-)axis 
    of beam and target particle in the overall CM frame for MSTV(31)$\ge$ 1.
\item[ PARV(32)] : (D=0. fm) determines 
    value of {\it minimum} impact parameter $b_{min}$, 
    transverse to the collision ($z$-)axis, such that
    $b_{min} = $PARV(32) for MSTV(32) = 0 or 1, 
    or $b_{min} = $PARV(32)$\cdot\sqrt{\sigma_{nd}/\pi}$ for MSTV(32)=2.
\item[ PARV(33)] : (D=1. fm) determines 
    value of {\it maximum} impact parameter $b_{max}$, 
    transverse to the collision ($z$-)axis, such that
    $b_{max} = $PARV(33) for MSTV(32) = 0 or 1, 
    or $b_{max} = $PARV(33)$\cdot\sqrt{\sigma_{nd}/\pi}$ for MSTV(32)=2.
\item[ PARV(34)] : (D=1.19 GeV$^{-1}$) width $w$ in exponential distribution 
    $\exp[-r/w]$ of partons' positions around its parent-hadron center-of-mass 
    (c.f. MSTV(33)).
    (Note: $w=1/\nu$, where $\nu=0.84$ GeV is the measured nucleon form factor.)
\item[ PARV(35)] : (D=2.38 GeV$^{-1}$) sigma $s$ in Gaussian distribution 
    $\exp[-r^2/(2\,s^2)]$ of 
    partons' positions around its parent-hadron center-of-mass (c.f. MSTV(33)).
    (Note: $s=2/\nu$, where $\nu=0.84$ GeV is measured nucleon form factor.)
\item[ PARV(36)] : (D=2.) cut-off for sampling initial partons' spatial 
    distributions 
    from exponential or Gaussian above, such that $r < a\,R_h$, where 
    $a=$PARV(36) and $R_h$ is the radius of the mother hadron.
\item[ PARV(37)] : (D=0.44 GeV) width $w$ in exponential distribution 
    $\exp[-k_\perp/w]$ of primordial $k_\perp$ distribution of initial partons 
    (c.f. MSTV(36)).
\item[ PARV(38)] : (D=0.44 GeV) sigma $s$ in Gaussian distribution 
    $\exp(-k_\perp^2/(2\,s^2)]$ of primordial $k_\perp$ distribution of 
    initial partons (c.f. MSTV(36)).
\item[ PARV(39)] : (D=2. GeV) upper cut-off for sampling primordial 
    $k_\perp$ from above exponential or Gaussian distribution.

\noindent
\sline

\end{description}
\medskip

\noindent
\lline

\noindent {\bf 3. Parton scatterings:}

\noindent
\lline

\begin{description}
\item[ MSTV(40)] : (D=3) master switch for parton collisions. \\
    {\bf = 0 :} generally switched off.
\\
    {\bf = 1 :} only primary-primary collisions, i.e. those involving 
    2 partons that both have not interacted before.
\\
    {\bf = 2 :} as = 1 plus primary-secondary collisions, i.e. those involving 
    2 partons, of which one has had already at least one previous interaction.
\\
    {\bf = 3 :} as = 2 plus secondary-secondary interactions, i.e.
    all multiple collisions included.
\item[ MSTV(41)] : (D=1) choice of (semi)hard QCD collision cross-sections.
\\
    {\bf = 1 :} standard $O(\alpha_s^2)$ perturbative cross-sections for 
    $2 \rightarrow 2$ and $2 \rightarrow 1$ processes.
\\
    {\bf = 2 :} alternative  of $2 \rightarrow n$ cross-sections 
    (not implemented yet!).
\item[ MSTV(42)] : (D=1) choice of $p_{0}$ 
    to separate {\it hard} and {\it soft} parton collisions, and to act
    as a regulator for the parton cross-sections $d\hat{\sigma}/dQ^2$,
    that are divergent 
    for vanishing momentum transfer (or invariant mass) $Q$.
\\
    {\bf = 0 :} fixed at 
    $p_0 = \max \left[ PARV(42), \max (VKIN(3),VKIN(5) \right]$.
\\
    {\bf = 1 :} energy and mass dependent, with  $p_0 = p_0(s,A)$, 
    according to the parametrization given by PARV(42), PARV(43).
\\
    {\bf = 2 :} initially, before the first event, set to 
    $p_0(s,A)$, c.f. =1, 
    but in subsequent events set equal to the average momentum transfer 
    of primary parton-parton collisions 
    $p_0 = \langle Q_{prim}\rangle$ accumulated by statistics.
\item[ MSTV(43)] : (D=2) $Q^2$-definition in $2 \rightarrow 2$ 
    parton collision processes;
    for $\hat s$-channel  or fusion processes, 
    $Q^2$ is always chosen to be $m^2=\hat s$.
    Note that the parton-parton collisions invariants are
    denoted as $\hat s = (p_1+p_2)^2$,
    $\hat t = (p_1-p_1')^2$, $\hat u = (p_1-p_2')^2$.
\\
    {\bf = 1 :} $Q^2 = 2 \hat{s} \hat{t} \hat{u}/(\hat{s}^2 + 
    \hat{t}^2 + \hat{u}^2)$.
\\
    {\bf = 2 :} $Q^2 = 0.5 \,(m_{\perp\,1}^2 + m_{\perp\,2}^2)$.
\\
    {\bf = 3 :} $Q^2 = \min (-\hat{t}, - \hat{u})$.
\\
    {\bf = 4 :} $Q^2 = \hat{s}$.
\item[ MSTV(44)] : (D=2) choice of scheme for parton collision-time estimate.
\\
    {\bf = 0 :} zero collision-time, i.e. instantaneous process. 
\\
    {\bf = 1 :} finite collision-time sampled from an exponential distribution
    $\exp(-x)$, where $x=t/\tau$, $t$ the time in the lab, and $\tau$
    the mean life-time of the parton in the lab, see PARV(44).
\\
    {\bf = 2 :} finite collision-time sampled from Gyulassy-Wang distribution
    $(2/\tau) \;x/(1+x^2)^2$, where $x=t/\tau$, $t$ the time in the lab, 
    and $\tau$ the mean life-time of the parton in the lab, see PARV(44).
\item[ MSTV(45)] : (D=1) selection of maximum allowed distance between two partons
    in their $cm$-frame in order to be able to collide (used for ruling
    out widely separated pairs to speed up simulation).
\\
    {\bf = 0 :} fixed at a value given by $r_{12\;max} =$ PARV(45).
\\
    {\bf = 1 :} variable at a value 
    $r_{12\;max} =$ PARV(45)/$(\sqrt{\hat{s}}/GeV)$.
\item[ MSTV(46)] : (D=0) choice of probability distribution W(x) from which 
    collision probability of 2 partons is sampled, where
    $x = r_{12}^{(sep)}/r_{12}^{(\hat{\sigma})}$, with $r_{12}^{(sep)}$ 
    the transverse  2-parton separation 
    at closest approach and $r_{12}^{(\hat{\sigma})}$ 
    the transverse radius of interaction as given by the parton-parton cross-section $\hat{\sigma}$.
\\
    {\bf = 0 :} flat distribution, $W(x) = \theta(a-x)$, with $a =$ PARV(46).
\\
    {\bf = 1 :} exponential distribution, $W(x) = \exp(-x/a)$, $a =$ PARV(46).
\item[ MSTV(47)] : (D=2) special treatment of `flavor excitation' by parton 
    collisions involving one or two primary hadronic or nuclear 
    (anti)quarks by requiring a minimum momentum transfer of the
    resolving parton with the heavy quark.
\\
    {\bf = 0 :} no special treatment of heavy quarks over light quarks.\\
    {\bf = 1 :} requirement of minimum momentum transfer for liberation of
    heavy quark out of initial parton distribution by scattering, where
    the minimum required momentum trasnfer 
    is given by PARV(47) times the mass of the heavy quark.
\\
    {\bf = 2 :} as = 1, but now both the struck heavy quark as well as its
    initial antiquark sibling are liberated (i.e. get on mass shell).
\item[ MSTV(48)] : (D=0) switch for including soft parton collisions according 
    to a phenomenological cross-section 
    $d\hat{\sigma}/dQ^2 \propto \alpha_s^2(p_0^2)/(Q^2 + \mu^2)$
    for soft parton collisions (c.f. eq. (\ref{sigs})), 
    where $p_0 =$ PARV(42), $\mu =$ PARV(48), and $Q^2 = p_\perp^2$
    for $\hat t, \hat u$ channel and
    $Q^2 = \hat s$ for $\hat s$ channel.
\\
    {\bf = 0 :} switched off.
\\
    {\bf = 1 :} soft collisions are treated complementary to hard collisions,
    i.e. occur only if $Q < p_{0}$.
\\
    {\bf = 2 :} soft collisions are treated supplementary to hard collisions,
    i.e. compete with the latter, whichever cross-section dominates.
\item[ MSTV(49)] : (D=1) switch to allow for $2 \rightarrow 1$
    parton fusion processes
    $a+b\rightarrow c^\ast$ in competition with $2\rightarrow 2$ 
    scattering processes $a+b\rightarrow c^\ast\rightarrow d+e$,
    such that for a given pair a,b either of those processes is selected
    according to the relative probability given by the respective 
    cross-sections and depending on the `life-time' of the particle $c^\ast$.
\\
    {\bf = 0 :} fusion processes switched off.
\\
    {\bf = 1 :} switched on, with relative weight determined by PARV(49).
\medskip

\noindent
\sline

\item[ PARV(41)] : (D=0.25 GeV) nominal $\Lambda$-value used in 
    running $\alpha_s$ for parton-parton collisions.
\item[ PARV(42)] : (D=0.0 GeV) cut-off for perturbative QCD cross-sections 
    that are divergent when the momentum transfer $Q\rightarrow 0$. Here the 
    2-body collsion scale $Q$ is taken equal to $p_\perp$, the
    transverse momentum 
    exchanged in a  parton scattering, or equal to $\hat s$ for
    annihilation or fusion processes. For MSTV(42)=0, the value 
    $p_{0} = a$ with $a =$ PARV(42) defines {\it hard} ($Q>p_{0}$) and 
    {\it soft} ($Q<p_{0})$ collisions. For MSTV(42)=1, see PARV(43). 
    {\it Note:} If PARV(42) $\le 0$, then it is set automatically upon
    initialization to = 2.0.
\item[ PARV(43)] : (D=0.0) parameter for effective $p_{0}$ value, 
    if MSTV(42)=1, 
    instead of $p_{0} = const$ for  MSTV(42)=0. The effective $p_{0}$ 
    assumes a dependence on the total collision energy $\sqrt{s}$ and 
    the mass of the (nuclear) collision system. It is parametrized as 
    $p_{0}(\sqrt{s},A,B) = \frac{a}{4} \cdot \left(E_h/GeV\right)^b$, 
    where $a =$ PARV(42), $b =$ PARV(43), and $E_h= 2 \sqrt{s}/(A+B)$ 
    with $A$ ($B$) the mass number of beam (target) particle, if 
    it is a hadron or nucleus.
    {\it Note:} If PARV(43) $\le 0$, then it is set automatically upon
    initialization to = 0.27 - 0.29.
\item[ PARV(44)] : (D=0.1) if MSTV(44)=1 or MSTV(44)=2, the 
    proportionality between 
    mean collision time $\tau$ of a 2-parton collision process, and the 
    associated transverse momentum generated (for $\hat{s}$-channel processes, the 
    invariant mass of the pair). It is parametrized as $\tau=\,c\,(1/Q)$,
    with $c =$ PARV(45) and $Q = p_\perp$ or $\hat{s}$.
\item[ PARV(45)] : (D=5. GeV$^{-1}$) maximum separation of two partons to be 
    able to
    interact through a 2-body collision. For MSTV(45)=0, it is fixed at
    $r_{12} =$ PARV(45), whereas for MSTV(45)=1, it is taken s-dependent as 
    $r_{12} =$ PARV(45)$/(\sqrt{\hat{s}}/GeV)$.
\item[ PARV(46)] : (D=1.0) parameter in the collision-probability distribution 
    $W(x)=\theta(a-x)$ (for MSTV(46)=0) and $W(x)=\exp(-x/a)$ for MSTV(46)=1,
    with $x = r_{12}^{(sep)}/r_{12}^{(\hat{\sigma})}$, $r_{12}^{(sep)}$
    the relative transverse parton 
    separation of closest approach and $r_{12}^{(\hat{\sigma})}$ 
    the transverse radius of interaction as given by the 
    parton-parton cross-section $\hat{\sigma}$.
\item[ PARV(47)] : (D=1.0) for `flavor excitation' processes where a primary
    hadronic or nuclear (anti)quark is struck out by a collision, an
    effective minimum resolution scale is required, depending on
    the flavor, such that the momentum transfer
      $Q^2 > a \cdot m_f$, with $a=$ PARV(47) and $m_f$ the flavor-dependent
     quark mass. This parameter acts in addition to the 
    constraint set by PARV(42) above.
\item[ PARV(48)] : (D=0.5 GeV) parameter in the phenomenological 
    cross-section
    $d\hat{\sigma}/dQ^2 \propto \alpha_s^2(p_0^2)/(Q^2 + \mu^2)$
    for soft parton collisions, 
    where $\mu =$ PARV(48), $p_0$ is given by PARV(42), and $Q^2 = p_\perp^2$
    for $\hat t, \hat u$ channel and
    $Q^2 = \hat s$ for $\hat s$ channel.
\item[ PARV(49)] : (D=10.) parametric strength of $2 \rightarrow 1$ 
    parton fusion processes $a+b\rightarrow c$ competing with
    $2\rightarrow 2$ scattering processes $a+b\rightarrow c+d$. It is  
    given by the ratio of the two processes which is proportional 
    to $1/(R^2 \hat{s})$, where $1/R^2 = $PARV(49).

\noindent
\sline

\end{description}
\medskip

\noindent
\lline

\noindent {\bf 4.  Space-like parton branchings:}

\noindent
\lline

\begin{description}
\item[ MSTV(50)] : (D=1) master switch for space-like branchings.
\\
    {\bf = 0 :} switched off
\\
    {\bf = 1 :} on.
\item[ MSTV(51)] : (D=3) level of coherence imposed on the space-like parton
    shower evolution.
\\
    {\bf = 1 :} none, i.e. neither $k^2$ values nor angles need be ordered.
\\
    {\bf = 2 :} $k^2$ values at branches are strictly ordered, increasing
        towards the hard interaction.
\\
    {\bf = 3 :} $k^2$ values and opening angles of emitted (on-mass-shell or
        time-like) partons are both strictly ordered, increasing
        towards the hard interaction.
\item[ MSTV(52)] : (D=2) structure of associated time-like parton evolution, 
    i.e.  showers initiated by emission off the incoming space-like partons.
\\
    {\bf = 0 :} no associated showers are allowed, i.e. emitted partons are
        put on mass-shell.
\\
    {\bf = 1 :} a shower may evolve, with maximum allowed time-like virtuality
        set by the phase space only.
\\
    {\bf = 2 :} a shower may evolve, with maximum allowed time-like virtuality
        set by phase space or by the scale $Q^2$, the virtuality of the 
        space-like parton created at the same vertex, whichever is the 
        stronger constraint.
\medskip

\noindent
\sline

\item[ PARV(51)] : (D=0.25 GeV)  $\Lambda$-value used in 
   $\alpha_s$ and in structure functions for space-like parton evolution.
\item[ PARV(52)] : (D=1.0 GeV) invariant mass cutoff $q_0$ of 
    space-like parton 
    showers, below which parton branchings re not evolved. For consistency
    it is taken as $q_{0} = \max(Q_0$,PARV(52)) where $Q_0 = $PARV(23) is
    the initial resolution scale in the hadron structure functions.
\item[ PARV(53)] : (D=4.) $Q^2$-scale of the hard scattering is multiplied by 
    PARV(53) to define the maximum parton virtuality allowed in 
    space-like branchings associated with the hard interaction. This does not 
    apply to $\hat{s}$-channel resonances, where the maximum virtuality is set 
    by $m^2=\hat s$.

\noindent
\sline
             
\end{description}
\medskip

\noindent
\lline

\noindent {\bf 5.  Time-like parton branchings:}

\noindent
\lline

\begin{description}
\item[ MSTV(60)] : (D=2) master switch for time-like branchings.
\\
    {\bf = 0 :} no branchings at all, i.e. switched off.
\\
    {\bf = 1 :} QCD type branchings of quarks and gluons.
\\
    {\bf = 2 :} also emission of photons off quarks; the photons are assumed
        on mass-shell.
\item[ MSTV(61)] : (D=2) branching mode for time-like parton evolution.
\\
    {\bf = 1 :} conventional branching, i.e. without angular ordering.
\\
    {\bf = 2 :} coherent branching, i.e. with angular ordering.
\item[ MSTV(62)] : not used.
\item[ MSTV(63)] : (D=1) choice of formation-time scheme for parton emission.
\\
    {\bf = 0 :} no formation time, i.e. instantanous emission. 
\\
    {\bf = 1 :} finite formation time sampled from an exponential distribution
    $\exp(-x)$, where $x=t/\tau$, $t$ 
    the time in the lab, and  $\tau$ the mean life-time 
    of the parton in the lab, see PARV(63).
\\
    {\bf = 2 :} finite formation time sampled from Gyulassy-Wang distribution
    $ (2/\tau) \;x/(1+x^2)^2$, where $x=t/\tau$, $t$ the time in the lab, 
    and $\tau$ the mean life-time of the parton in the lab, see PARV(63).
\item[ MSTV(64)] : not used.
\item[ MSTV(65)] : (D=0) selection of kinematics reconstruction for branchings
    initiated by a single off-shell parton.
\\
    {\bf = 0 :} conservation of energy and jet direction, but longitudinal 
    momentum is not conserved.
\\
    {\bf = 1 :} with full energy-momentum conservation at each branching, but
    the jet direction is not conserved due to recoil.
\medskip

\noindent
\sline

\item[ PARV(61)] : (D=0.29 GeV) $\Lambda$-value used in running $\alpha_s$ for
    time-like parton showers.
\item[ PARV(62)] : (D=1.0 GeV) invariant mass cutoff $\mu_0$ of time-like 
    parton evolution,
    below which partons are not assumed to radiate. Exception: for
    the case MSTV(84)=3, for reasons of consistency,
    the value is automatically 
    scaled  to 1.5 $\times$ PARV(62). 
\item[ PARV(63)] : (D=0.01) if MSTV(63)=1 or MSTV(63)=2, the proportionality 
    between mean-life time $\tau$ of an off-shell parton and its virtuality 
    $k^2$, is parametrized as $\tau = c\,(1/\sqrt{k^2})$, with $c =$ PARV(63). 
    (Is inversely related to PARV(65)).
\item[ PARV(64)] : (D=4.) $Q^2$-scale of the hard scattering is multiplied by 
    PARV(64) to define the maximum parton virtuality allowed in 
    time-like branchings associated with the hard interaction. This does not 
    apply to $\hat{s}$-channel resonances, where the maximum virtuality is set 
    by $m^2=\hat s$.
\item[ PARV(65)] : (D=4.) invariant virtuality $k^2$ 
    of off-shell partons is multiplied
    PARV(65) to define the maximum parton virtuality allowed in time-like 
    branchings of single parton decays. (Is inversely related to 
    PARV(63)).

\noindent
\sline
\end{description}
\medskip

\noindent
\lline

\noindent {\bf 6. Parton-cluster formation and cluster-hadron decay: }

\noindent
\lline

\begin{description}
\item[ MSTV(80)] : (D=1) master switch for hadronization of parton clusters.
\\
    {\bf = 0 :} off.
\\
    {\bf = 1 :} on.
\item[ MSTV(81)] : (D=1) choice of definition of 2-parton spatial separation 
    $r_{12}= r_1-r_2$ between two partons 1, 2, for measuring
    probability of pre-hadronic cluster formation of the two partons.
\\
    {\bf = 0 :} distance $r_12$ measured in the global frame of the collision 
     event.
\\
    {\bf = 1 :} distance $r_12$ measured in the center-of-mass frame of 1 and 2.
\\
    {\bf = 2 :} distance $r_12$ measured in the rest-frame of parton 1.
\item[ MSTV(82)] : (D=4) choice of definition of 2-parton momentum separation 
    $p_{12}$ of two partons 1, 2, for pre-hadronic cluster search.
\\
    {\bf = 1 :} three-momentum, i.e. 
    $p_{12}=\sqrt{p_{x\;12}^2+p_{y\;12}^2+p_{z\;12}^2}$.
\\
    {\bf = 2 :} transverse momentum, i.e. $p_{12}=\sqrt{p_{x\;12}^2+p_{y\;12}^2}$.
\\
    {\bf = 3 :} `pseudo-mass', i.e. $p_{12}=\sqrt{2\,E_1E_2\,(1-p_1p_2)/(m_1m_2)}$.
\\
    {\bf = 4 :} invariant mass', i.e. 
    $p_{12}=\sqrt{E_{12}^2-p_{x\;12}^2+p_{y\;12}^2+p_{z\;12}^2}$.
\item[ MSTV(83)] : (D=1) choice of probability distribution from which cluster
    formation of 2 partons is sampled. Simulates the conversion process as
    a `tunneling' of the partons through a potential barrier set by PARV(81) 
    and PARV(82), separating perturbative and non-perturbative regimes.
\\
    {\bf = 0 :} flat distribution, i.e. $\theta$-function.
\\
    {\bf = 1 :} exponential distribution.
\item[ MSTV(84)] : (D=3) level of including color correlations among pairs of 
     partons in the process of cluster formation.
\\
    {\bf = 0 :} none, except that clustering of two partons originating from the 
     same mother is vetoed.
\\
    {\bf = 1 :} probability of two cluster candidates being in a color singlet 
     state is sampled uniformly with factors 
     $C_{gg}=1/9, \;C_{q\bar q}=1/3, \;C_{gq}=1/3$.
\\
    {\bf = 2 :} exact matching of colors and anticolors of two partons 
     is required to give color singlet(s).
\\
    {\bf = 3 :} general case, where arbitrary color configuration of two partons
     is allowed, and color singlet formation is balanced by additional gluon
     emission(s) to conserve color locally at each vertex.
\item[ MSTV(85)] : (D=0) option allowing for diquark clustering, i.e. sequential
     clustering of two  (anti)quarks plus a third (anti)quark (not yet working
     properly!).
\\
    {\bf = 0 :} off.
\\
    {\bf = 1 :} on.
\item[ MSTV(87)] : (D=0) choice of effective invariant masses 
     of partons to ensure that
     kinematical thresholds for cluster-hadron decay can be satisfied.
\\
    {\bf = 0 :} `curent' masses are used, i.e. $m_f =$ (0.05, 0.01, 0.01, 0.2, 1.5, 
     5.0, 150) GeV for $f = g, d, u, s, c, b, t$, respectively.
\\
    {\bf = 1 :} `constituent' masses are used, i.e. $m_f =$ (0.5, 0.32, 0.32, 0.5,
     1.8, 5.2, 150) GeV for $f= g, d, u, s, c, b, t$, respectively.
\item[ MSTV(88)] : (D=0) choice of measure for rough distinction between  `slow' 
     and `fast' clusters in statistics accumulation of exogamous cluster 
     production with numerical value of boundary given by PARV(88).
\\
    {\bf = 0 :} energy $E_{cl}$.
\\
    {\bf = 1 :} 3-momentum $P_{cl}$
\\
    {\bf = 2 :} cluster velocity $\beta = P_{cl}/E_{cl}$.
\\
    {\bf = 3 :} Lorentz factor $\gamma = E_{cl}/M_{cl}$. 
\item[ MSTV(89)] : (D=1) smearing of cluster rapidities within
     an interval $| y | \le y_c$, where $y_c$ is determined by PARV(89).
\item[ MSTV(90)] : (D=1) master switch for hadronization of beam/target remnants.
\\
    {\bf = 0 :} off.
\\
    {\bf = 1 :} on.
\item[ MSTV(91)] : (D=1) switch for the decay of unstable hadrons, emerging either 
     directly from cluster-decays, or from previous unstable hadron decays,
\\
    {\bf = 0 :} off.
\\
    {\bf {\boldmath $\ge 1$} :} 
    the decay chain is iterated MSTV(91) times, but at most 5 times.
\item[ MSTV(92)] : (D=1) 
     switch for simulating {\it stopping effect} in
     nucleus-nucleus collisions of initial-state nucleons by redistributing 
     them  in transverse momentum $p_\perp$ according to
     a $1/(1+b \,p_\perp^2)^4$ distribution (approximately an exponential for
     $p_\perp \le 2 $ GeV), and in longitudinal momentum $p_z$ (or rapidity) 
     according to a Gaussian distribution $\exp[-P_{beam}^2/(2 c^2)]$.
     The parameters $b$ and $c$ are given by PARV(97) and PARV(98),
     respectively. Option MSTV(92) is inactive for collisions without nuclei.
\\
    {\bf = 0 :} off, i.e. beam nucleon distributions shaped analogous to proton-proton
     collisions.
\\
    {\bf = 1 :} beam protons and neutrons are redistributed.
\\
    {\bf = 2 :} as = 1, plus $\Delta$'s associated with  beam nucleon interactions 
                are included.
\\
    {\bf = 3 :} as = 2, plus $\pi$'s and $K$'s from beam nucleon interactions are 
                accounted.
\item[ MSTV(93)] : (D=0) option to distingiush  - in beam fragmentation 
     of colisions with nuclei - into
     `unwounded' and `wounded' beam nucleons. Former are those whose initial
     parton content has not altered throughout a collisions, latter are those
     who have been struck at least once in a hard parton collision.
     Option MSTV(93) is inactive for collisions without nuclei.
\\
    {\bf = 0 :} off, i.e. spectator partons are not reassembled to their initial
     mother nucleons, but all those non-interacted spectator partons that
     belong to an `unwoundsed' nucleon, are lumped together with the spectator
     partons from `wounded' nucleons and hadronize as usual via 
     the beam-fragmentation scheme.
\\
    {\bf = 1 :} on, i.e. for all those nucleons that have not lost a daughter
     parton through parton collisions, their initial parton content is recollected,
     and these `unwounded' nucleons appear in the final state with their
     initial energy-momentum, modulo some aquired Fermi motion.
     In this case, only the partons from `wounded' nucleons participate
     in the beam-fragmentation, specifically, only a fraction of all initial-state
     nucleons which is set by PARV(99) (see below).
\\
    {\bf = 2 :} as = 1, but now the reassembled nucleons undergo subsequent
			fragmentation and are broken up.
\item[ MSTV(94)] : (D=1)
     Switch to allow to cluster together excess pions from low-mass 
     parton recombinations, if their density gets above a critical value, 
     and to convert them either into energy transferred to the environment, 
     or to convert then into particle production (see MSTV(95) and MSTV(96)).
     Specifically, close-by pions lumped together in clusters of
                mass $\ge 1 - 5$ GeV and the energy-momentum of these clusters 
                is then transferred to the remaining particle system, or,
     alternatively, the energy-momentum of $\pi$  clusters 
                with mass $\ge 1-5$ GeV is converted
                into pair production of $\pi^\pm, \rho, \omega$.
\\
    {\bf = 0 :} off, i.e. no action taken.
\\
    {\bf = 1 :} on.
\item[ MSTV(95)] : (D=0)
     For MSTV(94) $>$ 0, the critical density above which pions may percolate
     to clusters is set by  MSTV(95). IF MSTV(95) $\le 0$, its value is set
     automatically upon initialization.
\item[ MSTV(96)] : (D=0)
     For MSTV(94) $>$ 0, the maximum mass of clusters of pions that can be
     percolated is set by  MSTV(96). IF MSTV(96) $\le 0$, its value is set
     automatically upon initialization.
\item[ MSTV(97)] : (D=0)
     Switch to select shape of $p_\perp$ spectrum of final-state hadrons emerging
     from fragmentation of beam/target clusters in collisions involving nuclei.
\\
    {\bf = 0 :} exponential distribution $\propto \exp(- b m_{\perp}]$, 
                $m_\perp = \sqrt{p_\perp^2+m^2}$.
\\
    {\bf = 1 :} power-law distribution $\propto 1/(m_{\perp}^2)^b$, 
                $m_\perp = \sqrt{p_\perp^2+m^2}$.
\item[ MSTV(98)] : (D=0)
     Switch to select shape of $y$ spectrum of final-state hadrons emerging
     from fragmentation of beam/target clusters in collisions involving nuclei.
     (Notice that in case MSTV(40) $>$ 0, the camel-like option is selected
     by default, in order to compensate the spectra of hadrons from partonic clusters
     and those from soft beam/target clusters.)
\\
    {\bf = 0 :} flat-top distribution, i.e. a central plateau  with Gaussian tails.
\\
    {\bf = 1 :} camel-like distribution, i.e. a central dip with Gaussian on either side.
\item[ MSTV(99)] : (D=0)
     Switch to adjust left-right imbalance of beam/target hadron fragments, if
     there is such occurrence.
\\
    {\bf = 0 :} off.
\\
    {\bf = 1 :} on.

\noindent
\sline

\item[ PARV(81)] : (D=3.6 GeV$^{-1}$) minimum separation of 2 partons, below which
     a cluster cannot be formed (denoted $R_\chi$ in Ref. \cite{ms37}).
\item[ PARV(82)] : (D=4. GeV$^{-1}$) critical separation of 2 partons that sets the
     space-time scale for cluster formation (denoted $R_{crt}$ 
     in Ref. \cite{ms37}).
     Is related to the average cluster size and typically about 10 \% 
     above $R_\chi$.
\item[ PARV(83)] : (D=1.0) `slope' of the exponential probability distribution 
     $W(R)=\theta(R-R_\chi)\exp(-$PARV(83)$R/R_{crt})$ for MSTV(83)$>$ 0, 
     describing
     the probability of 2 partons to form a pre-hadronic 
     cluster by tunneling through 
     the `potential barrier' marked by $R_\chi$ and $R_{crt}$, with average
     cluster size $R_{cl} =2 R_{crt}/$PARV(83).
\item[ PARV(84)] : (D=0.3 GeV) minimum invariant mass of 2 partons required to 
     form a cluster being necessarily above threshold for hadron decay.
\item[ PARV(85)] : (D=1000. GeV) maximum allowed invariant mass of 2 partons to form
     a cluster with mass $M_{cl}$, in order to provide option to suppress 
     production of  heavy clusters - in addition to the  requirement of PARV(81),PARV(82).
     Note: default corresponds to no suppression.
\item[ PARV(86)] : (D=5. GeV$^{-1}$) maximum allowed separation of 
     2 partons that are
     unambigous cluster candidates. Cluster formation is enforced above
     this value, in order to avoid unphysical separation. In this case
     the requirement $M_{cl}$ $<$ PARV(85) is overridden.
\item[ PARV(87)] : (D=1.) `slope' of exponential Hagedorn-type distribution for
     sampling the masses of clusters in their 2-body decays, given by
     $\exp(- a \cdot M_{cl}/M_0)$, where $a =$ PARV(87), 
     $M_{cl}$ is the cluster mass and 
     $M_0 = 0.2$ GeV is the `Hagedorn temperature'  (see Ref. \cite{ms37}).
\item[ PARV(88)] : (D=1. GeV) boundary for distinguishing `slow' and `fast' 
     clusters in statistics accumulation of exogamous cluster production
     (c.f. MSTV(88)).
\item[ PARV(89)] : (D=0.2) determines value of $y_c$ = PARV(89) $y_{beam}$
     for smearing of parton cluster rapidities within an 
     interval $| y | \le y_c$, if MSTV(89) $>$ 0.
\item[ PARV(90)] : (D=0.0) determines value of $y_c$ = PARV(90) $y_{beam}$
     for smearing of beam cluster rapidities within an 
     interval $| y | \le y_c$. If PARV(90) $<$ 0, it is ste automatically
     to pre-fitted value, otherwise, PARV(90) $>$ 0 as user input takes
     this as fixed value. 
\item[ PARV(91)] : (D=0.0) Parametric factor for charged multiplicity 
     from soft fragmentation of beam/target remnants that governs the magnitude
     of $N_{ch}(\sqrt{s})$ of hadrons emerging from spectator partons. 
     If PARV(91) $\le$ 0, then the value of PARV(91) is set automatically
     to pre-fitted value, otherwise, PARV(91) $>$ 0 as user input takes
     this as fixed value. (Note: applies only if if MSTV(90) = 1).
\item[ PARV(92)] : (D=0.0) Factor multiplying the energy of the remnant
     beam/target clusters that are formed from the spectator partons,
     and that are fragmented with the total energy E=PARV(92)$\cdot (E_{beam}+E_{target})$. If PARV(92) $\le 0$, then it is set automatically upon
     initialization.
\item[ PARV(93)] : (D=1.0) Nuclear dependence  in $eA$, $hA$ or $AB$ collisions 
     of the 
     charged multiplicity $N_{ch}$, resulting from soft fragmentation of nuclear 
     beam or/and target remnants of the original nuclei with mass number 
     $A\; (B)$ (case of hadron $h$ corresponds to $A=1$ or $B=1$). 
     It is parametrized as: 
     $N_{ch}^{(AB)}(\sqrt{s},A,B) = \left[(A+B)/2\right]^\beta \, 
     N_{ch}^{(p\bar p)}(\sqrt{s})$, where $\beta =$ PARV(93).
\item[ PARV(94)] : (D=15.GeV$^{-1}$) Time in the global collision frame when soft
     fragmentation starts to become gradually active (default corresponds to 35 fm.)
\item[ PARV(95)] : (D=1000.GeV$^{-1}$) The positions of particles produced 
     in fragmentation
     of both parton clusters and beam/target clusters, are assigned such 
     they become active only after an effective formation time 
     $\Delta t = PARV(95) \times \min(E/M^2,$PARV(95)) (default value corresponds 
     $\Delta t \le 200$ fm).
\item[ PARV(96)] : (D=40.) proportionality factor in the uncertainty relation 
     $\Delta t = PARV(96) \times E/M^2$, which determines the mean formation
     time of particles produced in soft beam/target fragmentation (c.f. PARV(95)).
\item[ PARV(97)] : (D=0.35 GeV) 
     parameter $b$ in  $1/(1+b \,p_\perp^2)^4$ distribution for simulating
     {\it stopping effect} in nucleus-nucleus collisions of protons and neutrons 
     (and optionally associated $\Delta$'s)  
     by redistributing them  in transverse momentum $p_\perp$ (c.f. MSTV(92)).
\item[ PARV(98)] : (D=0.14 GeV) 
     parameter $c$ in Gaussian distribution $\exp[-P_{beam}^2/(2 c^2)]$ for
     simulating {\it stopping effect} in nucleus-nucleus collisions 
      of protons and neutrons (and associated $\Delta$'s)  
     by redistributing them  in longitudinal momentum $p_z$ (c.f. MSTV(92)).
\item[ PARV(99)] : (D=0.) 
     Of all initial-state nucleons which 
     have not lost a daughter parton through parton collisions (those that remained
     `unwounded' nucleons) and  appear in the final state with their
     initial energy-momentum, a fraction $0 \le $PARV(93)$ \le 1$ 
     is fragmented via the usual beam/target fragmentation scheme,
      whereas the fraction 1$-$PARV(99) appears as original nucleons
     in the final state. Is inactive for MSTV(93) = 0.

\noindent
\sline
\end{description}
\medskip

\noindent
\lline

\noindent  {\bf 7.  `Afterburner' Hadron  cascade:}

\noindent
\lline

\begin{description}
\item[ MSTV(100)] : (D=3) switch to select various levels
of final-state hadron cascading in terms of scattering
and resonance production with decay of direction of primary
hadrons emerging from both, the parton cluster formation and
the beam/target fragmentation. (See additional control switches 
MSTV(101)-MSTV(103)).
\\
    {\bf = 0 :} switched off completely, i.e., no final-state hadron cascade.
\\
    {\bf = 1 :} switched on, but including {\it only}
                hadron cascading of hadrons from parton cluster decays
                and excluding re-interactions among hadrons from the beam/target
                fragmentation.
\\
    {\bf = 2 :} switched on, but in contrast to  = 1, now  {\it only}
                hadron cascading of hadrons from beam/target fragmentation
                are allowed ans  re-interactions among hadrons from the 
                parton cluster decays are excluded.
\\
    {\bf = 3 :} switched on in full, i.e., any hadron can re-interact
                irrespective to its origin.
\item[ MSTV(101)] : (D=1) Additional selection option in addition to MSTV(100)
that allows to include/exclude hadrons that have not become `real', i.e.,
that have not passed through their formation time.
\\
    {\bf = 0 :} only on-shell hadrons can interact, that is, those which
                have passed their formation time.
\\
    {\bf = 1 :} both on-shell hadrons and off-shell hadrons which
                have {\it not} passed their formation time yet, can 
                take part in the hadron cascading.
\\
\item[ MSTV(102)] : (D=3) switch to include/exclude interactions in which 
                    pre-hadronic clusters (not yet formed hadrons) are involved
                    (note: applies only if MSTV(101) = 1).
\\
    {\bf = 0 :} no pre-hadronic clusters can interact.
\\
    {\bf = 1 :} only pre-hadronic clusters from parton cluster formation
                may participate in the hadron cacade.
\\
    {\bf = 2 :} in contrast to = 1, now only clusters from beam/target
                fragmentation are included in the hadron cacade.
\\
    {\bf = 3 :} any pre-hadronic cluster can interact with other hadrons
                or clusters, irrespective of the origin.
\item[ MSTV(103)] : (D=2) Sets (for MSTV(102) $\ge 1$) the maximum number 
                 of preformed hadrons for which is
                 searched for in the event record in order to reconstruct
                 the original cluster from whose decay the hadrons emerged.
\item[ MSTV(105)] :  -- not used --
\medskip

\noindent
\sline

\item[ PARV(100)]{\bf - PARV(104)} : -- not used --
\item[ PARV(105)] : (D=5. GeV$^{-1}$) maximum separation of two hadronic
    particles to be able to
    interact through a 2-body collision. For MSTV(105)=0, it is fixed at
    $r_{12} =$ PARV(105), whereas for MSTV(105)=1, it is taken s-dependent as 
    $r_{12} =$ PARV(105)$/(\sqrt{{s}}/GeV)$.
\item[ PARV(106)] : (D=1.0) parameter in the collision-probability distribution 
    $W(x)=\theta(a-x)$ (for MSTV(106)=0) and $W(x)=\exp(-x/a)$ for MSTV(106)=1,
    with $x = r_{12}^{(sep)}/r_{12}^{({\sigma})}$, $r_{12}^{(sep)}$
    the relative transverse hadron 
    separation of closest approach and $r_{12}^{(\hat{\sigma})}$ 
    the transverse radius of interaction as given by the 
    hadron-hadron cross-section $\sigma$.
\medskip
\end{description}
                    
\noindent
\lline

\noindent  {\bf 8.  Other settings:}

\noindent
\lline

\begin{description}
\item[$\bullet$]
{\it Handling of input/output:}
\end{description}

\begin{description}
\item[ MSTV(110)] : (D=1) direction of program output.
\\
    {\bf = 0 :} all output is directed to standard output unit 6.
\\
    {\bf = 1 :} output is directed to units MSTV(111)-MSTV(113).
\item[ MSTV(111)] : (D=10) unit number to which general output is directed.
\item[ MSTV(112)] : (D=6) unit number for writing `on-line' information.
\item[ MSTV(113)] : (D=20) unit number for listing warnings and error messages.
\item[ MSTV(114)] : (D=0) switch for general information on unit MSTV(111), 
    which is directed automatically to file VNIRUN.DAT. Allows to
    select the amount of output provided on the global performance of a 
    simulation consisting of of a sample of collision events:
\\
    {\bf = 0 :} only minimal output concerning selected process and 
    main parameters is written out, as well as final summary of results.
\\
    {\bf = 1 :} as =0, plus a listing the 1st event in momentum and 
    position space.
\\
    {\bf = 2 :} as =1, plus listing of number and properties of elementary
    subprocesses that occurred during the simulation,
    written to files VNICOL.DAT, VNISBR.DAT and VNITBR.DAT
    for collisions, space-like and time-like branchings, respectively.
\\
    {\bf = 3 :} as =2, plus detailed event listing of particle record
    at the end of each single event, written to VNIRUN.DAT. 
    NOTE: for large particle systems,
    this slows down the program substantially, and moreover requires
    a lot of disc space if a large number of events are printed).
\item[ MSTV(115)] : (D=2) switch for `on-line' information on unit MSTV(112),
    i.e. writing of initialization and termination info, and  initial 
    and final entries for each event as the simulation goes on.
\\
    {\bf = 0 :} no `online output'.
\\
    {\bf = 1 :} only initialization and finalization info is printed.
\\
    {\bf = 2 :} 1-line info at beginning and at the end of each event
	to keep control of the program performance.
\\
    {\bf = 3 :} as = 2, plus listing of entries in  particle record
    after
    each single time-step in an event (useful for AB collisions).
\\
    {\bf = 4 :} as = 3, plus writing `standardized' output to file
        VNIOSC.DAT, according to the OSCAR (Open Standard Codes At RHIC)
        conventions established at RIKEN-BNL, June 1997.
\end{description}

\begin{description}
\item[$\bullet$]
{\it Miscellanous control switches for subroutines:}
\end{description}

\begin{description}
\item[ MSTV(120)] : (D=0) flag to indicate initialization of certain HERWIG 
    common blocks and default values that are necessary for using
    parts of the HERWIG program for the hadronization of clusters.
    Is set equal to 1 after first intitialization call. 
\item[ MSTV(121)] : (D=0) if MSTV(121) is set to 1 before a VNIROBO call, 
    the V vectors are reset (in the particle range to be rotated/boosted),
    i.e., they  are set to 0 before the rotation/boost. If MSTV(121) is
    equal to zero, the V vectors are not reset. MSTV(121) is inactive 
    during a VNIROBO call and is set back to 0 upon return.
\item[ MSTV(122)] : (D=0) specifies in a VNIROBO call the type of rotation of the 
    P, R, and V vectors. The rotation is clockwise (active rotation) for 
    MSTV(122)=0 and anticlockwise (passive rotation) otherwise. The 
    value of MSTV(122) is set back to 0 upon return.
\item[ MSTV(123)] : (D=0) specifies in a VNIROBO call for performing a boost,
    whether the vectors R are boosted (MSTV(123)=0) or not (MSTV(123)=1).
    The vectors P are always boosted. The value of MSTV(123) is set back
    to 0 upon return.
\item[ MSTV(124)] : (D=1) pointer to lowest entry in particle record 
     VNIREC1/VNIREC2 to be 
     included in data analysis routines VNIANA1-VNIANA4.
\item[ MSTV(125)] : (D=100000) pointer to highest entry in particle record 
     VNIREC1/VNIREC2 
     to be included in data analysis routines VNIANA1-VNIANA4.
\item[ MSTV(126)] : (D=20) number of lines in the beginning of the particle 
     record that are reserved for internal event history information. The 
     value should not be reduced, but can be increased if necessary.
\item[ MSTV(127)] : (D=1) in listing of the current state of the particle record
     by calling the subroutine VNILIST, it can be chosen between listing 
     either color/anticolor $C$ $A$ and the space time origin of a particle $I$
     according to the information contained  $V(I,5)$ explained above, 
     ( = 0 ), or, alternatively the mother $KMO$ and color flow information
     contained in $K(I,3)-K(I,5)$, ( = 1 ). All other listed quantities are 
     the same for either option.
\item[ MSTV(128)] : (D=0) specifies in calls to VNIEDIT and VNIROBO the classes
   of particles to be included.
   The default = 0 {\it ex}cludes all inactive or decayed particles with $K(I,1) \le 0$
   or $K(I,1) > 10$, whereas the setting = 1 {\it in}cludes any entry 
    listed in the particle record.  MSTV(128) is set back to 0 upon return.
\item[ MSTV(129)] : (D=0) specifies in calls to the `color-linking' routine
            VNIJCOL the way of encoding the color connections between partons
            in the components $J=4,5$ of the array $K(I,J)$. For very large number
            of entries ($\ge 35000$) in the record VNIREC, MSTV(129) is set automatically to
            = 1 in the hadronization process, in order to save memory space. 
            It is reset to = 0 afterwards.
\end{description}

\begin{description}
\item[$\bullet$]
{\it Handling of errors and warnings:}
\end{description}

\begin{description}
\item[ MSTV(130)] : (D=2) check on possible errors during program execution.
\\
    {\bf = 0 :} errors do not cause any immediate action.
\\
    {\bf = 1 :} possible errors are checked and in case of problem, it is
        excited from the subprogram, but the simulation is continued 
        from there on. For the first MSTV(101) errors a message is 
        printed; after that no messages appear.
\\
    {\bf = 2 :} possible errors are checked and in case of problem, the
        the simulation is forcibly terminated. For the first MSTV(101) 
        errors a message is printed; after that no messages appear.
\item[ MSTV(131)] : (D=10) max number of errors that are printed.
\item[ MSTV(132)] : (D=1) printing of warning messages.
\\
    {\bf = 0 :} no warnings are written.
\\
    {\bf = 1 :} first MSTV(103) warnings are printed, thereafter no warnings
        appear.
\item[ MSTV(133)] : (D=10) max number of warnings that are printed.
\end{description}

\begin{description}
\item[$\bullet$]
{\it Version, date of last change:}
\end{description}

\begin{description}
\item[ MSTV(140)] : (D=1) Print-out of VNI logo on first occasion; MSTV(140) 
       is reset to 0 afterwards.
\item[ MSTV(141)] : VNI version number.
\item[ MSTV(142)] : VNI subversion number.
\item[ MSTV(143)] : year of last change of VNI version
\item[ MSTV(144)] : month of last change of VNI version.
\item[ MSTV(145)] : day of last change of VNI version.

\end{description}
\medskip

    
\noindent
\lline
         
\noindent {\bf 9. Statistics, event study and data analysis:} 

\noindent
\lline

\begin{description}
\item[$\bullet$]
{\it  Particle record} VNIREC1/VNIREC2: {\it Choosing event history information.}
\end{description}

\begin{description}
\item[ MSTV(150)] : (D=0) specifies whether the complete particle
    history with all particles produced at any time during an event, 
    is kept throughout the simulation (rather inefficient for large 
    systems) or whether it is compressed at the end of each time step 
    (thereby loosing information on parent/child genealogy etc.).
\\
    {\bf = 0 :} particle record is greatly compressed after each time step,
                all `dead' particles are removed and genealogical history is lost.
\\
    {\bf = 1:} particle record is partly compressed, keeping some of the
               genelogical history.
\\
    {\bf = 2:} full particle history is kept throughout.
\end{description}

\begin{description}
\item[$\bullet$]
{\it Subroutine} VNIANA1: {\it General event statistics.}
\end{description}

\begin{description}
\item[ MSTV(151)] : (D=0) Statistics on initial parton state.
\item[ MSTV(152)] : (D=0) Number and momenta of produced particles.
\item[ MSTV(153)] : (D=0) Factorial moments.
\item[ MSTV(154)] : (D=0) Energy-energy correlations.
\item[ MSTV(155)] : (D=0) Decay channels.
\end{description}

\begin{description}
\item[$\bullet$]
{\it Subroutine} VNIANA2: {\it Statistics on hadronic observables.}
\end{description}

\begin{description}
\item[ MSTV(161)] : (D=0) time-evolution of flavor, energy and momentum composition.
\item[ MSTV(162)] : (D=0) Distributions of particles in $y=\ln(1/x)$.
\item[ MSTV(163)] : (D=0) Bose-Einstein correlation analysis.
\end{description}

\begin{description}
\item[$\bullet$]
{\it Subroutine} VNIANA3: {\it Statistics on pre-hadronic clusters.}
\end{description}

\begin{description}
\item[ MSTV(171)] : (D=0) Rapidity spectra of clusters $dN/dy$.
\item[ MSTV(172)] : (D=0) Longitudinal space-time spectra  $dN/dz$.
\item[ MSTV(173)] : (D=0) Transverse momentum spectra $(1/p_\perp)\;dN/dp_\perp$
\item[ MSTV(174)] : (D=0) Transverse space-time spectra $(1/r_\perp)\;dN/dr_\perp$.
\item[ MSTV(175)] : (D=0) Distributions of cluster sizes and masses.
\item[ MSTV(176)] : (D=0) Polarization profile of cluster density.
\end{description}

\begin{description}
\item[$\bullet$]
{\it Subroutine} VNIANA4: {\it Statistics on partons and produced hadrons.}
\end{description}

\begin{description}
\item[ MSTV(181)] : (D=0) Rapidity spectra of partons and hadrons $dN/dy$.
\item[ MSTV(182)] : (D=0) Longitudinal space-time spectra  $dN/dz$.
\item[ MSTV(183)] : (D=0) Transverse momentum spectra $(1/p_\perp)\;dN/dp_\perp$.
\item[ MSTV(184)] : (D=0) Transverse space-time spectra $(1/r_\perp)\;dN/dr_\perp$.
\end{description}


\noindent
\lline

\bigskip
\bigskip

\begin{center}
{\bf MSTW(200), PARW(200): generated quantities and statistics}
\end{center}
\bigskip   

\noindent
\lline

\noindent{\bf 1. General:}

\noindent
\lline

\begin{description}

\item[ MSTW(1)]  : total number of collision events to be generated.
\item[ MSTW(2)]  : total number of time steps per collision event.
\item[ MSTW(3)]  : physics process $A+B$ that is simulated. 
     The current available
     beam ($A$) and target ($B$) particles and the collision processes are the
     ones listed before in Table 1, with MSTW(3)=IPRO (see also selection switch
     MSTV(30)).
\item[ MSTW(4)]  : Overall Lorentz frame of reference for event siumulation.
\\
    = 1 : center-of-momentum frame (CMS) of beam and target particle.
\\
    = 2 : fixed-target frame (FIXT) with target particle at rest.
\\
    = 3 : user-defined frame (USER) with given {\it 3-momentum} of beam
          and target. Particles are assumed on the mass shell.
\\
    = 4 : user-defined frame (FOUR) with given {\it 4-momentum} 
         ( i.e., 3-momentum and energy) of beam
          and target particle. The particles need not to be on the mass shell.
\\
    = 5 : user-defined frame (FIVE) with given {\it 5-momentum} 
         ( i.e., 3-momentum, energy and mass) of beam
          and target particle. The particles need not to be on the mass shell,
          but 4-momentum and mass information must(!)  match. 
\item[ MSTW(5)] : KF flavour code for beam  particle $A$.
\item[ MSTW(6)] : KF flavour code for target  particle $B$.
\item[ MSTW(7)] : type of incoming beam particle $A$: 
      1 for lepton, 2 for hadron, and 3 for nucleus.
\item[ MSTW(8)] : type of incoming target particle $B$: 
     1 for lepton, 2 for hadron, and 3 for nucleus.
\item[ MSTW(9)] : combination of incoming beam and target particles.
\\
    {\bf = 1 :} lepton  on  lepton
\\
    {\bf = 2 :} lepton  on  hadron
\\
    {\bf = 3 :} lepton  on nucleus
\\
    {\bf = 4 :} hadron  on  lepton
\\
    {\bf = 5 :} hadron  on  hadron
\\
    {\bf = 6 :} hadron  on nucleus
\\
    {\bf = 7 :} nucleus on  lepton
\\
    {\bf = 8 :} nucleus on  hadron
\\
    {\bf = 9 :} nucleus on  nucleus
\item[ MSTW(10)] : performance flag for current event. Is  =0 at beginning, and 
    set =1 if during the evolution the event turns out to be rejectable,
    either due to unphysical kinematics, particle combinations, etc.,
    or due to numerical errors.
\item[ MSTW(11)] : current collision event.
\item[ MSTW(12)] : current time step in this collision event.
\item[ MSTW(13)] : current number of successful collision events,
     i.e. those that completed gracefully with MSTW(10)=0.
\item[ MSTW(14)] : counter for "non-diffractive" collisions events,
     i.e. those that involved at least one parton 
    collision (in  hadron or nucleus collisions).
\item[ MSTW(15)] :
Status of Lorentz transformation between different global Lorentz frames
in which the collision event is simulated  (which may be different from
the initially specified frame, c.f. MSTW(4)). The value of MSTW(15)
saves the last performed transformation:
\\
{\bf = 1 :} from global $cm$-frame to fixed-target or
user-specified frame;
\\
{\bf = 2 :} from global $cm$-frame to hadronic center-of-mass in
DIS (photon-hadron $cm$-frame);
\\
{\bf = -1 :} from fixed-target or user-specified frame
to global $cm$-frame;
\\
{\bf = -2 :} from  hadronic center-of-mass in
DIS (photon-hadron $cm$-frame) to global $cm$-frame.
\medskip

\noindent
\sline

\item[ PARW(1)] : time increment $TINC(I)$ in current time step $I$.
\item[ PARW(2)] : time $TIME(I)$ in current time step $I$.
\item[ PARW(3)] : local machine time (in CPU) passed when simulation started.
\item[ PARW(4)] : local machine time (in CPU) when simulation ended.
\item[ PARW(5)] : conversion factor CPU $\rightarrow$ seconds 
     (machine-dependent: for most processors, it is equal to 0.01, i.e. 
      1 sec = 100 CPU.
\item[ PARW(11)] : Total invariant $\sqrt{s} = E_{CM}$, i.e. the total CM energy
     of the collision system.
\item[ PARW(12)] : Invariant $s=E_{CM}^2$ mass-square of complete system.
\item[ PARW(13)] : CM energy per nucleon $E_{CM}/(A+B)$ (where 
     $A$ ($B$) is the mass number of beam (target), and $A+B$ is the total 
     number of nucleons in the system) in collisions involving nuclei. 
     Is equal to $E_{CM}$ for elementary particle- and hadronic collisions.
\item[ PARW(14)] : Invariant $s_{eff} =E_{CM}^2/(A+B)$ mass-square per nucleon 
     in collisions involving nuclei. 
     Is equal to $s$ for elementary particle- and hadronic collisions.
\item[ PARW(15)] : mass $M_A$ of beam particle.
\item[ PARW(16)] : mass $M_B$ of target particle.
\item[ PARW(17)] : longitudinal momentum $P_{z\,A}$
      of beam  particle in specified global frame.
\item[ PARW(18)] : longitudinal momentum $P_{z\,B}$
      of target  particle in specified global frame.
\item[ PARW(19)] : angle $\theta$ of rotation from CM frame to user-defined 
      frame (also, in $e^+e^-$ via $W^+W^-$, the rotation of the 
      $W$-pair along the $z$-axis).
\item[ PARW(20)] : azimuthal angle $\phi$ of rotation from CM frame to 
      user-defined frame,  corresponding to PARW(18).

\noindent
\sline

\end{description}
\medskip

\noindent
\lline

\noindent {\bf 2. Initial state of collision system:}

\noindent
\lline

\begin{description}
\item[ MSTW(21)] : number of neutrons of beam particle $A$.
\item[ MSTW(22)] : number of neutrons of target particle $B$.
\item[ MSTW(23)] : number of protons of beam particle $A$.
\item[ MSTW(24)] : number of protons of target particle $B$.
\item[ MSTW(25)] : number of $d$-valence quarks of beam particle $A$.
\item[ MSTW(26)] : number of $d$-valence quarks of target particle $B$.
\item[ MSTW(27)] : number of $u$-valence quarks of beam particle $A$.
\item[ MSTW(28)] : number of $u$-valence quarks of target particle $B$.
\item[ MSTW(29)] : number of $s$-valence quarks of beam particle $A$.
\item[ MSTW(30)] : number of $s$-valence quarks of target particle $B$.
\item[ MSTW(31)] : total number of initial state partons in collisions 
           involving one or more hadron or nucleus.
\item[ MSTW(32)] : number of initial gluons.
\item[ MSTW(33)] : $10^5\times$ number of $d$ quarks + number of $\bar{d}$ antiquarks.
\item[ MSTW(34)] : $10^5\times$ number of $u$ quarks + number of $\bar{u}$ antiquarks.
\item[ MSTW(35)] : $10^5\times$ number of $s$ quarks + number of $\bar{s}$ antiquarks.
\item[ MSTW(36)] : $10^5\times$ number of $c$ quarks + number of $\bar{c}$ antiquarks.
\item[ MSTW(37)] : $10^5\times$ number of $b$ quarks + number of $\bar{b}$ antiquarks.
\item[ MSTW(38)] : $10^5\times$ number of $t$ quarks + number of $\bar{t}$ antiquarks.
\medskip

\noindent
\sline

\item[ PARW(21)] : absolute value of velocity $\vec{\beta}_A$ of beam particle $A$.
\item[ PARW(22)] : absolute value of velocity $\vec{\beta}_B$ of target particle$B$.
\item[ PARW(23)] : Lorentz factor $\gamma_A$ of beam particle $A$.
\item[ PARW(24)] : Lorentz factor $\gamma_B$ of target particle $B$.
\item[ PARW(25)] : rapidity $Y_A$ of beam particle $A$.
\item[ PARW(26)] : rapidity $Y_B$ of target particle $B$.
\item[ PARW(27)] : radius $R_A$  (in 1/GeV) of beam particle in its restframe.
\item[ PARW(28)] : radius $R_B$ (in 1/GeV) of target particle in its restframe.
\item[ PARW(29)] : radius $R_n$ (in 1/GeV) of neutron within in nucleus.
\item[ PARW(30)] : radius $R_p$ (in 1/GeV) of proton within in nucleus.
\item[ PARW(31)] : accumulated average $Q_0$ of initial hadron (nucleus) parton 
    distributions (equals the average momentum scale of primary parton 
    collisions $\langle Q_{prim} \rangle$).
\item[ PARW(32)] : accumulated average momentum fraction $x_A$ of beam-side
    initial hadron (nucleus) parton distributions.
\item[ PARW(33)] : accumulated average momentum fraction $x_B$ of target-side
    initial hadron (nucleus) parton distributions.
\item[ PARW(34)] : 3 times the total charge of beam-side particle.
\item[ PARW(35)] : 3 times the total charge of target-side particle.
\item[ PARW(36)] : ratio $b/b_{max}$ of actual impact parameter
    $b_{min} \le b \le b_{max}$ to maximum allowed impact parameter
    of beam and target particles, if variable impact parameter is
    chosen  (c.f. MSTV(32) and PARV(32), PARV(33)).
\item[ PARW(37)] : azimuthal angle of colliding beam and target particles
    around the beam axis, if variable impact parameter is chosen.
\item[ PARW(38)] : sign of $r_x$-coordinate of beam-particle $A$ in global frame
    for variable impact parameter selection (= $-$ sign of target-particle $B$).
\item[ PARW(39)] : sign of $r_y$-coordinate of beam-particle $A$ in global frame
    for variable impact parameter selection (= $-$ sign of target-particle $B$).

\noindent
\sline

\end{description}
\medskip

\noindent
\lline

\noindent {\bf 3. Parton scatterings:}

\noindent
\lline

\begin{description}

\item[ MSTW(40)] : Total number of 2-parton collisions $a+b \rightarrow N$ 
                   ($N = 1,2$).
\item[ MSTW(41)] : Number of hard $2 \rightarrow 2$ collisions of $q+q, q+\bar{q}, 
    \bar{q}+\bar{q}$.
\item[ MSTW(42)] : Number of hard $2 \rightarrow 2$ collisions of $q+g, \bar{q}+g$.
\item[ MSTW(43)] : Number of hard $2 \rightarrow 2$ collisions of $g+g$.
\item[ MSTW(44)] : Number of soft $2 \rightarrow 2$ collisions of $q+q, q+\bar{q}, 
    \bar{q}+\bar{q}$.
\item[ MSTW(45)] : Number of soft $2 \rightarrow 2$ collisions of $q+g, \bar{q}+g$.
\item[ MSTW(46)] : Number of soft $2 \rightarrow 2$ collisions of $g+g$.
\item[ MSTW(47)] : Number of $2 \rightarrow 1$ fusions of $q+\bar{q}$.
\item[ MSTW(48)] : Number of $2 \rightarrow 1$ fusions of $q+g, \bar{q}+g$.
\item[ MSTW(49)] : Number of $2 \rightarrow 1$ fusions of $g+g$.
\medskip

\noindent
\sline

\item[ PARW(40)] : total number of `primary' (i.e., first) 
     collisions among primary partons.
\item[ PARW(41)] : accumulated average $Q^2$ of the `primary' collisions.
\item[ PARW(42)] : accumulated average $\sqrt{\hat{s}}$ of hard $2 \rightarrow 2$ 
     collisions.
\item[ PARW(43)] : accumulated average rapidity $y^\ast=y_1^\ast-y_2^\ast$ of hard 
     $2\rightarrow 2$ collisions.
\item[ PARW(44)] : accumulated average $p_\perp$ of hard 
     $2\rightarrow 2$ collisions.
\item[ PARW(45)] : accumulated average $\sqrt{\hat{s}}$ of soft
     $2\rightarrow 2$ collisions.
\item[ PARW(46)] : accumulated average rapidity $y^\ast=y_1^\ast - y_2^\ast$ of soft
     $2\rightarrow 2$ collisions.
\item[ PARW(47)] : accumulated average $p_\perp$ of soft 
     $2\rightarrow 2$ collisions.
\item[ PARW(48)] : accumulated average $\sqrt{\hat{s}}$ of hard
     $2\rightarrow 1$ collisions.
\item[ PARW(49)] : accumulated average rapidity $y^\ast=y_1^\ast - y_2^\ast$ of hard 
     $2\rightarrow 1$ collisions.
\item[ PARW(50)] : integrated  2-parton cross-section $\hat{\sigma}(\hat{s})
    =\int dp_\perp^2 (d\hat{\sigma}(\hat{s},p_\perp^2)/dp_\perp^2)$.

\noindent
\sline

\end{description}
\medskip

\noindent
\lline

\noindent {\bf 4.  Space-like parton branchings:}

\noindent
\lline

\begin{description}

\item[ MSTW(50)] : Total number of space-like branchings $a \rightarrow b+c$.
\item[ MSTW(51)] : Number of space-like processes  $CMshower \rightarrow b+c$.
\item[ MSTW(52)] : Number of space-like processes  
     $q (\bar{q}) \rightarrow q (\bar{q})+g$.
\item[ MSTW(53)] : Number of space-like processes  $ g  \rightarrow   g+g$.
\item[ MSTW(54)] : Number of space-like processes  $ g   \rightarrow  q+\bar{q}$.
\item[ MSTW(58)] : Number of space-like processes  
     $q/q~ \rightarrow q (\bar{q}) +\gamma$.

\medskip

\noindent
\sline

\item[ PARW(50)] : average $\sqrt{Q^2}$ of starting scale for 
   space-like branching evolution.
\item[ PARW(51)] : average $\sqrt{-q^2}$ of virtuality of branching particle.
\item[ PARW(52)] : average value of $x$ in $x \rightarrow x'x''$ branching.
\item[ PARW(53)] : average fraction $z = x'/x$ of space-like branchings.
\item[ PARW(54)] : average relative transverse momentum $q_\perp$ of 
     space-like branchings.
\item[ PARW(55)] : average longitudinal momentum $q_z$ of branching particle.
\item[ PARW(56)] : average energy of $q^0$ branching particle.
\item[ PARW(57)] : average invariant mass $\sqrt{|q^2|}$ of branching particle.

\noindent
\sline
             
\end{description}
\medskip

             
\noindent
\lline

\noindent {\bf 5.  Time-like parton branchings:}

\noindent
\lline

\begin{description}

\item[ MSTW(60)] : Total number of time-like branchings $a \rightarrow b+c$.
\item[ MSTW(61)] : Number of time-like processes  $CMshower \rightarrow b+c$.
\item[ MSTW(62)] : Number of time-like processes  
    $q (\bar{q}) \rightarrow q (\bar{q})+g$.
\item[ MSTW(63)] : Number of time-like processes   $g  \rightarrow   g+g$.
\item[ MSTW(64)] : Number of time-like processes   $g  \rightarrow  q+\bar{q}$.
\item[ MSTW(68)] : Number of time-like processes  
    $q (\bar{q}) \rightarrow q (\bar{q}) + \gamma$.
\medskip

\noindent
\sline

\item[ PARW(60)] : average $\sqrt{Q^2}$ of starting scale for 
time-like branching evolution.
\item[ PARW(61)] : average $\sqrt{k^2}$ of virtuality of branching particle.
\item[ PARW(62)] : average value of $x$ in $x \rightarrow x'x''$ branching.
\item[ PARW(63)] : average energy fraction $z = x'/x$ of time-like branchings.
\item[ PARW(64)] : average relative  transverse momentum $k_\perp$ 
    of time-like branchings.
\item[ PARW(65)] : average longitudinal momentum $k_z$ of branching particle.
\item[ PARW(66)] : average energy of branching particle.
\item[ PARW(67)] : average invariant mass $\sqrt{m^2}$ of branching particle.

\noindent
\sline
\end{description}
\medskip

\noindent
\lline

\noindent {\bf 6. Parton-cluster formation and cluster-hadron decay: }

\noindent
\lline

\begin{description}

\item[ MSTW(80)] : Total number of 2-parton cluster formations.
\item[ MSTW(81)] : Total number of $gg$ clusterings, $gg \rightarrow CC + X$, 
\     where $X$ is either `nothing', or $g$, or $gg$.
\item[ MSTW(82)] : Number of processes, $gg \rightarrow CC + g$.
\item[ MSTW(83)] : Number of processes, $gg \rightarrow CC + gg$.
\item[ MSTW(84)] : Total number of $q\bar{q}$ clusterings, 
     $q\bar{q} \rightarrow CC + X$, where $X$ is either `nothing', 
     or $g$, or $gg$.
\item[ MSTW(85)] : Number of processes, $q\bar{q} \rightarrow CC + g$.
\item[ MSTW(86)] : Number of processes, $q\bar{q} \rightarrow CC + gg$.
\item[ MSTW(87)] : Total number of $gq\;(g\bar{q})$ clusterings, 
     $gq \rightarrow C + X$, where $X$ is either $q$ or $gq$.
\item[ MSTW(88)] : Number of processes, $gq \rightarrow CC + gq$.
\item[ MSTW(89)] : Total number of $(qq)q \rightarrow CC$, 
     $(\bar{q}\bar{q})\bar{q} \rightarrow CC$ clusterings.
\item[ MSTW(90)]{\bf - MSTW(99)} : Same as MSTW(80) - MSTW(89), but now for the 
      corresponding numbers of `exogamously' produced clusters, with 
      `exogamy index' $e_{cl} = (e_i+e_j)/2$ unequal to 0 or 1 
       (see Ref. \cite{ms40}).
\item[ MSTW(100)] : Total number of `fast' clusters, classified according to
      the choices of MSTV(88)] and PARV(88).
\item[ MSTW(101)] : Total number of `slow' clusters, the fraction complimentary
      to the `fast' one.
\item[ MSTW(110)] : Total number of primary "neutral" particles per event, 
      produced directly by cluster-hadron decays.
\item[ MSTW(111)] : Number of primary leptons and gauge bosons.
\item[ MSTW(112)] : Number of primary light mesons.
\item[ MSTW(113)] : Number of primary strange mesons.
\item[ MSTW(114)] : Number of primary charm and bottom mesons.
\item[ MSTW(115)] : Number of primary tensor mesons.
\item[ MSTW(116)] : Number of primary light baryons.
\item[ MSTW(117)] : Number of primary strange baryons.
\item[ MSTW(118)] : Number of primary charm and bottom baryons.
\item[ MSTW(119)] : Number of other particles.
\item[ MSTW(120)]{\bf - MSTW(129)} : Same as MSTW(110) - MSTW(119), but now for 
      charged particles only.
\item[ MSTW(130)]{\bf - MSTW(139)} : Same as MSTW(110) - MSTW(119), but now for 
      the numbers of secondary neutral particles per event (i.e. 
      those produced by decays of unstable particles).
\item[ MSTW(140)]{\bf - MSTW(149)} : Same as MSTW(130) - MSTW(139), but now for 
      charged particles only.
\medskip

\noindent
\sline

\item[ PARW(81)] : accumulated average space-time separation of clustered 
     partons.
\item[ PARW(82)] : accumulated average energy-momentum separation of clustered 
     partons.
\item[ PARW(83)] : minimum  encountered separation of any two clustered
     partons with respect to chosen space-time measure (c.f. MSTV(81)).
\item[ PARW(84)] : maximum encountered separation of any two clustered
     partons with respect to chosen space-time measure (c.f. MSTV(81)).
\item[ PARW(85)] : minimum encountered separation of any two clustered
     partons with respect to chosen energy-momentum measure (c.f. MSTV(82)).
\item[ PARW(86)] : maximum encountered separation of any two clustered
     partons with respect to chosen energy-momentum measure (c.f. MSTV(82)).
\item[ PARW(91)]{\bf - PARW(99)} : Gives the `exo(endo)-gamy distribution' 
     (see Ref. \cite{ms40}) of clusters 
     formed from partons with `exogamy' index $e_i, e_j$, such that 
     $e_{cl} = (e_i+e_j)/2$. Inital beam/target particles have $e = 0 (1)$, so that 
     $e_{cl}$ lies in the interval [0,1], which is binned as 
     $0, 0.05, 0.2, 0.3, 0.45,0.55, 0.7, 0.8, 0.95, 1$, with 
     PARW(90)-PARW(99) giving the numbers of clusters 
     in the bins. Note: only those clusters are counted that are classified as
     `fast' (c.f. MSTW(88)).
\item[ PARW(101)]{\bf - PARW(109)} : Same as PARW(91)-PARW(99), but for `slow' clusters
     only. The sums PARW(91)+PARW(101), etc., hence give the total numbers.
\item[ PARW(110)]{\bf - PARW(149)} :  not used.

\noindent
\sline
\end{description}
\medskip

                    
\noindent
\lline

\noindent  {\bf 7.  `Afterburner' Hadron  cascade:}

\noindent
\lline

\begin{description}

\item[ MSTW(150)] : Total number of hadronic collisions involving 
                    clusters and/or hadrons.
\item[ MSTW(151)] : Number of collisions between two clusters $C+C$,
                    where only {\it light} quark flavors are involved.
\item[ MSTW(152)] : Number of $C + C$ collisions where {\it strange} quark 
                    flavor is involved.
\item[ MSTW(153)] : Number of $C + C$ collisions where {\it charm} quark 
                    flavor is involved.
\item[ MSTW(154)] : Number of $C + C$ collisions where {\it bottom} quark 
                    flavors is involved.
\item[ MSTW(155)] : -- not used --
\item[ MSTW(156)]{\bf - MSTW(160)} : Same as MSTW(151)-MSTW(155), 
                    but for cluster-meson collisions $C+M1$, where
                    $M1$ are light mesons only.
\item[ MSTW(161)]{\bf - MSTW(165)} : Same as MSTW(156)-MSTW(160), 
                    but for cluster-meson collisions $C+M2$, where
                    $M2$ are heavy or excited mesons.
\item[ MSTW(166)]{\bf - MSTW(170)} : Same as MSTW(161)-MSTW(165), 
                    but for cluster-baryon collisions $C+B$, where
                    $B$ is any type of baryon.
\item[ MSTW(171)]{\bf - MSTW(175)} : Same as MSTW(166)-MSTW(170), 
                    but for meson-meson collisions $M1+M1$, where
                    $M1$ {\it and} $M2$ are light mesons.
\item[ MSTW(176)]{\bf - MSTW(180)} : Same as MSTW(171)-MSTW(175), 
                    but for meson-meson collisions $M1+M2$, where
                    $M1$ is a light meson and $M2$ an excited or heavy meson.
\item[ MSTW(181)]{\bf - MSTW(185)} : Same as MSTW(176)-MSTW(180), 
                    but for meson-baryon collisions $M1+B$, where
                    $M1$ is a light meson and $B$ any type of baryon.
\item[ MSTW(186)]{\bf - MSTW(190)} : Same as MSTW(181)-MSTW(185), 
                    but for baryon-baryon collisions $B+B$, where
                    both $B$ can be any type of baryon.
\item[ MSTW(191)]{\bf - MSTW(195)} : Same as MSTW(186)-MSTW(190), 
                    but for baryon-meson collisions $B+M2$, where
                    $B$ is any type of baryon and $M2$ an excited or heavy meson.
\item[ MSTW(196)]{\bf - MSTW(200)} : Same as MSTW(191)-MSTW(195), 
                    but for meson-meson collisions $M2+M2$, where
                    both $M2$ are excited or heavy mesons.
\medskip

\noindent
\sline
\item[ PARW(150)] : $\langle \, \sqrt{s}\,\rangle$ of {\it all}
                    cluster/hadron collisions.
\item[ PARW(151)] : $\langle \, \sqrt{s}\,\rangle$ of cluster/hadron collisions
                    involving only {\it light} flavors.
\item[ PARW(152)] : $\langle \, \sqrt{s}\,\rangle$ of cluster/hadron collisions
                    involving {\it strange} flavor.
\item[ PARW(153)] : $\langle \, \sqrt{-t}\,\rangle$ of cluster/hadron collisions
                    involving  {\it charm} flavor.
\item[ PARW(154)] : $\langle \, \sqrt{-t}\,\rangle$ of cluster/hadron collisions
                    involving  {\it bottom} flavor.
\item[ PARW(155)]{\bf - PARW(159)} : -- not used --
\item[ PARW(160)]{\bf - PARW(164)} : $\langle \, \sqrt{-t}\,\rangle$ 
                    of cluster/ hadron collisions in paralell to
\item[ PARW(165)]{\bf - PARW(169)} : -- not used --
\item[ PARW(170)]{\bf - PARW(174)} : $\langle \, \sqrt{-u}\,\rangle$ 
                    of cluster/ hadron collisions in paralell to
                    PARW(150)-PARW(154).
\item[ PARW(175)]{\bf - PARW(179)} : -- not used --
\item[ PARW(180)]{\bf - PARW(184)} : $\langle \, n_{prod}\,\rangle$ 
                    of particles produced in cluster/ hadron collisions 
                    in paralell to
                    PARW(150)-PARW(154).
\item[ PARW(185)]{\bf - PARW(190)} : -- not used --
\item[ PARW(191)] : Total number of `unwounded' nucleons at the end of 
		    {\it current event} of collision with nuclei.
\item[ PARW(192)] : Number of `unwounded' nucleons at the end of 
		    {\it  current event} which are {\it within} the transverse
		    area of nuclear overlap.
\item[ PARW(193)] : Number of `unwounded' nucleons at the end of 
		    {\it  current event} which are {\it outside} the transverse
		    area of nuclear overlap.
\item[ PARW(194)]  {\bf - PARW(196} : Numbers of `unwounded' nucleons 
		    {\it accumulated over all events} in collisions with nuclei,
		    in paralell to PARV(191) - PARV(193).
\item[ PARW(197)]  {\bf - PARW(200)} : -- not used --

\noindent
\sline

\end{description}

\noindent
\lline
\bigskip
\bigskip

\subsection{Kinematics cuts and selection of subprocesses}  
\bigskip

  The commonblock VNIKIN contains the arrays VKIN and MSUB, which may be used 
to impose specific kinematics cuts and to switch on/off certain
 subprocesses, respectively. The notation used in the following 
is that 'hat' denotes internal variables in a partonic interaction, 
while '*' is for variables in the global CM frame of the system as-a-whole. 
All dimensionful quantities are in units of GeV or GeV$^{-1}$.

  Kinematics cuts can be set by the user before the VNIXRUN call. Most of 
the cuts are directly related to the kinematical variables used in the 
evolution of the system in each event, and in the event data analysis. 
Except for the entries VKIN(7) - VKIN(12), the assignments follow the
convention of PYTHIA, in order to allow an easy interfacing.

  Note, that in the current version of VNI, only the values 
VKIN(7)-VKIN(12) are actively used. All other quantities are related to 
parton collision processes involving one or more partons, which are only
of relevance in reactions with hadrons (nuclei) in the initial state.
As stated before, the latter are presently worked out, but not yet
included in the package.

    
\begin{verbatim}
            COMMON/VNIKIN/VKIN(200),MSUB(200),ISET(200)
\end{verbatim}

\noindent
{\it Purpose}: To impose kinematics cuts on the particles' interactions
         and their space-time evolution.  Also, to allow the user to 
         run the program with a desired subset of processes.

\begin{description}
\item[ VKIN(1),]{\bf VKIN(2) :} (D=2.,-1. GeV) range of allowed 
     $\hat{m} = \sqrt{\hat{s}}$ values. 
     If VKIN(2) $<$ 0, the upper limit is inactive.
\item[ VKIN(3),]{\bf VKIN(4) :} (D=2.,-1. GeV) range of allowed 
     (D=0.,-1.) range of allowed $\hat{p}_\perp$ values for
    parton collisions, with transverse momentum 
    $\hat{p}_\perp$ is defined
    in the rest-frame of the 2-body interaction. If VKIN(4) $<$ 0,
    the upper limit is inactive. For processes which are singular
    in the limit $\hat{p}_\perp\rightarrow 0$ (see VKIN(6)), VKIN(5) provides
    an additional constraint. The VKIN(3) and VKIN(4) limits can
    also be used in $2 \rightarrow 1 \rightarrow 2$ processes. 
    Here, however, the product
    masses are not known and hence assumed vanishing in the event
    selection. The actual $\hat{p}_\perp$ range for massive products is thus
    shifted downwards with respect to the nominal one.

\item[ VKIN(5) :] 
 (D=0.5 GeV) lower cutoff on $\hat{p}_\perp$ values, in addition to the
    VKIN(3) cut above, for processes which are singular in the
    limit $\hat{p}_\perp \rightarrow 0$ (see VKIN(6)).
\item[ VKIN(6) :] 
 (D=1. GeV)  2-body parton collision  processes, which do not proceed only
    via an intermediate resonance (i.e. are $2 \rightarrow 1 \rightarrow 2$ 
    processes),
    are classified as singular in the limit $\hat{p}_\perp\rightarrow 0$,
    if either
    or both of the two final state products has a mass $m$ $<$ VKIN(6).
\\
\item[ VKIN(7),]{\bf VKIN(8) :}
   (D=-100.,100.) range of allowed particle rapidities
    $y^\ast$ in the CM frame of the overall system.
\item[ VKIN(9),]{\bf VKIN(10) :} 
    (D=0.,1000.) range (in GeV) of allowed particle
    transverse momenta $k_\perp^\ast$ perpendicular to the 
    $z$-axis in the global 
    CM frame of the event.
\item[ VKIN(11),]{\bf VKIN(12) :}
    (D=-1000.,1000) range (in GeV$^{-1}$) of allowed 
    longitudinal positions $z^\ast$ of particles from the CM of the global
    collision system.
\item[ VKIN(13),]{\bf VKIN(14) :} 
    (D=0.,1000.) range (in GeV$^{-1}$) of allowed 
    transverse distances $r_\perp^\ast$ perpendicular to the $z$-axis in the CM 
    frame of the global collision system.
\\
\item[ VKIN(21),]{\bf VKIN(22) :}
    (D=0.,1.) range of allowed $x_A$ (Bjorken-$x$) values for the parton
    of beam-side $A$ that enters  a parton collision.
\item[ VKIN(23),]{\bf VKIN(24) :}
     (D=0.,1.) range of allowed $x_B$ (Bjorken-$x$) values for the parton
    of target-side $B$ that enters  a parton collision.
\item[ VKIN(25),]{\bf VKIN(26) :} 
    (D=-1.,1.) range of allowed Feynman $x_F$ values, 
    where $x_F = x_A-x_B$.
\item[ VKIN(27),]{\bf VKIN(28) :}
    (D=-1.,1.) range of allowed $\cos(\hat{\theta})$ values
    in a  $2\rightarrow 2$ parton collision, where $\hat{\theta}$ 
    is the scattering
    angle in the rest-frame of the 2-body collision.
\\
\item[ VKIN(31),]{\bf VKIN(32) :}
    (D=2.,-1.) range of allowed $\hat{m}'$ values, where
    $\hat{m}'$ is the mass of the complete three- or four-body final state
    in $2 \rightarrow 3$ or $2 \rightarrow 4$ processes 
    (while $\hat{m}$ (without a prime), constrained in VKIN(1)
    and VKIN(2), here corresponds to the one- or two-body central
    system). If VKIN(32) $<$ 0, the upper limit is inactive.
\item[ VKIN(35),]{\bf VKIN(36) :}
    (D=0.,-1.) range of allowed $\mbox{abs}(\hat{t}) = -\hat{t}$
    values in $2 \rightarrow 2$ processes. 
    Note that for deep inelastic scattering
    this is nothing but the $Q^2$ scale of the photon vertex, 
    in the limit that initial and
    final state radiation is neglected. If VKIN(36) $<$ 0, the upper
    limit is inactive.
\item[ VKIN(37),]{\bf VKIN(38) :}
    (D=0.,-1.) range of allowed $\mbox{abs}(\hat{u}) = - \hat{u}$
    values in $2 \rightarrow 2$ processes. If VKIN(38) $<$ 0, 
    the upper limit is inactive.
\\
\item[ VKIN(39) -]{\bf VKIN(100) :}
currently not used.
\\
\item[ VKIN(101) -]{\bf VKIN(200) :} 
     reserved for internal use of storing kinematical
    variables of an event.
\\
\\
\item[ MSUB(ISUB) :]  array to be filled 
    when MSTV(14) = 1 (for MSTV(14) = 0, the
    array MSUB is initialized in VNIXRIN automatically) to choose which 
    subset of subprocesses to include in the generation. The ordering 
    follows the ISUB code given before in Table 3.
\\
{ = 0 :} the subprocess ISUB is {\it excluded}.
\\
{ = 1 :} the subprocess ISUB is {\it included}.
\\
\item[ ISET(ISUB) :] gives the type of kinematical variable selection scheme
    used for subprocess ISUB.
\\
{\bf =  1 :} 2 $\rightarrow$ 1 processes (parton fusion processes).
\\
{\bf =  2 :} 2 $\rightarrow$ 2 processes (hard and soft parton scattering processes).
\\
{\bf =  3 :} 1 $\rightarrow$ 2 processes (parton branching processes).
\\
{\bf = -1 :} process  is not yet implemented, but space is reserved.
\\
{\bf = -2 :} process  is not defined.
\end{description}
\bigskip
\bigskip

\subsection{Instructions on how to use the program}

  There are a variety of example programs available from the website
  http://rhic.phys.columbia.edu/rhic/vni.   
  Below described is a very simple example, {\it  vnixple.f}, disguised 
as a generic skeleton program that calls the steering routines VNIXRIN, 
VNIXRUN and VNIXFIN (c.f. Sec. 3.4) on a black-box-level.
This example 
generates 1000 $p\bar{p}$ events at $\sqrt{s} = 546$ GeV.
Each event is evolved for  a time range 
of 0 - 30 fm in the $p\bar{p}$ center-of-mass frame (the global CM frame). 


\begin{verbatim}
      PROGRAM VNIXPLE

C...Purpose: example for p+p~ collisions at CERN collider with 546 GeV.
      INCLUDE 'vni1.inc'
      CHARACTER*(8) FRAME,BEAM,TARGET


C...Required user input:
C.....number of events and final time (in fm).
      NEVT  = 1000
      TFIN  =   30.
C.....colliding particles, global Lorentz frame, and total energy.
      BEAM  = 'p+      '
      TARGET= 'p-      '
      FRAME = 'cms '
      WIN   = 546.

C.....Specify your change of default parameters here:
C     ..........


C...Initialize simulation procedure.
      CALL VNIXRIN(NEVT,TFIN,FRAME,BEAM,TARGET,WIN)


C...Begin loop over of events.
      DO 500 IEVT=1,NEVT

C.....Generate collision event.
      CALL VNIXRUN(IEVT,TFIN)

C.....Analysis of event as-a whole.
C     ..........

C...End loop over events.
  500 CONTINUE


C...Finish up simulation procedure.
      CALL VNIXFIN()


      END
\end{verbatim}

This example is rather self-explanatory, and will become
further clear with the following subsections. The "$\ldots$"
indicate where one can interface on a black-box-level with the program 
as-a-whole
without needing to dig deeper into its subroutine structure. In particular, one
can insert here histograming (either using the included
portable package "vnibook.f", or any other package such  as 
the HBOOK or PAW of the CERN library).
Note also the pre-programmed histogram options 
discussed in Sec. 3.7 (switches MSTV(150)-MSTV(200)) and Appendix B (subroutines VNIANA1-VNIANA4).

In the above example only the final state at the end of each event
is collected. If the user is interested in the actual
space-time development of the collision events, he/she can access this by
inserting in the loop over events the   following  extension:

\begin{verbatim}

C...Generate collision event; analyze evolution every 5 fm.
      TSTEP=5.
      NTIM=20

C.....Begin loop over time intervals.
      DO 400 ITIM=1,NTIM
      TRUN=ITIM*TSTEP
      IF(TRUN.GT.TFIN.OR.ITIM.EQ.NTIM) TRUN=TFIN
      CALL VNIXRUN(IEVT,TRUN)

C.....Analysis of time development.
C     ..........

C.....End loop over time intervals.
  400 IF(TRUN.EQ.TFIN) GOTO 450
  450 CONTINUE      

\end{verbatim}

\noindent
{\bf Remarks:}
\begin{description}
\item[(i)]
      In case one desires to interface with the PYTHIA program,
       one should write a little program with common block

\begin{verbatim}
       COMMON/LUJETS/N,K(4000,5),P(4000,5),V(4000,5) 
\end{verbatim}

       and which copies the variables $N, K, P$, and $V$ of the particle 
       record VNIREC1/VNIREC2 directly.  
       Be aware however, that VNIREC1/VNIREC2 is dimensioned 
       for 100000 entries, whereas LUJETS has space for only 4000 
       entries.
\\
\item[(ii)]
       In case one desires to interface with the HERWIG program, one 
       should write a little program with common block

\begin{verbatim}
       PARAMETER (NMXHEP=2000) 
      COMMON/HEPEVT/NEVHEP,NHEP,ISTHEP(NMXHEP),IDHEP(NMXHEP), 
     &JMOHEP(2,NMXHEP),JDAHEP(2,NMXHEP),PHEP(5,NMXHEP),VHEP(4,NMXHEP) 
\end{verbatim}

       and which calls the routine HWHEPC($MCONV,IP,NP$) with $MCONV = 1$,
       $IP = 1$,and $NP = N$. Again, watch the dimensions, 100000 versus 2000.
\end{description}
\bigskip
\bigskip
\bigskip

\subsection{Example for a typical collision event}

  As a result of running the above example program {\it vnixple.f},
a number of files are created in the directory, all of which are called 
VNI???.DAT, where "???" are either 3 letters, or 3 digits.
The latter are data files that contain event statistics and other information
on the performance of a run. 
(see Sec. 3.7, switches MSTV(150)-MSTV(200) and Appendix B  for further options).
By default only  a few files are created:
\begin{description}
\item
VNIRUN.DAT containing
the main summary of a simulation with print-out of results,
\item
VNICOL.DAT containing a detailed  summary of all parton collisions that occurred,
\item
VNITIM.DAT containing the generated time-step grid and its real-time
translation, 
\item
VNIRLU.DAT containing the status of the random number generator,
\item
VNIERR.DAT containing error messages in case problems occur.
\end{description}

Most important is the file VNIRUN.DAT, which
gives a summary of what
occurred during a simulation on the parton cascade level, the cluster 
formation level and the hadron decay level.
One can obtain a listing of an  event by simply calling VNILIST(1)
at the end or during an event,
which gives a first impression
(for a complete listing see Appendix D): 
\begin{description}
\item[(i)]
The event record begins with the beam/target particles, here $p$ and $\bar{p}$,
which are then transformed to the center-of-mass frame (here it coincides
with the global frame of reference). 
\item[(ii)]
Then follow the
three valence quarks of $p$, and after that its intrinsic  gluons and
seaquark-pairs. After that the initial-sate valence quarks and gluons and
seaquarks of $\bar{p}$ are listed.
\item[(iii)]
Then comes the history of the cascade evolution, including
hard scatterings with shower (bremsstrahlung), subsequent
cluster-formation of the materialized partons
and hadron decay of these parton clusters.
\item[(iv)]
Finally, all non-materialized (non-interacted) partons
are recombined to a beam- and a target cluster that
undergo a soft fragmentation, again via cluster-hadron decay.
\item[(v)]
At the very end the sum of charge, three-momentum, energy, and
the total invariant mass of the collision system is printed.
\end{description}
\bigskip
\bigskip

\begin{center}
Example of the particle record for a $p\bar{p}$ collision event
at $\sqrt{s} = 546$ GeV.
\end{center}

\begin{verbatim}

                  Particle listing (summary: momentum space P in GeV)
                  event no.    1  time step  200  at time  34.75 fm

    I particle     KS     KF  C A     P_x      P_y      P_z       E        M

    1 (p+)         15   2212  0 0      .000     .000  272.998  273.000     .938
    2 (p~-)        15  -2212  0 0      .000     .000 -272.998  273.000     .938
 ==============================================================================
    3 (p+)         15   2212  0 0      .000     .000  272.998  273.000     .938
 ==============================================================================
    4 (p~-)        15  -2212  0 0      .000     .000 -272.998  273.000     .938
 ==============================================================================
    5 (u)          13      2  1 0      .650     .048    1.786    2.319     .006
    6 (d)          13      1  1 0     -.593    -.165    2.092    2.587     .010
    7 (d~)         13     -1  0 3     -.135    -.452    6.724    6.886    -.490
    8 (d)          13      1  3 0      .413    -.211     .408     .935    -.484
    9 (g)          14     21  1 1     1.751     .556    2.232    3.398    -.682
      .
      .
 ==============================================================================
   48 (d~)         13     -1  0 2      .526    -.150  -16.659   15.157     .010
   49 (u~)         13     -2  0 3     -.121    -.114   -4.941    4.106    -.174
   50 (u)          13      2  3 0     -.252    -.044   -1.576     .935    -.249
   51 (g)          13     21  3 2     -.110     .229  -12.733   11.449    -.234
      .
      .
 ==============================================================================
      .
      .
   83 (g)          14     21  2 1     2.052    2.858    1.999    1.045   -3.909
   84 (d)          14      1  1 0      .000     .000   -8.008    8.008     .000
   85 (d)          14      1  2 0      .000     .000    8.008    8.008     .000
   86 (CMshower)   11     94  0 0     2.661    3.728   -3.710    8.110    5.571
   87 (d)          14      1  2 0     1.128     .073   -1.364    3.504    3.023
   88 (cluster)    11     91  0 0     1.116    1.615   -2.687    3.412     .750
   89 (cluster)    11     91  0 0     -.450    -.605    -.228    1.213     .922
   90 (cluster)    11     91  0 0    -1.528    -.662    1.381    2.181     .280
   91 (cluster)    11     91  0 0     1.252     .689   -1.272    3.136    2.484
   92 (cluster)    11     91  0 0      .110    -.650    5.219    5.796    2.433
   93 pi0           6    111  0 0      .660     .523    -.880    1.171     .135
   94 pi0           6    111  0 0      .461    1.090   -2.094    2.377     .135
   95 rho-          6   -213  0 0     -.334    -.562    -.332    1.068     .769
   96 pi0           6    111  0 0     -.110    -.046    -.182     .186     .135
   97 pi-           6   -211  0 0     -.678    -.312     .548    1.066     .140
   98 pi+           6    211  0 0     -.845    -.352     .547    1.203     .140
    .
    .
 ==============================================================================
  109 (Beam-REM)   11     92  0 0     -.315     .737  274.428  281.871   70.164
  110 (Targ-REM)   11     93  0 0     -.907   -1.587 -276.508  247.044 -116.181
  111 (CMF)        15    100  0 0    -1.222    -.850   -2.080  528.915  528.909
 ==============================================================================
  112 (cluster)    11     91  0 0     -.315     .737  274.428  281.871   70.164
  113 (cluster)    11     91  0 0     -.907   -1.587 -276.508  247.044 -116.181
  114 (Beam-REM)   11     92  0 0      .411     .648  127.485  127.543    3.762
  115 (a_1-)       17 -20213  0 0      .493     .228   19.200   19.250    1.275
  116 (Delta+)     17   2214  0 0     -.081     .420  108.286  108.293    1.232
  117 pi0           7    111  0 0      .490     .124    4.473    4.831     .135
  118 (rho-)       17   -213  0 0      .005     .102   14.584   14.605     .769
  119 pi0           7    111  0 0     -.160     .015   35.475   37.042     .135
  120 p+            7   2212  0 0      .085     .402   72.525   75.579     .938
  121 pi0           7    111  0 0     -.122    -.268    8.519    9.014     .135
  122 pi-           7   -211  0 0      .133     .367    5.779    6.174     .140
  123 (Targ-REM)   11     93  0 0    -1.252    -.510  -63.766   63.825    2.384
  124 (Delta~-)    17  -2214  0 0     -.779    -.303  -21.201   21.253    1.232
  125 (eta)        17    221  0 0     -.473    -.208  -42.565   42.572     .549
  126 pi0           7    111  0 0     -.072    -.254   -3.205    3.199     .135
  127 p~-           7  -2212  0 0     -.701    -.052  -18.281   18.903     .938
  128 pi0           7    111  0 0     -.179     .069  -12.277   12.622     .135
  129 pi0           7    111  0 0     -.060    -.060  -12.901   13.269     .135
  130 pi0           7    111  0 0     -.225    -.220  -17.816   18.383     .135
    .
    .
    .
  336 (rho+)       17    213  0 0     -.263     .404   -5.474    5.549     .769
  337 (omega)      17    223  0 0      .067    -.237  -31.485   31.496     .783
  338 pi0           7    111  0 0     -.342     .522   -5.018    5.113     .135
  339 pi+           7    211  0 0      .085    -.120    -.742     .657     .140
  340 pi0           7    111  0 0     -.309     .029  -10.513   10.790     .135
 ==============================================================================
                  sum:   .00           .000     .000     .000  546.000  546.000

\end{verbatim}

More information on 
pre-programmed 
print-out of results concerning for particle distributions, time evolution, 
observables, etc., can be switched on with switches MSTW(151) - MSTW(200), as 
explained in the previous Sec. 3.7.
\bigskip
\bigskip
\bigskip
\bigskip

\noindent {\bf ACKNOWLEDGEMENTS}
\medskip

The present program is, with respect to both its
physics and computational aspects, the product of several
years of gradual experience and development,
during which I profited from many colleagues, most of all
from John Ellis, Miklos Gyulassy, and Berndt M\"uller.
Thank you!

This work was supported in part by the D.O.E under contract no.
DE-AC02-76H00016.
\bigskip

\newpage

\newpage
\appendix

\section{Further physics routines}
\label{sec:appa}

Aside from the three main routines VNIXRIN, VNIXRUN, and VNIXFIN,
the physics routines described 
below form the heart of the program. In general,  
for each of the included collision processes of Table 1,
there is an initialization 
routine VNI??IN that sets up the initial state, and an evolution 
routine VNIEV?? that carries out the time evolution of the particle 
distributions in phase-space, starting from the initial state. 
The routines VNICOLL, VNISBRA, VNITBRA, etc., therein perform the 
perturbative parton evolution in terms of 
parton collisions, space-like branchings, 
and time-like branchings, respectively. 
The routines VNICHAD, VNIBHAD, etc., deal with the cluster hadronization. 
These routines are universal and independent of the process under consideration.

\begin{verbatim}
SUBROUTINE VNICOL2(NCOL)
\end{verbatim}

\noindent
{\it Purpose}: to test for all possible pairings of `alive' partons, whether a 
    collision is allowed, and, if it is allowed, to evaluate the quantities 
    characterizing the 2-body collision at the parton level according to the 
    cross-section and to choose one of the possible subprocesses (channels).

\begin{description}
\item[{\boldmath $NCOL$} :] 
      total number of found collisions.
\end{description}
\medskip

\begin{verbatim}
SUBROUTINE VNICOLL(ISUB,Q2,IP1,IP2,IP3,IP4)
\end{verbatim}

{\it Purpose}: to perform a 2-body parton collision, find outgoing 
flavors	and colors and calculate  
kinematics and  parton-parton collision.

\begin{description}
\item[{\boldmath $ISUB$} :] 
      Subprocess of 2-body collision (c.f. Table 3).
\item[{\boldmath $Q2$} :] 
      Momentum transfer scale of subprocess (c.f. MSTV(43), Sec. 3.7).
\item[{\boldmath $IP1$,}]{\boldmath $IP2$} : 
      Line numbers of paerons entering the interaction vertex.
\item[{\boldmath $IP3$,}]{\boldmath $IP4$} : 
      Line numbers of particles emerging from the interaction vertex.
      For $2\rightarrow 1$ processes $IP4$ is equal to zero.
\end{description}
\medskip

\begin{verbatim}
SUBROUTINE VNISBRA(IP1,IP2,QMAX)
\end{verbatim}

\noindent
{\it Purpose}: to generate space-like parton evolution, by backward tracing of
    branching processes down to some initial resolution scale associated
    with a non-perturbative cut-off. The performance is regulated by 
    the switches MSTV(50)-MSTV(52) and parameters PARV(51)-PARV(53) 
    (c.f. Sec. 3.7). 

\begin{description}
\item[{\boldmath $IP1, $}]{\bf {\boldmath $IP2$} :} 
    partons in lines $IP1$ and $IP2$ in the commonblocks VNIREC1/VNIREC2,
    are evolved by space-like parton branchings from initial mass scale 
    QMAX. If one of the particles is a non-radiating one, then only
    one branch is evolved, and the spectator particle role is then only
    to take up recoil due to emitted particles by the radiating one.
\item[{\boldmath $QMAX$} :]
    initial mass scale, from which the two particles are evolved
    from $Q = QMAX$ 
    backwards to lower $Q < QMAX$ until the cut-off $Q=q_0 \equiv $PARV(52)
    is reached.
\end{description}
\medskip

\begin{verbatim}
SUBROUTINE VNITBRA(IP1,IP2,QMAX)
\end{verbatim}

\noindent
{\it Purpose}:
    to generate time-like parton evolution, by coherent branching
    processes within a finite time interval, $q \rightarrow qg$, 
    $q \rightarrow \gamma$, $g \rightarrow gg$, $g \rightarrow q\bar q$.
    The performance is regulated by the switches 
    MSTV(61)-MSTV(65) and parameters PARV(61)-PARV(65) (c.f. Sec. 3.7). 

\begin{description}
\item[{\boldmath $IP1 > 0$}]{\bf {\boldmath $IP2 = 0$} :} 
    generate a time-like parton branching for the parton
    in line $IP1$ in commonblocks VNIREC1/VNIREC2, with maximum allowed mass
    $QMAX$. 
\item[{\boldmath $IP1 > 0$}]{\bf {\boldmath $IP2 > 0$} :} 
    generate time-like parton branchings for the two
    partons in lines $IP1$ and $IP2$ in the commonblocks VNIREC1/VNIREC2,
    with maximum allowed mass for each parton $QMAX$.
\item[{\boldmath $QMAX$} :]
    the maximum allowed mass of a radiating parton, i.e. the
    starting value for the subsequent evolution. (In addition, the
    mass of a single parton may not exceed its energy, the mass of a
    parton in a system may not exceed the invariant mass of the
    system.)
\end{description}
\medskip

\begin{verbatim}
SUBROUTINE VNICHAD(I1,I2)
\end{verbatim}

\noindent
{\it Purpose}: to search at a given point of time the 
particle record for possible clustering of materialized
partons,  which are those which have been struck out from the initial state
or have been produced as secondaries by 
parton collisions or radiative processes.
    It consists of checking for each possible pair of two partons, the 
    invariant space-time distance between them, their flavor and origin, 
    the matching of color and anticolor, their common invariant mass, 
    etc.. If a pair satisfies all necessary conditions for the formation
    of one or two color singlet clusters, then they are combined and
    converted into a composite cluster with internal flavor content and
    invariant mass determined by the partons, which then decays either
    into a pair of color singlet clusters, or into a singlet cluster plus
    an additional gluon or quark.

\begin{description}
\item[{\boldmath $I1, \, I2$} :]
    first and last entry of particle record VNIREC1/VNIREC2 in between the
    cluster formation is carried out by
    looping over all particles $I$ with $I1 \le I \le I2$.
\end{description}
\medskip

\begin{verbatim}
SUBROUTINE VNICFRA(I1,I2)
\end{verbatim}

\noindent
{\it Purpose}: to fragment color singlet clusters with given 
    $q\bar q$ substructure
    into final state hadrons. For each cluster its invariant mass and
    flavor content are checked, and it is determined whether it decays
    isotropically into a pair of hadrons, or converts directly into a
    single hadron of corresponding flavor. The hadron decay is described
    by a phase-space model, in which the probability for forming a 
    certain hadron state is given by the density of hadronic states with 
    mass below the cluster mass. Once a hadron state is chosen, it either
    propagates on undisturbed if it is a stable particle, or it decays 
    further according to the particle data tables, if it is a resonance.

\begin{description}
\item[{\boldmath $I1 > 0$}  :] 
    entry of particle record VNIREC1/VNIREC2 from which analysis starts.
\item[{\boldmath $I2 > I2$} :] 
    entry of particle record up to which analysis extends.
\end{description}
\medskip

\begin{verbatim}
SUBROUTINE VNIBHAD(I1,I2)
\end{verbatim}

\noindent
{\it Purpose}:
    to perform beam cluster formation and hadronization. The
    remnant partons of beam/target particles, those that originate from
    the partonic coherent initial state in collisions involving hadrons 
    or nuclei and have not interacted with decohereing effect. These
    primary partons are reassembled to clusters along the beam axis,
    which are then fragmented based on the minimum bias model of the
    UA5 collaboration, such that color, flavor and energy-momentum of 
    the collision system as-a-whole are conserved.

\begin{description}
\item[{\boldmath $I1, \, I2$} :] 
    first and last entry of particle record VNIREC1/VNIREC2 in between the
    soft beam/target fragmentation is carried out by
    looping over all particles $I$ with $I1 \le I \le I2$.
\end{description}
\medskip

\newpage 

\begin{verbatim}
SUBROUTINE VNIBFRA(I1,I2)
\end{verbatim}

\noindent
{\it Purpose}: to hadronize beam/target clusters consisting of remnant
    beam partons,  using parts of the HERWIG 5.7 program. Beam remnants
    are those that emerge from the coherent partonic initial state 
    without having undergone any decohering interaction, and carry the
    remaining energy, flavor and color of the system. The beam remnants
    are lumped to beam clusters which are fragmented similarly as the
    clusters arising from the parton cascade evolution.
\begin{description}
\item[{\boldmath $I1 > 0$}  :] 
    entry of particle record VNIREC1/VNIREC2 from which analysis starts.
\item[{\boldmath $I2 > I1$} :] 
    entry of particle record up to which analysis extends.
\end{description}
\medskip

\begin{verbatim}
SUBROUTINE VNIPADC(I1,I2,NUST) 
\end{verbatim}

\noindent
{\it Purpose}: to decay  unstable particles resulting from cluster decays
    to hadrons. Out of these primary hadrons, those that are unstable 
    resonances are decayed according to the Particle Data, which
    gives lower mass hadrons plus possibly gauge bosons and leptons.
    These secondary decay products may decay further.
	
\begin{description}
\item[{\boldmath $I1 > 0$}  :] 
     entry of particle record VNIREC1/VNIREC2 from which analysis starts.
\item[{\boldmath $I2 > I1$} :] 
     entry of particle record up to which analysis extends.
\item[{\boldmath $NUST$}    :] 
      returns number of decay products that are still unstable.
\end{description}
\bigskip
\bigskip

\newpage

\section{Event analysis routines} 
\label{sec:appb}

  The following routines collect, analyze and print out information
on the result of a simulation, and give the user immediate access
to calculated particle data, spectra, and other observables. 
\begin{itemize}
\item
The routine VNILIST provides a useful tool to obtain a
quick overall view of the state of the particle record at any time during the 
simulation of one or many events, and (in parts) the space-time history 
of the system. 
\item
 The routine VNIEDIT allows to edit the particle record, e.g. to
compress, or to exclude certain particle species from the record. It is
used frequently during the simulation procedure to clean up used but 
further on unecessary particle information.
\item
 Via the functions KVNI and PVNI, values of some frequently appearing
variables may be obtained more easily. 
\item
 The routines VNIANAL and VNIANA1 - VNIANA4 carry out various data 
accumulation duties, and print out tables at the end of the simulation, 
which are automatically directed to files named VNI???.DAT where ??? is 
a 3-digit number, specified below.
\item
 Finally, the six routines VNISPHE, VNITHRU, VNICLUS, VNICELL, 
VNIJMAS and VNIFOWO 
allow the possibility to find some global event shape 
properties as discussed in detail in the PYTHIA documentation \cite{jetset}.
\medskip

\item
The common block VNIANA allows to access the bin sizes, and bin ranges
of histograms in the  routines VNIANA1 - VNIANA4.

\begin{verbatim}
                COMMON/VNIANA/MANA(50),PANA(50)
\end{verbatim}

\noindent
{\it Purpose}:  to provide data for variable ranges, bin sizes, etc., 
                for histograms and tables created by analysis routines 
                VNIANA1 - VNIANA4.
\medskip

\begin{description}
\item[ MANA(1)]{\bf - MANA(10)} : 
      (D=1,21,41,61,91,121,151,181,221,301) Sequential
      numbers of time steps, at which a snapshot analysis during the
      space-time evolution of the collision system may be performed
      in routines VNIANA4 - VNIANA4.
\item[ MANA(11)] : (D=40) Number of bins for energy distribution in
      $\ln(1/x)$ (where $x$ is the energy fraction) in routine VNIANA2.
\item[ MANA(12)] : not used.
\item[ MANA(13)] : (D=40) Number of bins for pair mass distribution of 
      Bose-Einstein correlations in routine VNIANA2.
\item[ MANA(14)-MANA(20)] : not used.
\item[ MANA(21)] : (D=20) Number of bins for cluster $y$-distribution in 
      routine VNIANA3. 
\item[ MANA(22)] : not used.
\item[ MANA(23)] : (D=20) Number of bins for cluster $r_z$-distribution in 
      routine VNIANA3. 
\item[ MANA(24)] : not used.
\item[ MANA(25)] : (D=20) Number of bins for cluster $p_\perp$-distribution in 
      routine VNIANA3. 
\item[ MANA(26)] : not used.
\item[ MANA(27)] : (D=20) Number of bins for cluster $r_\perp$-distribution in 
      routine VNIANA3. 
\item[ MANA(28)]{\bf -MANA(30)} : not used.
\item[ MANA(31)] : (D=20) Number of bins for parton/hadron $y$-distribution 
      in routine VNIANA4. 
\item[ MANA(32)] : not used.
\item[ MANA(33)] : (D=20) Number of bins for parton/hadron $r_z$-distribution 
      in routine VNIANA4. 
\item[ MANA(34)] : not used.
\item[ MANA(35)] : (D=20) Number of bins for parton/hadron $p_\perp$-distribution
      in routine VNIANA4. 
\item[ MANA(36)] : not used.
\item[ MANA(37)] : (D=20) Number of bins for parton/hadron $r_\perp$-distribution 
      in routine VNIANA4. 
\item[ MANA(38)]{\bf - MANA(50)} : not used.
\medskip

\item[ PANA(1)]{\bf - PANA(10)} : not used.

\item[ PANA(1)],{\bf PANA(12)} : 
      (D=0.,6.) Lower and upper bound for energy distribution in $\ln(1/x)$ 
      (where $x$ is the energy fraction of particles), in routine VNIANA2.
\item[ PANA(13)],{\bf PANA(14)} : (D=0.,3. GeV) Lower and upper bound 
     for pair mass 
     distribution of Bose-Einstein correlations in routine VNIANA2.
\item[ PANA(15)]{\bf - PANA(20)} : not used.

\item[ PANA(21)],{\bf PANA(22)} : (D=-8.,8.) Lower and upper bound for cluster 
      $y$-distribution in routine VNIANA3.
\item[ PANA(23)],{\bf PANA(24)} : (D=-40.,40. GeV$^{-1}$) 
      Lower and upper bound for cluster 
      $r_z$-distribution in routine VNIANA3.
\item[ PANA(25)],{\bf PANA(26)} : (D=0.,10. GeV) 
      Lower and upper bound for cluster 
      $p_\perp$-distribution in routine VNIANA3.
\item[ PANA(27)],{\bf PANA(28)} : (D=0.,40. GeV$^{-1}$) 
      Lower and upper bound for cluster 
      $r_\perp$-distribution in routine VNIANA3.
\item[ PANA(29)]{\bf - PANA(30)} : not used.

\item[ PANA(31)],{\bf PANA(32)} : (D=-8.,8.) 
      Lower and upper bound for parton/hadron
      $y$-distribution in routine VNIANA4.
\item[ PANA(33)],{\bf PANA(34)} : (D=-100.,100. GeV$^{-1}$) 
      Lower and upper bound for 
      parton/hadron $r_z$-distribution in routine VNIANA4.
\item[ PANA(35)],{\bf PANA(36)} : (D=0.,10. GeV) 
      Lower and upper bound for parton/hadron 
       $p_\perp$-distribution in routine VNIANA4.
\item[ PANA(37)],{\bf PANA(38)} : (D=0.,100. GeV$^{-1}$) 
      Lower and upper bound for parton/hadron $r_\perp$-distribution 
      in routine VNIANA4.
\item[ PANA(39)]{\bf - PANA(50)} : not used.

\end{description}

\end{itemize}
\bigskip

\begin{verbatim}
SUBROUTINE VNILIST(MSEL)
\end{verbatim}

\noindent
{\it Purpose}: to list the current state of the particle record during the
     evolution of an event, or certain specified excerpts of it. Also,
     to list parton or particle data, or current parameter values.

\begin{description}
\item[{\boldmath $MSEL$}  :] determines what is to be listed.
\\
{\bf =  0 :} writes a logo with program version number and last date of 
        change; is mostly for internal use.
\\
{\bf =  1 :} gives a simple list of current event record, in an 80 column
        format suitable for viewing directly on the computer terminal.
        For each entry, the following information is given: the entry
        number I, the parton/particle name, the status code $KS$ (K(I,1)), 
        the flavour code $KF$ (K(I,2)), the color and anticolor, $C$ (L(I,1)),
        and $A$ (L(I,2)), and the three-momentum, energy and mass 
        (P(I,1)-P(I,5)). A final line contains information on total charge, 
        momentum, energy and invariant mass.
\\
{\bf =  2 :} gives a more extensive list of the current event record, in a
        132 column format, suitable for printers or workstations.
        For each entry, the following information is given: the entry
        number I, the parton/particle name (with padding as described
        for $MSEL = 1$), the status code $KS$ (K(I,1)), the flavour code
        $KF$ (K(I,2)), the color and anticolor, $C$ (L(I,1)) and $A$ (L(I,2)), 
        the origin of the particle in a 3 digit $N1 N2 N3$, where $N1$ is
        the event and $N2$ the time step in this event, when the particle
        was first produced, and $N3$ is the number of the entry in the 
        particle record that the particle occupied when it first appeared.
        This labeling allows a unique specification of its `historical'
        origin. Finally, it follows the three-momentum, energy and mass 
        (P(I,1) - P(I,5)).  A final line contains information on total
        charge, momentum, energy and invariant mass. 
\\
{\bf =  3 :} gives the same basic listing as = 2, but with an additional
        line for each entry containing information on production vertex
        position and time (V(I,1) - V(I,4)).
\\
{\bf = -1 :} gives the format of listing as = 1, but instead of momentum, 
        energy and mass, it lists now the particle's position in the 
        overall reference frame, (R(I,1) - R(I,3)), its momentum-rapidity 
        (Y(P)) and its transverse momentum (PT) with respect to the 
        $z$-axis, which is usually the jet-axis, or beam axis. 
\\
{\bf = -2 :} gives the format of listing as = 2, but as for = -1, instead
        of momentum, energy and mass, now the particle's positions,
        rapidities and transverse momenta in the overall Lorentz frame
        are listed. As for = 2, additional information on the `historical'
        origin of the particle is listed.
\\
{\bf = -3 :} gives the format of listing as = 3, again, instead of momentum, 
        energy and mass, now the particle's positions, rapidities and 
        transverse momenta are listed.
\\
{\bf = 11 :} provides a simple list of all parton/particle codes defined
        in the program, with $KF$ code and corresponding particle name.
        The list is grouped by particle kind, and only within each group
        in ascending order.
\\
{\bf = 12 :} provides a list of all parton/particle and decay data used in
        the program. Each parton/particle code is represented by one
        line containing $KF$ flavour code, $KC$ compressed code, particle
        name, antiparticle name (where appropriate), electrical and
        color charge, mass, resonance width and maximum broadening, 
        average invariant lifetime and whether the particle is considered 
        stable or not.
\\
{\bf = 13 :} gives a list of current parameter values for MSTV, PARV, and for
        MSTW and PARW. This is useful to keep check of which default values 
        were changed in a given run.
\\
{\bf = 14 :} gives a list of time grid and size of time slizes used in the 
        current run.
\end{description}
\medskip

\begin{verbatim}
SUBROUTINE VNIEDIT(MEDIT)
\end{verbatim}

\noindent
{\it Purpose}: to exclude unstable or undetectable partons/particles from the
    event record. One may also use VNIEDIT to store spare copies of
    events (specifically initial parton configuration) that can be
    recalled to allow e.g. different fragmentation schemes to be run
    through with one and the same parton configuration. Finally, an
    event which has been analysed with VNISPHE, VNITHRU or VNICLUS
    (see below) may be rotated to align the event axis with
    the $z$-direction.

\begin{description}
\item[{\boldmath $MEDIT$}  :] tells which action is to be taken.
\\
{\bf = 0 :} empty ($KS = 0$) and documentation ($KS > 20$) lines are removed.
        The partons/particles remaining are compressed in the beginning of
        the VNIREC1 and VNIREC2  commonblocks 
	and the $N$ value is updated accordingly.
        The event history is lost, so that information stored in K(I,3),
        K(I,4) and K(I,5) is no longer relevant.
\\
{\bf = 1 :} as = 0, but in addition all partons/particles that have
        fragmented/decayed ($KS > 10$) are removed.
\\
{\bf = 2 :} as = 1, but also all neutrinos and unknown particles (i.e.
        compressed code $KC = 0$) are removed.
\\
{\bf = 3 :} as = 2, but also all uncharged, color neutral particles are
        removed, leaving only charged, stable particles (and
        unfragmented partons, if fragmentation has not been performed).
\\
{\bf = 5 :} as = 0, but also all partons which have branched or been
        rearranged in a parton shower and all particles which have
        decayed are removed, leaving only the fragmenting parton
        configuration and the final state particles.
\\
{\bf = 11 :} remove lines with K(I,1) $<$ 0. Update event history
        information (in K(I,3) - K(I,5)) to refer to remaining entries.
\\
{\bf = 12 :} remove lines with K(I,1) = 0. Update event history
        information (in K(I,3) - K(I,5)) to refer to remaining entries.
\\
{\bf = 13 :} remove lines with K(I,1) = 11, 12 or 15, except for any line
        with K(I,2) = 94. Update event history information (in K(I,3) -
        K(I,5)) to refer to remaining entries. In particular, try to
        trace origin of daughters, for which the mother is decayed,
        back to entries not deleted.
\\
{\bf = 14 :} remove lines with K(I,1) = 13 or 14, and also any line with
        K(I,2) = 94. Update event history information (in K(I,3) -
        K(I,5)) to refer to remaining entries. In particular, try to
        trace origin of rearranged partons back through the parton shower
        history to the shower initiator.
\\
{\bf = 15 :} remove lines with K(I,1) $>$ 20. Update event history
        information (in K(I,3) - K(I,5)) to refer to remaining entries.
\\
{\bf = 21 :} all partons/particles in current event record are stored
        (as a spare copy) in bottom of commonblocks VNIREC1/VNIREC2.
\\
{\bf = 22 :} partons/particles stored in bottom of event record with = 21
        are placed in beginning of record again, overwriting previous
        information there (so that e.g. a different fragmentation scheme
        can be used on the same partons). Since the copy at bottom is
        unaffected, repeated calls with = 22 can be made.
\\
{\bf = 23 :} primary partons/particles in the beginning of event record
        are marked as not fragmented or decayed, and number of entries
        N is updated accordingly. Is simple substitute for = 21 plus = 22
        when no fragmentation/decay products precede any of the
        original partons/particles.
\\
{\bf = 31 :} rotate largest axis, determined by VNISPHE, VNITHRU or VNICLUS,
        to sit along the $z$-direction, and the second largest axis into
        the $xz$ plane. For VNICLUS it can be further specified to $+z$ axis
        and $xz$ plane with $x > 0$, respectively. Requires that one of
        these routines has been called before.
\\
{\bf = 32 :} mainly intended for VNISPHE and VNITHRU, this gives a further
        alignment of the event, in addition to the one implied by = 31.
        The "slim" jet, defined as the side ($z > 0$ or $z < 0$) with the
        smallest summed $p_\perp$ over square root of number of particles,
        is rotated into the $+z$ hemisphere. In the opposite hemisphere
        (now $z < 0$), the side of $x > 0$ and $x < 0$ which has the largest
        summed $p_z$ absolute is rotated into the $z < 0$, $x > 0$ quadrant.
        Requires that VNISPHE or VNITHRU has been called before.
\end{description}
\medskip

\begin{verbatim}
FUNCTION KVNI(I,J)
\end{verbatim}

\noindent
{\it Purpose:} to provide various integer-valued event data. Note that many
    of the options available (in particular $I > 0, J \ge 14$) which
    refer to event history will not work after a VNIEDIT call.

\begin{description}
\item[{\boldmath $I = 0, J = $} :] properties referring to the complete event.
\\
{\bf = 1 :} N, total number of lines in event record.
\\
{\bf = 2 :} total number of partons/particles remaining after
        fragmentation and decay.
\\
{\bf = 6 :} three times the total charge of remaining (stable) partons
        and particles.
\\
\item[{\boldmath $I > 0, J = $} :] 
properties referring to the entry in line no. $I$ of the event record.
\\
{\bf = 1 - 5 :} K(I,1) - K(I,5), i.e. parton/particle status KS, flavour
        code KF and origin/decay product/color-flow information.
\\
{\bf = 6 :} three times parton/particle charge.
\\
{\bf = 7 :} 1 for a remaining entry, 0 for a decayed/fragmented/
        documentation entry.
\\
{\bf = 8 :} $KF$ code (K(I,2)) for a remaining entry, 0 for a
        decayed/fragmented/documentation entry.
\\
{\bf = 9 :} $KF$ code (K(I,2)) for a parton (i.e. not color-neutral entry),
        0 for a particle.
\\
{\bf = 10 :} $KF$ code (K(I,2)) for a particle (i.e. color-neutral entry),
        0 for a parton.
\\
{\bf = 11 :} compressed flavour code $KC$.
\\
{\bf = 12 :} color information code, i.e. 0 for color neutral, 1 for
        color triplet, -1 for antitriplet and 2 for octet.
\\
{\bf = 13 :} flavour of "heaviest" quark or antiquark (i.e. with largest
        code) in hadron or diquark (including sign for antiquark),
        0 else.
\\
{\bf = 14 :} generation number. Beam particles or virtual exchange
        particles are generation 0, original partons/particles generation
        1 and then 1 is added for each step in the fragmentation/decay
        chain.
\\
{\bf = 15 :} line number of ancestor, i.e. predecessor in first generation
        (generation 0 entries are disregarded).
\\
{\bf = 16 :} rank of a hadron in a jet system it belongs to. 
        Rank denotes the
        ordering in flavour space, with hadrons containing the original
        flavour of the jet having rank 1, increasing by 1 for each step
        away in flavour ordering. All decay products inherit the rank
        of their parent. Whereas the meaning of a first-rank hadron
        in a quark jet is always well-defined, the definition of higher
        ranks is only meaningful for independently fragmenting quark
        jets. In other cases, rank refers to the ordering in the actual
        simulation, which may be of little interest.
\\
{\bf = 17 :} generation number after a collapse of a jet system into one
        particle, with 0 for an entry not coming from a collapse, and
        -1 for entry with unknown history. A particle formed in a
        collapse is generation 1, and then one is added in each decay
        step.
\\
{\bf = 18 :} number of decay/fragmentation products (only defined in a
        collective sense for fragmentation).
\\
{\bf = 19 :} origin of color for showering parton, 0 else.
\\
{\bf = 20 :} origin of anticolor for showering parton, 0 else.
\\
{\bf = 21 :} position of color daughter for showering parton, 0 else.
\\
{\bf = 22 :} position of anticolor daughter for showering parton, 0 else.
\\
{\bf = 23 :} number of 2-body collisions of parton/particle.
\\
{\bf = 24 :} number of space-like decays of parton/particle.
\\
{\bf = 25 :} number of time-like decays of parton/particle.
\\
{\bf = 27 :} entry of direct mother of parton/particle.
\\
{\bf = 28 :} timestep of production of parton/particle.
\\
{\bf = 29 :} index of generation in the cascade tree of parton/particle.
\end{description}
\medskip

\begin{verbatim}
FUNCTION PVNI(I,J)
\end{verbatim}

\noindent
{\it Purpose}: to provide various real-valued event data. Note that some
    of the options available ($I > 0, J = 20-25$) are primarily intended
    for studies of systems in their respective CM frame.

\begin{description}
\item[{\boldmath $I = 0, J = $} :] properties referring to the complete event.
\\
{\bf = 1 - 4 :} sum of $p_x, p_y, p_z$ and $E$, respectively, for the stable
        remaining entries.
\\
{\bf = 5 :} invariant mass of the stable remaining entries.
\\
{\bf = 6 :} sum of electric charge of the stable remaining entries.

\item[{\boldmath $I > 0, J = $} :]
    properties referring to the entry in line no. $I$ of the event
    record.
\\
{\bf = 1 - 5 :} P(I,1) - P(I,5), i.e. normally $p_x, p_y, p_z, E$ and $m$ for
        parton/particle.
\\
{\bf = 6 :} electric charge $q$.
\\
{\bf = 7 :} momentum squared $\vec{p}^2 = p_x^2 + p_y^2 + p_z^2$.
\\
{\bf = 8 :} momentum $p = \sqrt{\vec{p}^2}$.
\\
{\bf = 9 :} transverse momentum squared $p_\perp^2 = p_x^2 + p_y^2$.
\\
{\bf = 10 :} transverse momentum $p_\perp$.
\\
{\bf = 11 :} transverse mass squared $m_\perp^2 = m^2 + p_x^2 + p_y^2$.
\\
{\bf = 12 :} transverse mass $m_\perp$.
\\
{\bf = 13 - 14 :} polar angle $\theta$ in radians (between 0 and $\pi$) or
        degrees, respectively.
\\
{\bf = 15 - 16 :} azimuthal angle $\phi$ in radians (between $-\pi$ and $\pi$)
         or degrees, respectively.
\\
{\bf = 17 :} true rapidity $y = \frac{1}{2} \ln((E+p_z)/(E-p_z))$.
\\
{\bf = 18 :} rapidity $y_\pi$ obtained by assuming that the particle is a
        pion when calculating the energy $E$, to be used in the
        formula above, from the (assumed known) momentum $p$.
\\
{\bf = 19 :} pseudorapidity $\eta = \frac{1}{2} \ln((p+p_z)/(p-p_z))$.
\\
{\bf = 20 :} momentum fraction $x_p = 2p/W$, where $W$ is the total energy of
        initial parton/particle configuration.
\\
{\bf = 21 :} $x_F = 2p_z/W$ (Feynman-$x$ if system is studied in CM frame).
\\
{\bf = 22 :} $x_\perp = 2p_\perp/W$.
\\
{\bf = 23 :} $x_E = 2E/W$.
\\
{\bf = 24 :} $z_+ = (E+p_z)/W$.
\\
{\bf = 25 :} $z_- = (E-p_z)/W$.
\end{description}
\medskip

\begin{verbatim}
SUBROUTINE VNIANAL(MSEL)
\end{verbatim}

\noindent
{\it Purpose}: to administrate in each event, which of the available analysis 
     duties (described below) are carried out, and when during the time
     evolution of the particle system data samples are taken and accumula-
     ted to perform a statistical analysis at the end. It calls the 
     routines VNIANA1 - VNIANA4 to do this job, for the options that are 
     switched on by default or by the user. The switches and their effect
     are described in Sec. 3.7.  After the last event of a run, the 
     results are automatically printed out in tabular form to files 
     called VNI???.DAT, with numbers ???, corresponding to the selected 
     analysis options.

\begin{description}
\item[{\boldmath $MSEL$} :] gives the type of action to be taken.
\\
{\bf =  0 :} all arrays for data analysis in VNIANA1 - VNIANA4 are reset.
\\
{\bf =  1 :} data samples are accumulated, stored and analyzed in VNIANA1 - 
     VNIANA4.
\\
{\bf = 10 :} final results are written to output files called VNI.???.DAT,
     where `???' is a 3-digit number specified in the descriptions below.
\end{description}
\medskip

\begin{verbatim}
SUBROUTINE VNIANA1(MSEL)
\end{verbatim}

\noindent
{\it Purpose}: to provide a number of global analyses referring to time-
    independent quantities of events as-a-whole, as e.g. total charged 
    multiplicities, etc.. Is adopted from PYTHIA 
    and only slightly altered.
    Accumulated statistics is written out at the end of a run. When errors 
    are quoted, these refer to the uncertainty in the average value for the 
    event sample as a whole, rather than to the spread of the individual 
    events, i.e.  errors decrease like one over the square root of the 
    number of events analyzed.

\begin{description}
\item[{\boldmath $MSEL$} :]
    determines which action is to be taken. Generally, a last digit
    equal to 0 indicates that the statistics counters for this option
    is to be reset. Last digit 1 leads to an analysis of current event 
    with respect to the desired properties. The statistics accumulated 
    is output in tabular form with last digit 2, and written to disc.
\\
\\
{\bf = 10 :} statistics on parton multiplicity is reset.
\\
{\bf = 11 :} the parton content of the current event is analyzed,
        classified according to the flavour content of the hard
        interaction and the total number of partons.
\\
{\bf = 12 :} gives a table on parton multiplicity distribution, written
        to file  VNI110.DAT.
\\
\\
{\bf = 20 :} statistics on particle content is reset.
\\
{\bf = 21 :} the particle/parton content of the current event is analyzed,
        also for particles which have subsequently decayed and partons
        which have fragmented. Particles are subdivided into primary
        and secondary ones, the main principle being that primary
        particles are those produced directly, while secondary come from 
        decay of other particles. 
\\
{\bf = 22 :} gives a table of particle content in events, written to file
        VNI120.DAT.
\\        
\\
{\bf = 30 :} statistics on factorial moments is reset.
\\
{\bf = 31 :} analyzes the factorial moments of the multiplicity
        distribution in different bins of rapidity and azimuth.
\\
{\bf = 32 :} prints a table of the first four factorial moments for various 
        bins of pseudorapidity and azimuth, written to file VNI130.DAT. 
        The moments are properly normalized so that they would be unity 
        (up to statistical fluctuations) for uniform and uncorrelated particle
        production according to Poissonian statistics, but increasing
        for decreasing bin size in case of "intermittent" behaviour.
\\        
\\
{\bf = 40 :} statistics on energy-energy correlation is reset.
\\
{\bf = 41 :} the energy-energy correlation (EEC) of the current event is
        analyzed. Events are assumed given in their CM frame. The weight 
        assigned to a pair $i$ and $j$ is $2E_iE_j/W^2$, where $W$ 
        is the sum of 
        energies of all analyzed particles in the event. Energies are 
        determined from the momenta of particles and masses. Statistics is 
        accumulated for the relative angle $\theta_{ij}$, 
        ranging between 0 and 180 
        degrees, subdivided into 50 bins.
\\
{\bf = 42 :} prints a table of the energy-energy correlation (EEC) and
        its asymmetry (EECA), with errors, written to file VNI140.DAT.  
        The definition of errors is not unique. Each event is viewed as 
        one observation, i.e. an EEC and EECA distribution is obtained by 
        summing over all particle pairs of an event, and then the average 
        and spread of this event-distribution is calculated in the standard 
        fashion, with the quoted error containing a factor of 
        $1/\sqrt{\mbox{no. events}}$. 
\\
\\
{\bf = 50 :} statistics on complete final states is reset.
\\
{\bf = 51 :} analyzes the particle content of the final state of the
        current event record. During the course of the run, statistics
        is thus accumulated on how often different final states appear.
        Only final states with up to 8 particles are analyzed, and there
        is only reserved space for up to 200 different final states.
        Most high energy events have multiplicities far above 8, so the
        main use for this tool is to study the effective branching
        ratios obtained with a given decay model for e.g. charm or
        bottom hadrons. 
\\
{\bf = 52 :} gives a list of the (at most 200) channels with up to 8
        particles in the final state, with their relative branching
        ratio, written to file VNI150.DAT. The ordering is according to 
        multiplicity, and within each multiplicity according to an ascending 
        order of $KF$ codes.  The $KF$ codes of the particles belonging to a 
        given channel are given in descending order.
\end{description}
\medskip

\begin{verbatim}
SUBROUTINE VNIANA2(MSEL)
\end{verbatim}

\noindent
{\it Purpose}: to analyze the time evolution of particles yields, energy
     ad momentum composition. Also, to analyze final state hadron 
     spectra in the energy variable $\ln(1/x)$ and to
     model Bose-Einstein correlations.

\begin{description}
\item[{\boldmath $MSEL$} :]
 determines which action is to be taken, with the convention noted before. 
\\
\\
{\bf = 10 :} statistics on time evolution of particle numbers, energy 
        partitions, longitudinal momentum fractions and  average 
        transverse momentum is reset.
\\
{\bf = 11 :} the particle content is analyzed in each single time step 
        during an event, for the different particle species individually.
\\
{\bf = 12 :} gives a table on the obtained time evolution profiles, written 
        to file VNI210.DAT.
\\
\\
{\bf = 20 :} statistics on $y=ln(1/x)$ distribution of final hadrons is reset,
        where $x = 2 E/\sqrt{s}$ is the energy fraction of a particle.
\\
{\bf = 21 :} hadron content at the end of each event is binned in $y=\ln(1/x)$
        individually for all, and for charged, hadrons.
        Covered range in $y$ is given by PARW(161) and PARW(162).
\\
{\bf = 22 :} gives a table of the obtained $y=ln(1/x)$ spectra, written to
        file VNI220.DAT.
\\
\\
{\bf = 30 :} statistics on Bose-Einstein correlations of final state pions
        and kaons is reset.
\\
{\bf = 31 :} Bose-Einstein effects are simulated according to a simple
        parametrization, with an algorithm adopted from PYTHIA. The 
        distribution of pair masses is stored for both cases, without and 
        with Bose-Einstein correlations taken into account.
        Covered range in pair mass is given by PARW(163) and PARW(164).
\\
{\bf = 32 :} gives a table of the obtained pair mass spectrum of `like-sign'
        pions and kaons, without and with Bose-Einstein effects, written
        to file VNI230.DAT.
\end{description}
\medskip

\begin{verbatim}
SUBROUTINE VNIANA3(MSEL)
\end{verbatim}

\noindent
{\it Purpose}: to analyze the dynamics of pre-hadronic clusters formed from 
    partons during the parton-hadron conversion stage of the system.
    The cluster size- and mass-distributions, and momentum and spatial
    spectra are accumulated at various time steps during the time 
    evolution of the system, so that the time change of the spectra 
    gives a history of the cluster formation processes, locally in
    phase-space.

\begin{description}
\item[{\boldmath $MSEL$} :]
    determines which action is to be taken, with the convention noted
    before. The statistics accumulated is output in tabular form with
    bins, ideally covering the range of momenta and coordinates of relevant
    particles. Depending on the kinematics, however, the limits of this range
    may require an adjustment, and/or a finer resolution of bins may be
    necessary.
\\
\\
{\bf = 10 :} statistics on cluster sizes and their invariant mass spectrum is 
        reset.
\\
{\bf = 11 :}  
        the cluster content of the system is analyzed each single time step 
        during an an event, and cluster sizes and masses are stored in bins.
        Covered range of analysis is $0 \le R \le 1.5$ fm, and 
        $0 \le M \le 7$ GeV.
\\
{\bf = 12 :} gives a table on the obtained cluster size- and mass-spectra,
        written to file VNI310.DAT.
\\
\\
{\bf = 20 :} statistics on rapidity ($y$) spectra of clusters is reset.
\\
{\bf = 21 :} the cluster content at certain pre-selected time steps during an
        an event is analyzed. 
        Covered range in $y$ is given by PARW(171) and PARW(172).
\\
{\bf = 22 :} gives a table on the obtained $y$-spectra, written to file VNI320.DAT.
\\
\\
{\bf = 30 :} statistics on longitudinal ($r_z$) spatial distribitions of clusters
        is reset.
\\
{\bf = 31 :} the cluster positions at certain pre-selected time steps during an
        event are analyzed, with respect to the overall CM frame and to the 
        $z$-axis, usually the jet-axis, or beam-axis. 
        Covered range in $r_z$ is given by PARW(173) and PARW(174).
\\
{\bf = 32 :} gives a table on the obtained $r_z$-spectra, 
      written to file VNI330.DAT.
\\
\\
{\bf = 40 :} statistics on transverse momentum ($p_\perp$) 
      distribitions of clusters is reset.
\\
{\bf = 41 :} the cluster momenta $p_\perp$ 
       at certain pre-selected time steps during an
       an event are analyzed, transverse with respect to the z-axis. 
        Covered range in $p_\perp$ is given by PARW(175) and PARW(176).
\\
{\bf = 42 :} gives a table on the obtained $p_\perp$-spectra, 
       written to file VNI340.DAT.
\\
\\
{\bf = 50 :} statistics on transverse ($r_\perp$) 
       spatial distribitions of clusters is reset.
\\
{\bf = 51 :} the cluster content at certain pre-selected time steps during an
        an event is analyzed, with respect to the distance transverse to the 
        $z$-axis. 
        Covered range in $r_\perp$ is given by PARW(177) and PARW(178).
\\
{\bf = 52 :} gives a table on the obtained $r_\perp$-spectra, 
       written to file VNI350.DAT.
\end{description}
\medskip

\begin{verbatim}
SUBROUTINE VNIANA4(MSEL)
\end{verbatim}

\noindent
{\it Purpose}: to analyze momentum and spatial distributions of partons and
    hadrons at various time steps during the time evolution of the system, 
    so that the time change of the spectra gives a history of the dynamics 
    from the initial state all the way to the final state of hadrons.

\begin{description}
\item[{\boldmath $MSEL$} :]
    determines which action is to be taken, with the convention noted
    before. The statistics accumulated is output in tabular form with
    bins, ideally covering the range of momenta and coordinates of relevant
    particles. Depending on the kinematics, however, the limits of this range
    may require an extension. 
\\
\\
{\bf = 10 :} statistics on rapidity ($y$) spectra of partons and hadrons is reset.
\\
{\bf = 11 :} the particle content at certain pre-selected time steps during an
        an event is analyzed, for gluons, quarks, mesons and baryons seperately.
        Covered range in $y$ is given by PARW(181) and PARW(182).
\\
{\bf = 12 :} gives a table on the obtained $y$-spectra, 
        written to file VNI410.DAT.
\\
\\
{\bf = 20 :} statistics on longitudinal ($r_z$) spatial distribitions of 
        partons and hadrons is reset.
\\
{\bf = 21 :} the particle positions at certain pre-selected time steps during an
        event are analyzed, with respect to the overall CM frame and to the 
        $z$-axis, usually the jet-axis, or beam-axis. 
        Covered range in $r_z$ is given by PARW(183) and PARW(184).
\\
{\bf = 22 :} gives a table on the obtained $r_z$-spectra, 
       written to file VNI420.DAT.
\\
\\
{\bf = 30 :} statistics on transverse momentum ($p_\perp$) 
        distribitions of partons and hadrons is reset.
\\
{\bf = 31 :} the particle momenta $p_\perp$ at 
        certain pre-selected time steps during an
        an event are analyzed, transverse with respect to the $z$-axis.
        Covered range in $p_\perp$ is given by PARW(185) and PARW(186).
\\
{\bf = 32 :} gives a table on the obtained pT-spectra, 
        written to file VNI430.DAT.
\\
\\
{\bf = 40 :} statistics on transverse ($r_\perp$) spatial distribitions of 
        partons and 
        hadrons is reset.
\\
{\bf = 41 :} the particle content at certain pre-selected time steps 
        during an an event is analyzed, 
        with respect to the distance transverse to the 
        $z$-axis.
        Covered range in $r_\perp$ is given by PARW(187) and PARW(188).
\\
{\bf = 42 :} gives a table on the obtained $r_\perp$-spectra, 
        written to file VNI440.DAT.
\end{description}
\medskip

\begin{verbatim}
SUBROUTINE VNISPHE(SPH,APL)
\end{verbatim}

\noindent
{\it Purpose:} to diagonalize the momentum tensor, i.e. find the
    eigenvalues $\lambda_1 > \lambda_2 > \lambda_3$, with sum unity,
    and the corresponding eigenvectors.
    Momentum power dependence is given by PARU(41); default corresponds
    to sphericity, PARU(41)=1. gives measures linear in momenta.
    Which particles (or partons) are used in the analysis is determined
    by the MSTU(41) value.

\begin{description}
\item[{\boldmath $SPH$} :]
   $3(\lambda_2 + \lambda_3)/2$, i.e. sphericity (for PARU(41)=2.).
   Is returned  = -1, if analysis not performed because event contained less
        than two particles (or two exactly back-to-back particles, in
        which case the two transverse directions would be undefined).
\item[{\boldmath $APL$} :]
 $3/2 \lambda_3$, i.e. aplanarity (for PARU(41)=2.).
    Is returned = -1,  as $SPH = -1$
\item[Remark:]
    the lines $N+1$ through $N+3$ ($N-2$ through $N$ for MSTU(43) = 2) in
    VNIREC1/VNIREC2 will, after a call, contain the following information:
\\
    K($N+i,1) = 31$;
\\
    K($N+i,2) = 95$;
\\
    K($N+i,3) : i$, the axis number, $i=1,2,3$;
\\
    K($N+i,4)$, K($N+i,5) = 0$;
\\
    P($N+i,1)$ - P($N+i,3)$ : the $i$-th eigenvector, $x,y$ and $z$ components;
\\
    P($N+i,4) : \lambda_i$, the $i$-th eigenvalue;
\\
    P($N+i,5) = 0$;
\\
    V($N+i,1)$ - V($N+i,5) = 0$.
\\
    Also, the number of particles used in the analysis is given in
    MSTU(62).
\end{description}
\medskip

\begin{verbatim}
SUBROUTINE VNITHRU(THR,OBL)
\end{verbatim}

\noindent
{\it Purpose}: to find the thrust, major and minor axes and corresponding
    projected momentum quantities, in particular thrust and oblateness.
    The performance of the program is affected by MSTU(44), MSTU(45),
    PARU(42) and PARU(48). In particular, PARU(42) gives the momentum
    dependence, with the default value 1. corresponding to linear
    dependence. Which particles (or partons) are used in the analysis
    is determined by the MSTU(41) value.

\begin{description}
\item[{\boldmath $THR$} :]
    thrust (for PARU(42)=1.).
    Is returned = -1, if analysis cannot be not performed,
        because event contained less than two particles.
    Is returned = -2, if remaining space in VNIREC1/VNIREC2 
        (partly used as working area)
        is not large enough to allow analysis.
\item[{\boldmath $OBL$} :]
   oblateness (for PARU(42)=1.).
   Is returned = -1, respectively = -2,  as for $THR$.
\item[Remark:]
     the lines $N+1$ through $N+3$ ($N-2$ through $N$ for MSTU(43) = 2) in
    VNIREC1/VNIREC2 will, after a call, contain the following information:
\\
    K($N+i,1) = 31$;
\\
    K($N+i,2) = 96$;
\\
    K($N+i,3) : i$, the axis number, $i=1,2,3$;
\\
    K($N+i,4$), K($N+i,5) = 0$;
\\
    P($N+i,1$) - P($N+i,3)$ : the thrust, major and minor axis,
        respectively, for $i =$ 1, 2 and 3;
\\
    P($N+i,4)$ : corresponding thrust, major and minor value;
\\
    P($N+i,5) = 0$;
\\
    V($N+i,1$) - V($N+i,5) = 0$.
\\
    Also, the number of particles used in the analysis is given in
    MSTU(62).
\end{description}
\medskip

\begin{verbatim}
SUBROUTINE VNICLUS(NJET)
\end{verbatim}

\noindent
{\it Purpose}: to reconstruct an arbitrary number of jets using a cluster
    analysis method based on particle momenta. Two different distance
    measures are available, the traditional one roughly corresponding
    to relative transverse momentum, and  one based on the JADE method, 
    which roughly corresponds to invariant mass (in both cases with 
    some important modifications).  The choice is controlled by MSTU(46). 
    The distance scale $d_{join}$, above which two clusters may not be joined, 
    is normally given by PARU(44). In general, $d_{join}$ may be varied to 
     describe different "jet resolution powers". With the mass distance,
    PARU(44) can be used to set the absolute maximum cluster mass, or
    PARU(45) to set the scaled one, i.e. in $y = m^2/W^2$, where $W^2$ is
    the total invariant mass-squared of the particles being considered.
    It is possible to continue the cluster search from the configuration
    already found, with a new higher $d_{join}$ scale, by selecting MSTU(48)
    properly. In MSTU(47) one can also require a minimum number of jets
    to be reconstructed; combined with an artificially large $d_{join}$ this
    can be used to reconstruct a predetermined number of jets. Which
    particles (or partons) are used in the analysis is determined by the
    MSTU(41) value, whereas assumptions about particle masses are given
    by MSTU(42). The parameters PARU(43) and PARU(48) regulate more
    technical details (for events at high energies and large
    multiplicities, however, the choice of a larger PARU(43) may be
    necessary to obtain reasonable reconstruction times).

\begin{description}
\item[{\boldmath $NJET$} :]
    the number of clusters reconstructed.
    Is returned = -1, if analysis cannot performed, 
        because event contained less than
        MSTU(47) (normally 1) particles, or analysis failed to
        reconstruct the requested number of jets.
    Is returned = -2, if remaining space in VNIREC1/VNIREC2 
        (partly used as working area)
        is not large enough to allow analysis.
\item[Remark:] if the analysis does not fail, further information is found
    in MSTU(61) - MSTU(63) and PARU(61) - PARU(63). In particular,
    PARU(61) contains the invariant mass for the system analyzed, i.e.
    the number used in determining the denominator of $y = m^2/W^2$.
    PARU(62) gives the generalized thrust, i.e. the sum of (absolute
    values of) cluster momenta divided by the sum of particle momenta
    (roughly the same as multicity). PARU(63) gives the minimum
    distance d (in $p_\perp$ or $m$) between two clusters in the final cluster
    configuration, 0 in case of only one cluster. Further, the lines
    $N+1$ through $N+NJET$ ($N-NJET+1$ through $N$ for MSTU(43) = 2) in 
    VNIREC1/VNIREC2
    will, after a call, contain the following information:
\\
    K($N+i,1) = 31$;
\\
    K($N+i,2) = 97$;
\\
    K($N+i,3) : i$, the jet number, with the jets arranged in falling
        order of absolute momentum;
\\
    K($N+i,4) :$ the number of particles assigned to jet i;
\\
    K($N+i,5) = 0$;
\\
    P($N+i,1$) - P($N+i,5) :$ momentum, energy and invariant mass of jet i;
\\
    V($N+i,1$) - V($N+i,5) = 0$.
\\
    Also, for a particle which was used in the analysis, K($I,4) = i$,
    where $I$ is the particle number and $i$ the number of the jet it has
    ben assigned to. Undecayed particles not used then have K($I,4) = 0$.
    An exception is made for lines with K($I,1) = 3$ (which anyhow are not
    normally interesting for cluster search), where the color flow
    information stored in K($I,4)$ is left intact.
\end{description}
\medskip

\begin{verbatim}
SUBROUTINE VNICELL(NJET)
\end{verbatim}

\noindent
{\it Purpose}: to provide a simpler cluster routine more in line with
    what is currently used in the study of high-$p_\perp$ collider events.
    A detector is assumed to stretch in pseudorapidity between -PARU(51)
    and +PARU(51) and be segmented in MSTU(51) equally large $\eta$
    (pseudorapidity) bins and MSTU(52) $\phi$ (azimuthal) bins. Transverse
    energy $E_\perp$ for undecayed entries are summed up in each bin. For
    MSTU(53) nonzero, the energy is smeared by calorimetric resolution
    effects, cell by cell. This is done according to a Gaussian
    distribution; if MSTU(53) = 1 the standard deviation for the $E_\perp$ is
    PARU(55)$ \sqrt{E_\perp}$, if MSTU(53) = 2 the standard deviation for the
    $E$ is PARU(55)$ \sqrt{E}$, $E_\perp$ and $E$ expressed in GeV. 
    The Gaussian is
    cut off at 0 and at a factor PARU(56) times the correct $E_\perp$ or $E$.
    All bins with $E_\perp > $PARU(52) are taken to be possible initiators
    of jets, and are tried in falling $E_\perp$ sequence to check whether
    the total $E_\perp$ summed over cells no more distant than PARU(54) in
    $\sqrt{(\Delta\eta)^2 + (\Delta\phi)^2}$ exceeds PARU(53). If so, these
    cells define one jet, and are removed from further consideration.
    Contrary to VNICLUS, not all particles need be assigned to jets.
    Which particles (or partons) are used in the analysis is determined
    by the MSTU(41) value.

\begin{description}
\item[{\boldmath $NJET$} :]
    the number of jets reconstructed (may be 0).
    Is returned = -2, if remaining space in VNIREC1/VNIREC2
    (partly used as working area)
        is not large enough to allow analysis.
\item[Remark:] the lines $N+1$ through $N+NJET$ ($N-NJET+1$ through $N$ for
    MSTU(43) = 2) in VNIREC1/VNIREC2 will, after a call, contain the
    following information:
\\
    K($N+i,1) = 31$;
\\
    K($N+i,2) = 98$;
\\
    K($N+i,3) : i$, the jet number, with the jets arranged in falling
        order in $E_\perp$;
\\
    K($N+i,4) :$ the number of particles assigned to jet $i$;
\\
    K($N+i,5) = 0$;
\\
    V($N+i,1$) - V($N+i,5) = 0$;
\\
\\
    Further, for MSTU(54) = 1:
\\
    P($N+i,1$), P($N+i,2) =$ position in $\eta$ and $\phi$ of the center of the
        jet initiator cell, i.e. geometrical center of jet;
\\
    P($N+i,3), P($N+i,4) = position in $\eta$ and $\phi$ of the 
        $E_\perp$-weighted
        center of the jet, i.e. the center of gravity of the jet;
\\
    P($N+i,5) = \sum E_\perp$ of the jet;
\\
\\
    while for MSTU(54) = 2:
\\
    P($N+i,1$) - P($N+i,5) :$ the jet momentum vector, constructed from the
        summed $E_\perp$ and the $\eta$ and $\phi$ of the 
        $E_\perp$-weighted center of
        the jet as $(p_x, p_y, p_z, E, m) =
        E_\perp (\cos\phi, \sin\phi, \sinh\eta, \cosh\eta, 0)$;
\\
\\
    and for MSTU(54) = 3:
\\
    P($N+i,1$) - P($N+i,5) :$ the jet momentum vector, constructed by adding
        vectorially the momentum of each cell assigned to the jet,
        assuming that all the $E_\perp$ was deposited at the center of the
        cell, and with the jet mass in P($N+i,5)$ calculated from the
        summed $E$ and $p$ as $m^2 = E^2 - p_x^2 - p_y^2 - p_z^2$.
\\
\\
    Also, the number of particles used in the analysis is given in
    MSTU(62), and the number of cells hit in MSTU(63).
\end{description}
\medskip

\begin{verbatim}
SUBROUTINE VNIJMAS(PMH,PML)
\end{verbatim}

\noindent
{\it Purpose}: to reconstruct high and low jet mass of an event, by dividing
    the collision event into two hemispheres, transversely to the 
    sphericity axis. Then one particle at a time is reassigned to the 
    other hemisphere if that reduces the sum of the two jet masses-
    squared $m_H^2 + m_L^2$. The procedure is stopped when no further
    significant change (see PARU(48)) is obtained. Often, the original
    assignment is retained as it is. Which particles (or partons)
    are used in the analysis is determined by the MSTU(41) value, whereas
    assumptions about particle masses are given by MSTU(42).

\begin{description}
\item[{\boldmath $PMH$} :] heavy jet mass (in GeV).
    Is returned = -2m, if remaining space in VNIREC1/VNIREC2 
     (partly used as working area)
      is  not large enough to allow analysis.
\item[{\boldmath $PML$} :]  light jet mass (in GeV).
    Is returned 
    = -2, if remaining space in VNIREC1/VNIREC2 (partly used as working area)
        is not large enough to allow analysis.
\item[Remark:]  light jet mass (in GeV).
    After a successful call, MSTU(62) contains the number of
    particles used in the analysis, and PARU(61) the invariant mass of
    the system analyzed. The latter number is helpful in constructing
    scaled jet masses.
\end{description}
\medskip

\begin{verbatim}
SUBROUTINE VNIFOWO(H10,H20,H30,H40)
\end{verbatim}

\noindent
{\it Purpose}: to do an event analysis in terms of the Fox-Wolfram moments
    The moments $H_i$ are normalized to the lowest one, $H_0$.  Which 
    particles (or partons) are used in the analysis is determined by 
    the MSTU(41) value.

\begin{description}
\item[{\boldmath $H10$} :]
 $H_1/H_0$. Is = 0 if momentum is balanced.
\item[{\boldmath $H20$} :]
 $H_2/H_0$.
\item[{\boldmath $H30$} :]
 $H_3/H_0$.
\item[{\boldmath $H40$} :]
 $H_4/H_0$.
\item[Remark:]
    the number of particles used in the analysis is given in
    MSTU(62).
\end{description}
\bigskip
\bigskip

\newpage

\section{Further commonblocks}
\label{sec:appc}
\bigskip

The subroutines and commonblocks that one normally comes in direct
contact with have already been described. For the sake of completeness,
some  further commonblocks are here described, which might be of interest 
to a user that desires a more detailed insight.
\bigskip
\bigskip

\noindent
The common block VNISYS contains information  concerning
the type and status of the collision system $A+B$ defined by the
beam and target particles, c.f Sec. 3.4.

\begin{verbatim}
      COMMON/VNISYS/NSYS(2,5),KSYS(2,5),PSYS(2,5),RSYS(2,5),AB(500,0:10)
\end{verbatim}

\noindent
{\it Purpose}:   to store information and data concerning
beam and target particles. It is in particular of interest if one
desires to use the user-specified frames USER, FOUR or FIVE, in which case
the user has to give the initial momenta PSYS and coordinates RSYS
of the colliding particles by assigning values before the initialization.
Furthermore, for collisions involving nuclei, the array AB  provides information 
on how many initial-state nucleons have been `wounded' and how often
they have been struck. For collisions without nuclei, the array AB is empty!

\begin{description}
\item[ NSYS(I,1) :]  number of neutrons in beam/target particle $I=1,2$.
\item[ NSYS(I,2) :]  number of protons.
\item[ NSYS(I,3) :]  number of $d$ valence quarks.
\item[ NSYS(I,4) :]  number of $u$ valence quarks.
\item[ NSYS(I,5) :]  number of $s$ valence quarks.
\medskip

\item[ KSYS(I,1)-]{\bf KSYS(I,5) :} status code, particle id, origin of beam/target
                   particle $I=1,2$, in analogy to the array $K$ in the 
		   particle record VNIREC1/VNIREC2.
\item[ PSYS(I,1)-]{\bf PSYS(I,5) :} initial momentum, energy, mass of beam/target particle,
                   in analogy to the array $P$ in the particle record 
		   VNIREC1/VNIREC2.
\item[ RSYS(I,1)-]{\bf RSYS(I,5) :} initial $x, y, z$ position before collision of
                   of beam/target particle, in analogy to the array $R$ 
                   in the particle record VNIREC1/VNIREC2. 
		   (RSYS(I,4) and RSYS(I,5) are currently
                   not used).
\medskip
\item[ AB(IH,0) :] number of collisions of nucleon $IH$, that is, the number of times
          an initial-state parton belonging to that nucleon gets struck and liberated
          from its mother nucleon. Here $1 \le IH \le N_{nuc}(A)+N_{nuc}(B)$, with
          $A$ and/or $B$ being the initial beam and/or target nucleus.
          AB($IH$,0) = 0 initially for all $IH$, and  at the
          end allows to discriminate `wounded' ($\ge 1$)and `unwounded'(= 0) nucleons.
\item[ AB(IH,1) :] {\it initial} number of partons in nucleon $IH$, before collision.
\item[ AB(IH,2) :] {\it current} number of partons in nucleon $IH$, during time evolution.
                    For `unwounded' nucleons AB($IH$,2)=AB($IH$,1).
\item[ AB(IH,3) :] $KF$ particle id of nucleon $IH$ (either proton or neutron).
\item[ AB(IH,4) :] $KF$ particle id of mother nucleus to which nucleon $IH$ belongs.
\item[ AB(IH,5) :] {\it initial} $P_\perp$ of nucleon $IH$ (before collision) 
                   with respect to collision axis of beam/target. 
\item[ AB(IH,6) :] {\it initial} $P_z$ of nucleon $IH$ with respect to collision axis.
\item[ AB(IH,7) :] {\it initial} $E$ of nucleon $IH$ with respect to chosen Lorentz frame.
\item[ AB(IH,8) :] {\it current} $P_\perp$ of nucleon $IH$ (during time evolution of collision) 
                   with respect to collision axis of beam/target. 
\item[ AB(IH,9) :] {\it current} $P_z$ of nucleon $IH$ with respect to collision axis.
\item[ AB(IH,10):] {\it current} $E$ of nucleon $IH$ with respect to chosen Lorentz frame.
\end{description}
\bigskip
\bigskip
\bigskip

\noindent
The common block VNINIA  contains detailed information on the number
of parton interactions for each of the included sub-process channels.

\begin{verbatim}
      COMMON/VNINIA/NIA(5,0:10)
\end{verbatim}

\noindent
{\it Purpose}:   to store accumulated statistics on type and
                 number of parton interactions.

\begin{description}
\item[ NIA(1,0) :]  total number of hard parton collisions $a+b\rightarrow c+d$.
\begin{description}
\item[ NIA(1,1) :]  number of hard $q + q \rightarrow q + q$.
\item[ NIA(1,2) :]  number of hard $q + \bar q \rightarrow q + \bar q$.
\item[ NIA(1,3) :]  number of hard $q + \bar q \rightarrow g + g$.
\item[ NIA(1,4) :]  number of hard $q + \bar q \rightarrow g + \gamma$.
\item[ NIA(1,5) :]  number of hard $q + \bar q \rightarrow \gamma + \gamma$.
\item[ NIA(1,6) :]  number of hard $q + g \rightarrow q + g$.
\item[ NIA(1,7) :]  number of hard $q + g \rightarrow q + \gamma$.
\item[ NIA(1,8) :]  number of hard $g + g \rightarrow q + \bar q$.
\item[ NIA(1,9) :]  number of hard $g + g \rightarrow g + g$.
\end{description}
\item[ NIA(2,0) :]  total number of soft parton collisions $a+b\rightarrow c+d$.
\begin{description}
\item[ NIA(2,1) :]  number of soft $q + q \rightarrow q + q$.
\item[ NIA(2,2) :]  number of soft $q + \bar q \rightarrow q + \bar q$.
\item[ NIA(2,3) :]  number of soft $q + \bar q \rightarrow g + g$.
\item[ NIA(2,4) :]  number of soft $q + \bar q \rightarrow g + \gamma$.
\item[ NIA(2,5) :]  number of soft $q + \bar q \rightarrow \gamma + \gamma$.
\item[ NIA(2,6) :]  number of soft $q + g \rightarrow q + g$.
\item[ NIA(2,7) :]  number of soft $q + g \rightarrow q + \gamma$.
\item[ NIA(2,8) :]  number of soft $g + g \rightarrow q + \bar q$.
\item[ NIA(2,9) :]  number of soft $g + g \rightarrow g + g$.
\end{description}
\item[ NIA(3,0) :]  total number of parton fusions $a+b\rightarrow c^\ast$.
\begin{description}
\item[ NIA(3,1) :]  number of $q + \bar q \rightarrow g^\ast$.
\item[ NIA(3,2) :]  number of $q + g \rightarrow q^\ast$.
\item[ NIA(3,3) :]  number of $g + g \rightarrow g^\ast$.
\end{description}
\item[ NIA(4,0) :]  total number of space-like branchings $a\rightarrow b+c$.
\begin{description}
\item[ NIA(4,1) :]  number of $q \rightarrow q + g$.
\item[ NIA(4,2) :]  number of $g \rightarrow g + g$.
\item[ NIA(4,3) :]  number of $g \rightarrow q + \bar q$.
\item[ NIA(4,4) :]  number of $q \rightarrow q + \gamma$.
\end{description}
\item[ NIA(5,0) :]  total number of time-like branchings $a\rightarrow b+c$.
\begin{description}
\item[ NIA(5,1) :]  number of $q \rightarrow q + g$.
\item[ NIA(5,2) :]  number of $g \rightarrow g + g$.
\item[ NIA(5,3) :]  number of $g \rightarrow q + \bar q$.
\item[ NIA(5,4) :]  number of $q \rightarrow q + \gamma$.
\end{description}
\end{description}
\bigskip
\bigskip
\bigskip

\noindent
The common block VNITIM  contains the  
(upon initialization) pretabulated time increments and time slizes
used for following the time-evolution of an collision event.

\begin{verbatim}
      COMMON/VNITIM/TINC(0:2500),TIME(0:2500)
\end{verbatim}

\noindent
{\it Purpose}:   to store time increments TINC and time steps TIME (in 1/GeV).

\begin{description}
\item[ TINC(I) :]  $i$-th time increment between time steps $i-1$ and $i$,
                   where $i =$ MSTW(12) and $1\le i\le $ MSTW(2).
                   PARW(1) is assigned that increment value at a each time step
                   during a collision event.
\item[ TIME(I) :]  $i$-th time step TIME($i$) = TIME($i-1$) + TINC($i$),
                   where $i =$ MSTW(12) and $1\le i\le $ MSTW(2).
                   PARW(2) is assigned that time value at a each time step
                   during a collision event.
\end{description}
\bigskip
\bigskip
\bigskip

The common block VNIINT serves as a temporary storage of
quantities during the simulation of a specific parton collision plus its
associated branching evolution.
It is very similar to a corresponding array in the PYTHIA program.

\begin{verbatim}
                COMMON/VNIINT/MINT(400),VINT(400)
\end{verbatim}

\noindent
{\it Purpose}: to provide an intermediate storage for integer and real valued 
    variables as work area, used in the program to carry out the evolution 
    of a single partonic subsystem at a time, consisting of a parton 
    collision and its associated space-like (initial state) and time-like 
    (final state) branchings. These variables must not be changed by the 
    user.

\begin{description}
\item[ MINT(1) :]
     specifies the general type of partonic subprocess that has 
    occured, according to the ISUB code given in Table 3.
\item[ MINT(2) :] whenever MINT(1) (together with MINT(15) and MINT(16)) are
    not enough to specify the type of a $2 \rightarrow 2$
    parton collision uniquely, 
    MINT(2) provides an ordering of the different possibilities, related
    to different color-flow topologies. The different possibilities
    are discussed in Refs. \cite{bengt88}. With $i$ = MINT(15), $j$ = MINT(16)
    and $k$ = MINT(2), they are:
\begin{description}
  \item[ ]
    ISUB = 1, $i = j, q_i q_i \rightarrow q_i q_i$;
\\
        $k = 1$ : color configuration A.
\\
        $k = 2$ : color configuration B.
  \item[ ]
    ISUB = 1, $i = j, q_i q_j \rightarrow q_i q_j$;
\\
        $k = 1$ : only possibility.
  \item[ ]
    ISUB = 2, $q_i \bar{q}_i \rightarrow q_l \bar{q}_l$;
\\
        $k = 1$ : only possibility.
  \item[ ]
    ISUB = 3, $q_i \bar{q}_i \rightarrow g g$;
\\
        $k = 1$ : color configuration A.
\\
        $k = 2$ : color configuration B.
  \item[ ]
    ISUB = 6, $q_i g \rightarrow q_i g$;
\\
        $k = 1$ : color configuration A.
\\
        $k = 2$ : color configuration B.
  \item[ ]
    ISUB = 8, $g g \rightarrow q_l \bar{q}_l$;
\\
        $k = 1$ : color configuration A.
\\
        $k = 2$ : color configuration B.
  \item[ ]
    ISUB = 9, $g g \rightarrow g g$;
\\
        $k = 1$ : color configuration A.
\\
        $k = 2$ : colour configuration B.
\\
        $k = 3$ : color configuration C.
\end{description}
\item[ MINT(3) :] number of partons produced in the hard interactions, i.e.
    the number $n$ of the $2 \rightarrow n$ matrix elements used, 
     where currently
     only $n = 1$ or $n = 2$ processes are included.
\item[ MINT(4) :] number of documentation lines in commonblocks VNIREC1/VNIREC2
     that are reserved and used as work space for simulating a parton collision.
\item[ MINT(5) :] not used.
\item[ MINT(6) :] current frame of event as initialized, cf. MSTW(4).
\item[ MINT(7) :] line number for documentation of outgoing particle 1 from 
    a parton collision $2\rightarrow 2$ or $2\rightarrow 1$.
\item[ MINT(8) :] line number for documentation of outgoing particle 2 from 
    a parton collision $2 \rightarrow 2$. Is 0 for $2 \rightarrow 1$.
\item[ MINT(9) :] not used.
\item[ MINT(10) :] not used.
\item[ MINT(11) :] $KF$ flavour 
       code of initial beam (side 1) particle (c.f. MSTW(5)).
\item[ MINT(12) :] $KF$ flavour 
       code of initial target (side 1) particle (c.f. MSTW(6)).
\item[ MINT(13) :] $KF$ flavour 
    code of side 1 parton that initiates a space-like
    cascade (only for primary partons, is 0 else).
\item[ MINT(14) :] $KF$ flavour code of 
    side 2 parton that initiates a space-like
    cascade (only for primary partons, is 0 else).
\item[ MINT(15) :] $KF$ flavour code 
    for side 1 incoming parton to collision subprocess.
\item[ MINT(16) :] $KF$ flavour code 
     for side 2 incoming parton to collision subprocess.
\\
\item[ MINT(17) -]{\bf MINT(20) :} not used.
\\
\item[ MINT(21) :] $KF$ flavour code for 
     parton 3 outgoing from a parton collision.
\item[ MINT(22) :] $KF$ flavour code for 
     parton 4 outgoing from a parton collision (is 0
     for $2\rightarrow 1$ processes).
\item[ MINT(23) :] $KF$ flavour code for 
   parton 5 outgoing from a parton collision - not
   used currently (relevant only for $2 \rightarrow 3, 2\rightarrow 4$ 
   processes).
\item[ MINT(24) :] $KF$ flavour code for parton 6 
   outgoing from a parton collision - not
   used currently (relevant only for  $2 \rightarrow  4$ processes).
\\
\item[ MINT(25) -]{\bf MINT(30) :} not used.
\\
\item[ MINT(31) -]{\bf MINT(40) :} not used.
\\
\item[ MINT(41) :] type of initial beam particle; 1 for lepton and 2 for hadron
    and 3 for nucleus (c.f. MSTW(7)).
\item[ MINT(42) :] type of initial  target 
    particle; 1 for lepton and 2 for hadron
    and 3 for nucleus (c.f. MSTW(8)).
\item[ MINT(43) :] combination of 
    incoming beam and target particles (c.f. MSTW(9)).
\item[ MINT(44) :] as MINT(43), but a photon counts as a lepton.
\item[ MINT(45), ]{\bf MINT(46) :}
    structure of incoming beam and target particles.
\\
{\bf = 1 :} no internal structure, i.e. lepton carries full beam energy.
\\
{\bf = 2 :} defined with structure functions, i.e. hadrons and nuclei.
\item[ MINT(47) :] combination of incoming beam and target particle structure
    function types.
\\
{\bf = 1 :} no structure function either for beam or target.
\\
{\bf = 2 :} structure functions for target but not for beam.
\\
{\bf = 3 :} structure functions for beam but not for target.
\\
{\bf = 4 :} structure functions both for beam and target.
\item[ MINT(48) :] total number of subprocesses switched on.
\item[ MINT(49), ]{\bf MINT(50) :} not used.
\item[ MINT(51) :] 
    internal flag that the simulation of a 2-parton collision system,
    including space-like initial state and time-like final state cascades, 
    has failed cuts.
\\
{\bf = 0 :} no problem.
\\
{\bf = 1 :} failed and to be skipped.
\\
\item[ MINT(52) -]{\bf MINT(60) :} not used.
\item[ MINT(61) -]{\bf MINT(70) :} not used.
\item[ MINT(81), ]{\bf MINT(82) :} not used.
\item[ MINT(83) :] number of lines in the event record already filled by
    previously considered pileup events.
\item[ MINT(84) :] MINT(83) + MSTP(126), i.e. number of lines already filled
    by previously considered events plus number of lines to be kept
    free for event documentation.
\\
\\
\item[ VINT(1)  :] $E_{CM}$, CM energy.
\item[ VINT(2)  :] $s =E_{CM}^2$, mass-square of complete system.
\item[ VINT(3)  :] mass of beam particle.
\item[ VINT(4)  :] mass of target particle.
\item[ VINT(5)  :] momentum of beam (and target) particle in CM frame.
\item[ VINT(6)  :] angle $\Theta$ for rotation from CM frame 
    to fixed target frame.
\item[ VINT(7)  :] angle $\Phi$ for rotation from CM 
    frame to fixed target frame.
\item[ VINT(8)  :] $\beta_x$ velocity for boost from CM frame to fixed target frame.
\item[ VINT(9)  :] $\beta_y$ velocity for boost from CM frame to fixed target frame.
\item[ VINT(10) :] $\beta_z$ velocity for boost from CM frame to fixed target frame.
\item[ VINT(11) :] $\hat{\beta}_x$ 
     velocity for boost from overall CM frame to `microscopic'
     cm frame of a parton-parton collision subsystem.
\item[ VINT(12) :] $\hat{\beta}_y$ 
     velocity for boost from overall CM frame to parton-parton 
     cm frame.
\item[ VINT(13) :] $\hat{\beta}_z$ 
     velocity for boost from overall CM frame to parton-parton 
     cm frame.
\item[ VINT(14) :] angle $\hat{\theta}$ 
      for rotation from overall CM frame to `microscopic'
     cm frame of a parton-parton collision subsystem.
\item[ VINT(15) :] angle $\hat{\phi}$ 
     for rotation from overall CM frame to parton-parton
     cm frame.
\item[ VINT(16) :] Lorentz boost factor $\gamma$  
     to parton-parton collision cm frame.
\item[ VINT(17) :] Rapidity $\hat{y}_{cm}$ of 
    2-parton cm sytem with respect to global CM frame.
\item[ VINT(18) :] Transverse momentum $\hat{p}_{\perp\;cm}$ 
 of 2-parton cm sytem in global CM frame.
\item[ VINT(19) :] Longitudinal separation 
    $\Delta\hat{r}_z$  of 2 partons in their c.m. frame
     of collision, at the point of `closest approach'.
\item[ VINT(20) :] Transverse separation 
    $\Delta\hat{r}_\perp$  of 2 partons in their c.m. frame
     of collision, at the point of `closest approach'.
\item[ VINT(21) :] $\tau = x_1x_2 s$ with $x_{1,2}$ the 
     longitudinal momentum fractions (Bjorken-$x$) of 2 colliding partons.
\item[ VINT(22) :] $\hat{y}^\ast=1/2 \ln(x_1/x_2)$, 
     the `relative' rapidity of 2 colliding partons.
\item[ VINT(23) :] $\cos(\hat{\theta})$) (scattering angle) 
    in $2\rightarrow 2$ parton collision.
\item[ VINT(24) :] $\hat{\phi}$ (azimuthal angle) in 
    in $2\rightarrow 2$ parton collision.
\\
\item[ VINT(25) -]{\bf VINT(30) :} not used.
\\
\item[ VINT(31) :] $\hat{t}_l$, lower value of kinematic limit on 
     $\hat{t}$ in parton collision.
\item[ VINT(32) :] $\hat{t}_u$, upper  value of kinematic limit on 
     $\hat{t}$ in parton collision.
\item[ VINT(33) :] $|\hat{t}|_1$, upper absolute value of 
     limit of integration over $d\hat{t}$ of 2-parton cross-section.
\item[ VINT(34) :] $|\hat{t}|_2$, lower absolute value of 
     limit of integration over $d\hat{t}$ of 2-parton cross-section.
\item[ VINT(35) :] $\cos(\hat{\theta})_{max}$ for $\hat{\theta}\le 0$.
\item[ VINT(36) :] $\cos(\hat{\theta})_{max}$ for $\hat{\theta}\ge 0$.
\item[ VINT(37) :] integrated 2-parton cross-section 
     $\hat{\sigma}(\hat{s}) = \int d\hat{p}_\perp^2 
     (d\hat{\sigma}(\hat{s},\hat{p}_\perp^2)/d\hat{p}_\perp^2$.
\item[ VINT(38) :] differential 2-parton cross-section 
     $d\hat{\sigma}(\hat{s},\hat{p}_\perp^2)/d\hat{p}_\perp^2$.
\item[ VINT(39) :] characteristic proper collision time 
     $\tau_col(p_\perp^2) \propto 1/p_\perp$ in the 2-body $c.m.$-frame.
\item[ VINT(40), ] not used.
\item[ VINT(41) :] the momentum fraction 
     $x_1$ taken by parton 1 at the 2-parton vertex.
\item[ VINT(42) :] the momentum fraction 
     $x_2$ taken by parton 2 at the 2-parton vertex.
\item[ VINT(43) :] $\hat{m} = \sqrt{\hat{s}}$, 
     mass of 2-parton collision subsystem.
\item[ VINT(44) :] $\hat{s}$ of the collision subprocess 
     ($2 \rightarrow 2, 2 \rightarrow 1$).
\item[ VINT(45) :] $\hat{t}$ of the collision subprocess 
     ($2 \rightarrow 2, 2 \rightarrow 1$).
\item[ VINT(46) :] $\hat{u}$ of the collision subprocess 
     ($2 \rightarrow 2, 2 \rightarrow 1$).
\item[ VINT(47) :] $\hat{p}_\perp$ of the collision subprocess, 
    i.e., transverse momentum evaluated in the rest frame of the collision.
\item[ VINT(48) :] $\hat{p}_\perp^2$ of the collision subprocess; see VINT(47).
\item[ VINT(49) :] $\hat{p}_{\perp\, min}$ of process, 
    i.e. PARV(42), or max(VKIN(3),VKIN(5)), 
    depending on which is larger, and whether the process is singular in 
    $\hat{p}_\perp\rightarrow 0$, or not.
\item[ VINT(50) :] not used.
\item[ VINT(51) :] $Q$ of the hard subprocess. 
    The precise definition is process-dependent, 
    see MSTV(43).
\item[ VINT(52) :] $Q^2$ of the hard subprocess; see VINT(51).
\item[ VINT(53) :] $Q$ of the outer hard-scattering subprocess.
    Agrees with VINT(51) for a $2 \to 1$ or $2 \to 2$ process.
\item[ VINT(54) :] $Q^2$ of the outer hard-scattering subprocess;
    see VINT(53).
\item[ VINT(55) :] $Q$ scale used as maximum virtuality in parton
    showers. Is equal to VINT(53), except for 
    deep-inelastic-scattering processes.
\item[ VINT(56) :] $Q^2$ scale in parton showers; see VINT(55).
\item[ VINT(57) :] $\alphaem$ value of hard process.
\item[ VINT(58) :] $\alphas$ value of hard process.
\item[ VINT(59) :] $\sin\hat{\theta}$ (cf. VINT(23)); used for
    improved numerical precision in elastic and diffractive scattering.
\item[ VINT(61),]{\bf VINT(62) :} nominal $m^2$ values, i.e. without
    initial-state radiation effects, for the two partons entering
    the hard interaction.
\item[ VINT(63),]{\bf  VINT(64) :} nominal $m^2$ values, i.e. without 
    final-state radiation effects, for the two (or one) partons/particles
    leaving the hard interaction.
\item[ VINT(65) :] $\hat{p}_{init}$, i.e. common nominal absolute
    momentum of the two partons entering the hard interaction, in their
    rest frame.
\item[ VINT(66) :] $\hat{p}_{fin}$, i.e. common nominal absolute
   momentum of the two partons leaving the hard interaction, in their
   rest frame.
\item[ VINT(67) ]{\bf - VINT(100) :} not used.
\item[ VINT(101) ]{\bf - VINT(105) :} $\theta$, $\varphi$ and
    \mbox{{\boldmath $\beta$}} for rotation and boost
    from c.m. frame to hadronic c.m. frame of a lepton-hadron 
    or lepton-nucleus event.
\item[ VINT(106) ]{\bf - VINT(110) :} $\theta$, $\varphi$ and
    \mbox{{\boldmath $\beta$}} for rotation and boost
    from c.m. frame to frame of equal nucleon-nucleon velocity
    in the case of a hadron-nucleus or an asymmetric nucleus-nucleus event
    with $A\ne B$.
\end{description}
\bigskip
\bigskip
\bigskip

\newpage
\noindent
The common block VNICOL contains accumulated data on parton collisions
during a beam/target event.

\begin{verbatim}
              COMMON/VNICOL/MCOL(1000,0:10),VCOL(1000,0:10)
\end{verbatim}

\noindent
{\it Purpose}: 
    to provide an intermediate storage for a minimal set of integer and 
    real valued variables for all detected parton collisions that occur
    during a single timestep during an event. The first index $I$ in the arrays
    MCOL and VCOL labels the sequence number of a collision and the second
    index $J$ is a pointer to interface with the above 
    MINT and VINT arrays wherein
    the calculated quantities are written to.

\begin{description}
\item[ MCOL(I,1) :] 
line number in VNIREC1/VNIREC2 of parton 1 entering the collision subprocess.
\item[ MCOL(I,2) :] 
line number in VNIREC1/VNIREC2 of parton 2 entering the collision subprocess.
\item[ MCOL(I,3) :] = MINT(1)
\item[ MCOL(I,4) :] = MINT(2)
\item[ MCOL(I,5) :] = MINT(3)
\item[ MCOL(I,6) :] = MINT(15)
\item[ MCOL(I,7) :] = MINT(16)
\item[ MCOL(I,8) :] = MINT(21)
\item[ MCOL(I,9) :] = MINT(22)
\item[ MCOL(I,10):] not used.
\\
\\
\item[ VCOL(I,1) :] = VINT(23)
\item[ VCOL(I,2) :] = VINT(24)
\item[ VCOL(I,3) :] = VINT(26)
\item[ VCOL(I,4) :] = VINT(41)
\item[ VCOL(I,5) :] = VINT(42)
\item[ VCOL(I,6) :] = VINT(44)
\item[ VCOL(I,7) :] = VINT(45)
\item[ VCOL(I,8) :] = VINT(46)
\item[ VCOL(I,9) :] = VINT(48)
\item[ VCOL(I,10):] = VINT(52)
\end{description}
\bigskip
\bigskip

\newpage

\section{Sample output}
\label{sec:appd}

\begin{center}
{\bf 1. General information on the simulation as-a-whole}
\end{center}
\smallskip

The following output stems
simulating 500 $p\bar{p}$ collisions at
CERN collider energy $\sqrt{s} = 900$ GeV. 
Unless switched of by setting $MSTV(114)=0$, this type of
output information is
provided automatically and written to 
the file VNIRUN.DAT at the end of a simulation. The output
includes a summary of the initialization overall collision system,
the gluon and quark content of the initial state, 
the type and number of occured interaction processes during
the parton-cascade development, the number and type of
produced pre-hadronic clusters from partons, and
the numbers of resulting final-state hadrons.
\bigskip

\begin{center}

\end{center}
\bigskip
\bigskip

\newpage

\begin{center}
{\bf 2. A typical event listing}
\end{center}
\smallskip

The event listing below is part of the standard output
obtained for 500 $p\bar{p}$ collisions at
CERN collider energy $\sqrt{s} = 900$ GeV. 
The listing, which is  is written to the file VNIRUN.DAT,
embodies the complete history of the first event
from the initial beam particles $p$ and $\bar{p}$,
their original gluon and quark content, the newly 
produced partons by hard scattering and radiation during
the parton-cascade development, the formed pre-hadronic clusters 
and resulting final state particles. The latter
emerge from both the parton clusters that result from
parton-cascade products, and the  beam/target clusters
that consist of the remnant initial partons
which have not participated in the parton cascades.
As explained in Sec. 3.6,
each entry contains name, status, flavor, and origin of a particle,
as well as it's three-momentum, energy and mass. Particles in
brackets are inactive, decayed or fragmented particles. 
The final line of the listing shows the total charge, three-momentum,
anargy and invariant mass of the particle system, given by
the sums over all active, stable particles.
\vspace{-0.5cm}

\begin{center}
{\small
\begin{verbatim}

                  Particle listing (summary: momentum space P in GeV)
                  event no.    1  time step  200  at time  34.75 fm

    I particle     KS     KF   KMO    P_x      P_y      P_z       E        M

    1 (p+)         15   2212     0     .000     .000  449.999  450.000     .938
    2 (p~-)        15  -2212     0     .000     .000 -449.999  450.000     .938
 ==============================================================================
    3 (p+)         15   2212     1     .000     .000  449.999  450.000     .938
 ==============================================================================
    4 (p~-)        15  -2212     2     .000     .000 -449.999  450.000     .938
 ==============================================================================
    5 (u)          13      2     3    6.468    -.283   67.618   67.928     .006
    6 (d)          13      1     3   -2.033    7.351   87.053   87.387     .010
    7 (g)          13     21     3     .135    -.110    6.530    6.530    -.175
    8 (g)          13     21     3     .410     .205    2.324    2.324    -.459
    9 (g)          13     21     3    -.009     .066    3.136    3.136    -.067
   10 (g)          13     21     3    -.098     .153   14.160   14.160    -.184
   11 (g)          13     21     3    -.094     .012    2.851    2.851    -.095
   12 (g)          13     21     3    -.159    -.130    4.587    4.587    -.205
   13 (g)          13     21     3     .745    -.058    5.911    5.910    -.747
   14 (g)          13     21     3     .015     .208    1.636    1.636    -.208
   15 (g)          13     21     3    -.002     .082    1.964    1.964    -.082
   16 (g)          13     21     3    -.133    -.088    1.742    1.742    -.159
   17 (g)          13     21     3    -.067    -.580    4.587    4.587    -.584
   18 (g)          13     21     3    -.343     .022    1.636    1.636    -.344
   19 (u~)         13     -2     3    -.355    -.198     .988     .988    -.406
   20 (u)          13      2     3     .132     .153     .988     .988    -.202
   21 (g)          13     21     3     .007     .003    2.992    2.992    -.010
   22 (g)          13     21     3    -.154     .440    3.746    3.746    -.466
   23 (g)          13     21     3    -.038     .415   17.137   17.137    -.418
   24 (g)          13     21     3     .054    -.036    2.714    2.714    -.065
   25 (g)          13     21     3     .054     .037    1.338    1.338    -.065
   26 (u~)         13     -2     3    -.406     .103   49.084   49.083    -.431
   27 (u)          13      2     3    -.268     .098    1.245    1.245    -.285
   28 (g)          13     21     3    -.009    -.030    1.742    1.742    -.031
   29 (s~)         13     -3     3     .082     .062    9.842    9.842    -.225
   30 (s)          13      3     3     .314     .241    1.070    1.070    -.443
   31 (s~)         13     -3     3    -.008    -.001    7.634    7.634    -.200
   32 (s)          13      3     3     .016    -.105     .988     .988    -.226
   33 (d)          13      1     3    -.029     .227    1.851    1.851    -.229
   34 (d~)         13     -1     3    -.030    -.025     .988     .988    -.040
   35 (g)          13     21     3    -.015     .008    1.636    1.636    -.017
   36 (g)          13     21     3     .218    -.061    6.744    6.744    -.227
   37 (g)          13     21     3     .147    -.042    2.581    2.581    -.153
   38 (g)          13     21     3     .100    -.031    2.450    2.450    -.105
   39 (s~)         13     -3     3    -.080     .236    1.245    1.245    -.319
   40 (s)          13      3     3    -.057     .207     .988     .988    -.292
   41 (g)          13     21     3    -.044    -.095    2.992    2.992    -.105
   42 (u)          13      2     3    -.343    -.011    1.433    1.433    -.343
   43 (u~)         13     -2     3     .058     .067     .988     .988    -.089
   44 (g)          13     21     3     .084    -.195    1.245    1.245    -.212
   45 (g)          13     21     3     .048     .017    1.533    1.533    -.051
   46 (g)          13     21     3    -.008     .022    2.200    2.200    -.024
   47 (g)          13     21     3    -.019     .610    4.765    4.765    -.611
   48 (g)          13     21     3     .032    -.050    2.081    2.081    -.060
   49 (u~)         13     -2     3    -.119    -.036    1.636    1.636    -.125
   50 (u)          13      2     3    -.144    -.129     .988     .988    -.193
   51 (d)          13      1     3    -.225    -.119    2.581    2.581    -.255
   52 (d~)         13     -1     3     .209    -.049     .988     .988    -.215
   53 (g)          13     21     3     .017     .334   50.278   50.278    -.351
   54 (d~)         13     -1     3    -.127     .001    1.156    1.156    -.127
   55 (d)          13      1     3     .182     .007     .988     .988    -.182
   56 (g)          13     21     3    -.026    -.076   18.553   18.553    -.089
   57 (g)          13     21     3     .095     .035    3.746    3.746    -.101
   58 (g)          13     21     3    -.248    -.167    4.240    4.240    -.299
   59 (g)          13     21     3     .035     .128    1.338    1.338    -.133
   60 (g)          13     21     3    -.170     .029    6.320    6.320    -.173
   61 (g)          13     21     3     .552     .019   14.477   14.477    -.553
   62 (g)          13     21     3    -.178    -.282    2.324    2.324    -.334
   63 (d~)         13     -1     3    -.210     .044    2.200    2.200    -.214
   64 (d)          13      1     3     .012    -.049     .988     .988    -.051
   65 (u~)         13     -2     3    -.056    -.557    2.714    2.714    -.560
   66 (u)          13      2     3    -.137    -.062     .988     .988    -.151
   67 (g)          13     21     3     .067    -.147    2.851    2.851    -.161
   68 (g)          13     21     3    -.409    -.114   10.915   10.915    -.425
   69 (g)          13     21     3     .157    -.266    6.961    6.961    -.310
   70 (g)          13     21     3    -.363    -.114    5.133    5.133    -.380
   71 (g)          13     21     3     .110     .285    8.338    8.338    -.306
   72 (u)          13      2     3     .093     .272    6.320    6.320    -.288
   73 (u~)         13     -2     3    -.561     .058     .988     .988    -.564
   74 (s)          13      3     3    -.252    -.121    7.406    7.406    -.344
   75 (s~)         13     -3     3    -.078     .016     .988     .988    -.214
   76 (d)          13      1     3    -.095     .011    4.072    4.072    -.097
   77 (d~)         13     -1     3     .013     .005     .988     .988    -.017
   78 (g)          13     21     3    -.315     .023    5.711    5.711    -.317
   79 (u~)         13     -2     3     .376    -.047    1.245    1.245    -.379
   80 (u)          13      2     3    -.339     .142     .988     .988    -.368
   81 (g)          13     21     3     .009     .110    1.964    1.964    -.111
   82 (g)          13     21     3    -.143    -.143   12.337   12.337    -.203
   83 (d~)         13     -1     3     .279     .015    2.200    2.200    -.280
   84 (d)          13      1     3    -.093    -.020     .988     .988    -.096
   85 (g)          13     21     3    -.118     .159    4.587    4.587    -.199
   86 (u~)         13     -2     3    -.005    -.196    1.245    1.245    -.196
   87 (u)          13      2     3    -.012    -.449     .988     .988    -.449
   88 (u)          13      2     3     .203     .190    4.947    4.947    -.278
   89 (u~)         13     -2     3    -.086    -.351     .988     .988    -.361
   90 (g)          13     21     3    -.041    -.083    4.240    4.240    -.093
   91 (g)          13     21     3    -.038     .260    3.434    3.434    -.262
   92 (g)          14     21   206    -.149     .025    5.467    5.467    -.147
   93 (g)          13     21     3     .001    -.168    7.406    7.406    -.169
   94 (d~)         13     -1     3     .095     .010    1.636    1.636    -.096
   95 (d)          13      1     3     .029     .014     .988     .988    -.033
   96 (u~)         13     -2     4     .538    6.772 -468.861  468.910     .006
 ==============================================================================
   97 (d~)         13     -1     4     .239     .502 -143.968  143.969     .010
   98 (u~)         13     -2     4     .050     .087   -1.742    1.742    -.101
   99 (u)          13      2     4     .149    -.244    -.988     .988    -.286
  100 (g)          13     21     4    -.204     .198   -2.324    2.324    -.284
  101 (g)          13     21     4     .021     .084   -1.433    1.433    -.087
  102 (g)          13     21     4    -.078    -.026   -2.324    2.324    -.082
  103 (g)          13     21     4     .001     .009   -4.412    4.412    -.013
  104 (u)          13      2     4     .080     .216   -4.240    4.240    -.230
  105 (u~)         13     -2     4     .126     .018    -.988     .988    -.127
  106 (u)          13      2     4    -.178    -.106   -1.156    1.156    -.208
  107 (u~)         13     -2     4    -.008     .354    -.988     .988    -.354
  108 (g)          13     21     4     .243     .021   -4.072    4.072    -.244
  109 (g)          13     21     4    -.186    -.081   -1.433    1.433    -.203
  110 (g)          13     21     4    -.140    -.036   -7.634    7.634    -.145
  111 (g)          13     21     4    -.242    -.221   -3.588    3.588    -.328
  112 (d)          13      1     4     .129     .089   -4.072    4.072    -.157
  113 (d~)         13     -1     4    -.060     .039    -.988     .988    -.072
  114 (u)          13      2     4    -.080    0.000   -4.947    4.947    -.081
  115 (u~)         13     -2     4    -.029    -.118    -.988     .988    -.122
  116 (g)          13     21     4    -.225     .015  -51.488   51.488    -.249
  117 (g)          13     21     4     .274    -.045   -1.851    1.851    -.278
  118 (g)          13     21     4     .064    -.019   -6.530    6.530    -.068
  119 (g)          13     21     4     .304    -.007   -5.322    5.322    -.305
  120 (g)          13     21     4    0.000     .433   -2.200    2.200    -.433
  121 (g)          13     21     4    -.009    -.028  -10.105   10.105    -.036
  122 (u~)         13     -2     4     .044    -.575   -8.100    8.100    -.577
  123 (u)          13      2     4    -.047     .452    -.988     .988    -.454
  124 (g)          13     21     4     .064     .004   -2.081    2.081    -.065
  125 (g)          13     21     4    -.119     .074   -4.240    4.240    -.140
  126 (g)          13     21     4     .001    -.027   -2.324    2.324    -.027
  127 (g)          13     21     4     .035    -.102   -9.582    9.582    -.110
  128 (u~)         13     -2     4     .204    -.155   -2.200    2.200    -.256
  129 (u)          13      2     4     .128    -.267    -.988     .988    -.296
  130 (s~)         13     -3     4    -.010    -.039  -23.146   23.146    -.209
  131 (s)          13      3     4    -.383    -.083   -1.156    1.156    -.439
  132 (g)          13     21     4    -.041     .164   -4.765    4.765    -.169
  133 (d)          13      1     4     .090    -.176    -.988     .988    -.198
  134 (d~)         13     -1     4     .091    -.003    -.988     .988    -.092
  135 (s)          13      3     4     .162     .150   -3.434    3.434    -.297
  136 (s~)         13     -3     4    -.025     .084    -.988     .988    -.218
  137 (d)          13      1     4     .018     .090  -11.758   11.758    -.096
  138 (d~)         13     -1     4     .033    -.095    -.988     .988    -.101
  139 (g)          14     21   207     .149    -.025   -5.468    5.470    -.065
  140 (g)          13     21     4    -.197     .040   -1.636    1.636    -.201
  141 (u)          13      2     4     .105     .015   -1.338    1.338    -.106
  142 (u~)         13     -2     4     .091     .232    -.988     .988    -.249
  143 (g)          13     21     4    -.023    -.024   -4.765    4.765    -.035
  144 (g)          13     21     4    -.101    -.226   -7.865    7.865    -.248
  145 (d~)         13     -1     4     .001     .029    -.988     .988    -.031
  146 (d)          13      1     4    -.368     .013    -.988     .988    -.369
  147 (u)          13      2     4    -.037    -.162  -14.477   14.477    -.169
  148 (u~)         13     -2     4    -.036     .378    -.988     .988    -.380
  149 (g)          13     21     4    -.255    -.163   -1.070    1.070    -.303
  150 (u)          13      2     4     .178    -.143   -1.533    1.533    -.229
  151 (u~)         13     -2     4     .010     .247    -.988     .988    -.247
  152 (g)          13     21     4    -.169    -.036  -33.894   33.894    -.187
  153 (g)          13     21     4    -.070    -.382  -15.448   15.448    -.389
  154 (g)          13     21     4    -.005     .047   -1.851    1.851    -.048
  155 (g)          13     21     4    -.299     .012   -1.636    1.636    -.299
  156 (d~)         13     -1     4    -.013     .200   -1.245    1.245    -.201
  157 (d)          13      1     4     .078     .115    -.988     .988    -.139
  158 (g)          13     21     4     .161     .004   -1.533    1.533    -.161
  159 (g)          13     21     4     .055    -.005   -1.742    1.742    -.055
  160 (g)          13     21     4     .001    -.032   -1.338    1.338    -.032
  161 (u)          13      2     4     .170    -.191  -11.192   11.192    -.257
  162 (u~)         13     -2     4    -.263    -.035    -.988     .988    -.266
  163 (g)          13     21     4     .065    -.013   -2.581    2.581    -.067
  164 (g)          13     21     4    -.086    -.321   -6.530    6.530    -.332
  165 (d)          13      1     4     .029    -.140   -1.245    1.245    -.143
  166 (d~)         13     -1     4    -.162    -.100    -.988     .988    -.191
  167 (g)          13     21     4    -.039    -.325   -2.081    2.081    -.328
  168 (d~)         13     -1     4    -.091     .017   -1.338    1.338    -.093
  169 (d)          13      1     4    -.042    -.771    -.988     .988    -.772
  170 (g)          13     21     4     .206     .254   -1.533    1.533    -.328
  171 (g)          13     21     4    -.061     .332   -2.200    2.200    -.337
  172 (d)          13      1     4     .074     .105   -6.961    6.961    -.130
  173 (d~)         13     -1     4     .050    -.015    -.988     .988    -.053
  174 (s~)         13     -3     4    -.012    -.001   -1.533    1.533    -.199
  175 (s)          13      3     4    -.105     .060    -.988     .988    -.233
  176 (g)          13     21     4    -.470    -.003   -2.200    2.200    -.470
  177 (g)          13     21     4     .081    -.082   -1.338    1.338    -.115
  178 (u~)         13     -2     4     .053    -.060   -2.324    2.324    -.080
  179 (u)          13      2     4     .043     .004    -.988     .988    -.044
  180 (u)          13      2     4     .008     .105   -5.711    5.711    -.107
  181 (u~)         13     -2     4    -.027     .029    -.988     .988    -.040
  182 (g)          13     21     4     .105     .007   -1.245    1.245    -.105
  183 (g)          13     21     4    -.029     .063   -4.412    4.412    -.070
  184 (g)          13     21     4    -.044    -.104   -2.992    2.992    -.113
  185 (g)          13     21     4    -.005     .003   -1.338    1.338    -.007
  186 (d~)         13     -1     4     .003     .128   -1.964    1.964    -.129
  187 (d)          13      1     4    -.271     .071    -.988     .988    -.281
  188 (d~)         13     -1     4     .045     .236   -1.338    1.338    -.240
  189 (d)          13      1     4    -.009    -.141    -.988     .988    -.142
  190 (g)          13     21     4    -.148    -.079   -1.742    1.742    -.168
  191 (g)          13     21     4     .066    -.088   -6.320    6.320    -.111
  192 (g)          13     21     4    -.093     .334   -4.765    4.765    -.347
  193 (g)          13     21     4     .259     .034   -3.434    3.434    -.262
  194 (u)          13      2     4     .025     .168  -21.558   21.558    -.176
  195 (u~)         13     -2     4    -.037     .036    -.988     .988    -.052
  196 (g)          14     21    92     .141     .040    3.588    3.588    -.147
  197 (g)          14     21   139     .062     .001   -8.338    8.338    -.065
 ==============================================================================
  198 !g!          21     21   200     .141     .040    3.588    3.588    -.147
  199 !g!          21     21   201     .062     .001   -8.338    8.338    -.065
  200 !g!          21     21     0   -5.188     .107   -3.853    6.463     .000
  201 !g!          21     21     0    5.392    -.066    -.897    5.466     .000
  202 (g)          14     21     0     .141     .040    3.588    3.588    -.147
  203 (g)          14     21     0     .062     .001   -8.338    8.338    -.065
  204 (g)          14     21   200   -5.188     .107   -3.853    6.463     .000
  205 (g)          14     21   201    5.392    -.066    -.897    5.466     .000
  206 (g)          14     21     0     .000     .000    5.470    5.470    -.000
  207 (g)          14     21     0     .000     .000   -5.470    5.470    -.000
  208 (CMshower)   11     94   204     .204     .041   -4.750   11.929   10.941
  209 (g)          14     21   208    3.659    -.030   -2.183    7.624    6.322
  210 (cluster)    11     91   209   -2.497    1.963   -4.249    5.660    1.973
  211 (cluster)    11     91   209    -.843     .635   -1.481    2.437    1.623
  212 pi+           6    211   210     .627     .196     .847    1.081     .140
  213 (K*_10)      16  20313   210    -.463     .736   -1.950    2.646    1.563
  214 (phi)        16    333   211    -.743     .632    -.926    1.688    1.020
  215 K-            6   -321   211    0.000    0.000     .565     .751     .494
  216 (K*0)        16    313   215    0.000    0.000   -1.227    1.536     .924
  217 pi0           6    111   215    -.237    -.351     .038     .446     .135
  218 K-            6   -321   218    -.001    0.000     .373     .619     .494
  219 K+            6    321   218    0.000    0.000     .098     .503     .494
  220 K0            6    311   218    0.000    0.000     .093     .506     .498
  221 pi0           6    111   218    -.142     .010     .243     .312     .135
  222 (g)          13     21   209    2.096   -1.634    1.690    3.150     .010
  223 (Beam-REM)   11     92   209    7.456    2.673  668.286  671.476  498.322
  224 (Targ-REM)   11     93   209   -1.241    7.624-1060.622 1060.682  960.492
  225 (CMF)        15    100   223    6.215   10.297 -392.336 1732.158 1687.097
 ==============================================================================
  226 (cluster)    11     91   223    7.456    2.673  668.286  671.476  498.322
  227 (cluster)    11     91   224   -1.241    7.624-1060.622 1060.682  960.492
  228 (Beam-REM)   11     92   226    3.106    1.329  413.755  413.777    2.559
  229 (K*0)        17    313   228    2.034     .486  173.846  173.861     .892
  230 (Sigma0)     17   3212   228    1.071     .843  239.909  239.916    1.192
  231 pi0           7    111     0     .672     .201  220.546  220.547     .135
  232 K0            7    311     0    -.801    -.560 -295.110  295.112     .498
  233 ((gamma))    -7     22   234    0.000     .045    6.088    6.089     .000
  234 Lambda0       7   3122     0    -.317    -.704   93.191   93.200    1.116
  235 (Targ-REM)   11     93   226   -1.050    5.146 -773.796  773.817    2.117
  236 (Delta~-)    17  -2214   235    -.580    2.264 -368.320  368.329    1.232
  237 (rho+)       17    213   235    -.470    2.883 -405.476  405.487     .769
  238 pi0           7    111     0    -.021     .284   20.497   20.500     .135
  239 p~-           7  -2212     0     .229     .387   59.692   59.701     .938
  240 pi0           7    111     0    -.101     .147    6.308    6.312     .135
  241 pi+           7    211     0    -.298    -.690   11.165   11.191     .140
  242 (Beam-REM)   11     92   226    3.084     .532  228.653  228.682    1.886
  243 (rho+)       17    213   242     .631     .097   81.099   81.105     .769
  244 (rho-)       17   -213   242    2.453     .435  147.554  147.577     .769
  245 pi0           7    111     0    1.086    -.372   13.180   13.231     .135
  246 pi+           7    211     0     .670     .010    6.556    6.592     .140
  247 pi0           7    111     0    -.310     .098    1.379    1.423     .135
  248 pi-           7   -211     0     .317    -.273    1.674    1.731     .140
  249 (Beam-REM)   11     92   226     .350    1.923   41.821   41.947    2.589
  250 pi0           7    111     0     .338    -.061     .993    1.059     .135
  251 (f_2)        17    225   249     .251    2.340   38.840   38.932    1.273
  252 pi-           7   -211     0     .173     .234     .476     .575     .140
  253 pi+           7    211     0    -.126     .040     .015     .193     .140
  254 (Beam-REM)   11     92   226     .848     .209    2.363    2.994    1.619
  255 (rho+)       17    213   254     .405     .603     .830    1.345     .769
  256 pi-           7   -211     0     .311     .020     .030     .343     .140
  257 pi0           7    111     0    -.133     .085     .042     .212     .135
  258 pi+           7    211     0    -.142     .273     .267     .431     .140
  259 (Targ-REM)   11     93   226    -.645     .059   -4.906    6.079    3.532
  260 (a_20)       17    115   259    -.785    -.697   -3.994    4.335    1.318
  261 (f_2)        17    225   259     .140     .756    -.912    1.744    1.273
  262 pi+           7    211     0     .159    -.214     .257     .395     .140
  263 (rho-)       17   -213   260   -1.001    -.566   -3.481    3.746     .769
  264 pi-           7   -211     0    -.222     .216     .622     .709     .140
  265 pi+           7    211     0     .508    -.089    1.136    1.256     .140
  266 (pi0)        17    111   263    -.206     .103   -1.282    1.309     .135
  267 pi-           7   -211     0    -.131    -.273     .880     .941     .140
  268 (Targ-REM)   11     93   226     .334    -.009  -13.547   13.730    2.207
  269 (K*+)        17    323   268     .543    -.031  -10.390   10.442     .892
  270 (K*~0)       17   -313   268    -.209     .022   -3.157    3.288     .892
  271 pi+           7    211     0    -.937     .548   -3.593    3.756     .140
  272 K0            7    311     0    -.641     .364   -6.401    6.463     .498
  273 pi+           7    211     0     .032    -.219   -2.566    2.580     .140
  274 K-            7   -321     0     .135     .027   -5.088    5.113     .494
  275 (Targ-REM)   11     93   226     .342    -.955  -20.126   20.346    2.810
  276 pi-           7   -211     0     .063     .297   -4.395    4.407     .140
  277 (f_2)        17    225   275     .518    -.135  -18.057   18.110    1.273
  278 pi-           7   -211     0    -.897     .386  -16.770   16.799     .140
  279 pi+           7    211     0     .078     .064   -6.170    6.172     .140
  280 (Targ-REM)   11     93   226     .011    2.337 -277.032  277.048    1.826
  281 (omega)      17    223   280    -.282    1.678 -154.896  154.908     .783
  282 (rho0)       17    113   280     .293     .659 -122.136  122.141     .769
  283 ((gamma))    -7     22   285    -.076     .029  -16.715   16.715     .000
  284 pi0           7    111     0     .272     .281  -27.169   27.172     .135
  285 pi-           7   -211     0    -.065    -.210  -17.101   17.103     .140
  286 pi+           7    211     0    -.015    -.300  -48.046   48.047     .140
 ==============================================================================
                  sum:   .00           .000     .000    0.000  900.000  900.000

\end{verbatim}
}
\end{center}

\newpage

\section{Summary list of subroutines}
\label{sec:appe}
\bigskip
                                                                 
\begin{center}
{\small
\begin{verbatim}

C***********************************************************************
C                                                                      *
C                                                                      *
C  List of subprograms grouped according to general program structure  *
C                                                                      *
C            (S = subroutine, F = function, B = block data)            *
C                                                                      *
C                                                                      *
C        PART 1: Main steering routines and general utilities          *
C                                                                      *
C  B   VNIDATA  to give default values to particle and decay data      *
C  S   VNIXRIN  to initialize the overall run and simulation procedure *
C  S   VNIXRUN  to steer the simulation of a single collision event    *
C  S   VNIXFIN  to finish up the overall run and simulation procedure  *
C  S   VNICBIN  to interface parameters sets between commonblocks      *
C  S   VNILOGO  to write program header                                *
C  S   VNILIST  to list event record, particle, or parameter values    *
C  S   VNIEDIT  to perform global manipulations on the event record.   *
C  S   VNICOPY  to save/restore copy of commonblocks & particle record *
C  F   KVNI     to provide integer-valued event information            *
C  F   PVNI     to provide real-valued event information               *
C  F   RLU      to provide a random number generator                   *
C  S   RLUGET   to save the state of the random number generator       *
C  S   RLUSET   to set the state of the random number generator        *
C  S   VNIERRM  to write error messages and abort faulty run           *
C  S   VNIFILE  to open and close output files                         *
C  S   VNIDUMP  to dump(read) state of all commonblocks to(from) file  *
C                                                                      *
C                                                                      *
C    PART 2: Initialization routines (global and process dependent)    *
C                                                                      *
C  S   VNICSIN  to initialize collision system and kinematics          *
C  S   VNIABXS  to give parametrized cross-sections for beam/target    *
C  S   VNITMIN  to initialize time slizes and form of time increment   *
C  S   VNI2JIN  to set up an initial 2-jet in l+l- -> gamma/Z0 -> qq~  *
C  S   VNI2JFL  to select flavour for initially produced qq~ 2-jet     *
C  S   VNI2JPY  to get energy-momentum of initial 2-jet using PYTHIA   *
C  S   VNI4JIN  to set up an initial 4-jet in l+l- -> W+W- -> qq~ q'q'~*
C  S   VNI4JFL  to initialize W+ and W- decay vertices and the 4 jets  *
C  S   VNI4JPY  to get energy-momentum of W+W- and 4 jets using PYTHIA *
C  S   VNILHIN  to set up initial system for l- + h -> l- + jet(s) + X *
C  S   VNILAIN  to set up initial system for l- + A -> l- + jet(s) + X *
C  S   VNIXDIS  to generate the variables of DIS photon-hadron(nucleus)*
C  F   VNIGAQX  to give elementary photon-quark cross-sections         *
C  S   VNIGAQQ  to give qq~ fluctuation of photon from VDM             *
C  S   VNIHHIN  to set up initial system for h  + h'->  jets + X       *
C  S   VNIHAIN  to set up initial system for h  + A ->  jets + X       *
C  S   VNIABIN  to set up initial system for A  + A'->  jets + X       *
C  S   VNIBTIN  to set up initial state of hadronic or nuclear beams   *
C  S   VNINPDT  to tabulate nuclear neutron/proton distribution        *
C  S   VNIRPIN  to initialize single hadrons, or nucleons in nucleus   *
C  S   VNIQGDT  to tabulate distributions in x and flavor of partons   *
C  F   VNIEFFQ  to give parametrization of eff. parton resolution Q(x) *
C  F   VNIQGSH  to give simple parametrization of parton shadowing     *
C  S   VNIVQKF  to extract the valence quark content of a hadron       *
C  S   VNIQGXF  to initialize x and flavor of partons in parent hadron *
C  S   VNIQGCS  to arrange a set of partons as color singlet hadron    *
C  S   VNIQGIN  to distribute initialized partons inside parent hadron *
C  S   VNIPRKT  to give primordial kT to initial partons               *
C  S   VNIXTIN  to find spatial extension of initial particle system   *
C                                                                      *
C                                                                      *
C            PART 3: Particle structure function package               *
C                                                                      *
C  - portable, self-contained program unit with interface to PDFLIB -  *
C                                                                      *
C  S   VNISTFU  calls structure functions: photon, pi+, n, p, hyperon  *
C  S   VNISTGA  gives photon structure function                        *
C  S   VNISTAG  gives anomalous part of photon structure function      *
C  S   VNISTGS  parametrized parton distributions in a virtual photon  *
C  S   VNISTPI  gives pi+ structure function                           *
C  S   VNISTFL  gives proton structure function at small x and/or Q^2  *
C  S   VNISTPR  gives proton structure function                        *
C  S   VNIFCTQ  CTEQ2'94 parametrization of parton distributions       *
C  S   VNIFGRV  GRV'94 parametrizations of parton distributions        *
C  F   FV       Function for GRV parametrizations                      *
C  F   FW       Function for GRV parametrizations                      *
C  F   FWS      Function for GRV parametrizations                      *
C  F   FHS      Function for GRV parametrizations                      *
C  F   VNIHFTH  gives threshold factor for heavy flavour production    *
C  F   VNIDILN  gives Dilogarithm function                             *
C  F   VNIGAMM  gives ordinary Gamma function                          *
C                                                                      *
C                                                                      *
C     PART 4: Main routines for cascade evolution and hadronization    *
C                                                                      *
C  S   VNIEVND  to administer general duties at the end of an event    *
C  S   VNIEVOK  to decide whether event is o.k. (accept or regenerate) *
C  S   VNIEV2J  to perform evolution for l+ + l- -> 2 jets -> hadrons  *
C  S   VNIEV4J  to perform evolution for l+ + l- -> 4 jets -> hadrons  *
C  S   VNIEVLH  to perform evolution for l- + h  ->  jets -> hadrons   *
C  S   VNIEVLA  to perform evolution for l- + A  ->  jets -> hadrons   *
C  S   VNIEVAB  to perform evolution for A  + B ->   jets -> hadrons   *
C  S   VNICOLP  to find 2-parton collisions and their kinematics       *
C  S   VNICOLX  to give cross-sections for 2 -> 2 and 2 -> 1 collisions*
C  S   VNICOLI  to set up a found 2-parton collision subsystem         *
C  S   VNICOLL  to find outgoing flavors and type of parton collision  *
C  S   VNISBRA  to generate space-like parton branchings               *
C  S   VNITBRA  to generate time-like parton branchings                *
C  S   VNICHAD  to organize parton cluster formation and hadronization *
C  S   VNICFOR  to form one or two cluster(s) from a pair of partons   *
C  S   VNICREC  to reconnect cluster-partons according to color flow   *
C  S   VNICFRA  to fragment found parton clusters into hadrons         *
C  S   VNIBHAD  to do beam remnant cluster formation and hadronization *
C  S   VNIBFRA  to fragment beam/target remnant clusters into hadrons  *
C  S   VNIBMPS  to give phase-space dist. of hadrons from cluster decay* 
C  S   VNIBMMO  to balance beam/target side momentum of hadrons        *
C  S   VNIPADC  to decay unstable particles according to particle data *
C  S   VNIPAKF  to give a hadronic particle from a flavor pair         *
C  S   VNINSPE  to recollect/fragment partons of spectator nucleons    *
C  S   VNISTPG  to simulate baryon stopping in nuclear A+B collisions  *
C  S   VNIPICL  to cluster pions together if above critical density    *
C  S   VNICOLH  to find & perform 2-body collisions of clusters/hadrons*
C  S   VNIHACL  to reconstruct cluster from its hadronic decay products*
C                                                                      *
C                                                                      *
C                    PART 5: Auxiliary routines I:                     *
C                                                                      *
C               - parton cascades & cluster formation -                *
C                                                                      *
C  F   VNIALEM  to give the alpha_electromagnetic value                *
C  F   VNIALPS  to give the alpha_strong value                         *
C  F   VNIANGL  to give the angle from known x and y components        *
C  F   INVCHGE  to give three times the electric charge                *
C  F   INVCOMP  to compress standard KF flavour code to internal KC    *
C  S   VNICOLB  to boost/rotate 2-body collision subsystem along p_z   *
C  S   VNIDEC2  to generate 2-body decay                               *
C  S   VNIFLAV  to give flavor composition & baryon number of system   *
C  S   VNIFRAM  to perform boosts between different frames             *
C  S   VNIKLIM  to find kinematic limits in 2-parton collsions         *
C  S   VNICCOL  to connect color indices in 2-parton collisions        *
C  S   VNIJCOL  to connect  partons with color flow indices            *
C  S   VNILCOL  to label color indices according to color flow tracing *
C  F   VNIMASS  to give the mass of a particle or parton               *
C  F   VNIMTHR  to give `current' or `constituent' masses for partons  *
C  S   VNINAME  to give the name of a particle or parton               *
C  F   VNIGORI  to give the origin of a particle in space-time history *
C  F   VNISORI  to sey the origin of a particle in space-time history  *
C  S   VNIONSH  to put final-state off-shell hadrons on mass-shell     *
C  S   VNIPROP  to propagate a given number of particles               *
C  S   VNIPSEA  to generate  a hadron pair from the Dirac `sea'        *
C  S   VNIQSEA  to generate a q or q~, or a qq~-pair from the `sea'    *
C  S   VNIROBO  to rotate and/or boost a given number of particles     *
C  S   VNIRSCL  to rescale particle momenta for numerical precision    *
C  F   VNIRCAM  to generate random number from a `camel' distribution  *
C  F   VNIRMTP  to generate random number from a power-law distribution*
C  F   VNIRMTX  to generate random number from 2-dim exp. distribution * 
C  F   VNIRPLT  to generate random number from flat-top distribution   *
C  F   VNIRGAU  to generate random number from Gaussian distribution   *
C  S   VNISMRP  to smear out classical particle positions and momenta  *
C  S   VNISORT  to count and sort particles                            *
C  S   VNIWIDT  to calculate width of particle resonances              *
C                                                                      *
C                                                                      * 
C                    PART 6: Auxiliary routines II:                    * 
C                                                                      * 
C          - cluster-hadron decay and beam/target fragmentation -      * 
C                                                                      * 
C  S   HWHEPC  to convert VNI event record to or from standard HEP     *
C  S   HWLIST  to list event record in HEP format                      *
C  S   HWIGIN  to set parameters of HERWIG                             *
C  S   HWUINC  to compute constants and lookup tables                  *
C  S   HWUINE  to perform re-initialization of a new event             *
C  S   HWURES  to load common blocks with particle data                *
C  S   HWUIDT  to translate particle identifiers                       *
C  S   HWCCUT  to cut up heavy cluster into two clusters               *
C  S   HWCFLA  to to set up flavors for cluster decay                  *
C  S   HWCHAD  to hadronize prepared beam/target clusters              *
C  S   HWDHAD  to decay unstable hadrons from beam/target remnants     *
C  S   HWDTWO  to generate two-body decay 0->1+2                       *
C  S   HWDTHR  to generate three-body decay 0->1+2+3                   *
C  F   HWDPWT  to give matrix element squared for phase-space decay    *
C  F   HWDWWT  to give matrix element squared for V_A weak decay       *
C  F   HWMNBI  to compute negative binomial probability                *
C  S   HWMLPS  to generate cylindrical phase-space                     *
C  S   HWMULT  to pick charged multiplicity of beam/target diffraction *
C  F   HWUPCM  to compute c.m. momentum for masses in 2-body decay     *
C  S   HWULOB  to transform momenta from rest frame into lab frame     *
C  S   HWULOF  to transform momenta from lab frame into rest frame     *
C  S   HWUMAS  to put mass in 5th component of momentum vector         *
C  S   HWUROB  to rotate vectors by inverse of rotation matrix R       *
C  S   HWUROT  to perform rotation with rotation matrix R              *
C  S   HWUSOR  to sort n-tupel A(N) into ascending order               *
C  F   HWUSQR  to give square root with sign retention                 *
C  S   HWUSTA  to make a particle type stable                          *
C  S   HWVDIF  to give vector difference                               *
C  F   HWVDOT  to give vector dot product                              *
C  S   HWVEQU  to give vector equality                                 *
C  S   HWVSCA  to give product of vector times scalar                  *
C  S   HWVSUM  to give vector sum                                      *
C  S   HWVZRO  to give zero vector                                     *
C  F   HWRGEN  to provide random number generator                      *
C  S   HWRAZM  to randomly rotate 2-vector pT                          *
C  F   HWREXP  random number from 1-dim. exponential distribution      *
C  F   HWREXT  random number from 2-dim. exponential distribution      *
C  F   HWRGAU  random number from 2-dim Gaussian distribution          *
C  F   HWRINT  random integer in finite interval                       *
C  F   HWRLOG  random logical                                          *
C  S   HWRPOW  random number from power-law distribution               *
C  F   HWRPOT  random number from 2-dim power-law distribution         *
C  F   HWRCAM  random number from `camel' distribution                 *
C  F   HWRUNG  random number from flat-top distribution                *
C  F   HWRUNI  uniform random in finite interval                       *
C  S   HWUTIM  dummy routine                                           *
C  S   HWWARN  dummy routine                                           *
C  B   HWUDAT  to provide HERWIG common blocks with data               *
C                                                                      * 
C                                                                      *
C                   PART 7: Integrated package WIZ                     *
C                                                                      *
C                    - hadron / cluster cascade -                      *
C                                                                      *
C       All components of WIZ were developed and put together by the   *
C                 "Wizard of Poquott"    Ron Longacre                  *
C                                                                      *
C                                                                      *
C  S   WIZDATA  to give default values to switches & parameters of WIZ * 
C  S   WIZCOLL  to perform cluster / hadron collisions in a time step  *
C  S   WIZADQM  to get cross section using the additive quark model    *
C  S   WIZSIGM  to get cross section using the parametrized exp. data  *
C  S   WIZMMXS  to evaluate total meson-meson interaction cross-section*
C  S   WIZNNXS  to evaluate total nucleon-nucleon cross-section        *
C  S   WIZCNQU  to determine baryon, strange, charm and bottom number  *
C  S   WIZCHCK  to check energy-momentum, charge, flavor conservation  *
C  S   WIZXSEC  to set up collsion of two (pre)hadronic particles      *
C  S   WIZYSEC  to perform collision of two (pre-)hadronic particles   *
C                                                                      *
C                                                                      *
C  S   ANNIL    to annihilate a baryon and antibaryon to a meson pair  *
C  S   MAKBBAR  to make (create) a baryon-antibaryon from a meson pair *
C  S   TNICES   to generate outgoing particles of (pre-)hadronic coll. *
C  S   ROTIT    to rotate coordinate system for boosts                 *
C  S   RBIAS    to generate minimum bias event or beam jets            *
C  S   FXQUOD   to determine strangeness number in HEP id code         *
C  S   FXCSBY   to determine charm number in HEP id code               *
C  S   CHNGID   to perform isospin rotation for minimum bias events    *
C                                                                      *
C  S   SCATS    to do low-energy coll.'s - meson-nucleon, Delta-N      *
C  S   SCATKN   to do low-energy coll.'s - K-N                         *
C  S   SKATH    to do low-energy coll.'s - K-Lambda, K-Sigma           *
C  S   SWAVE    to rotate colliding particle momenta creating a S-wave *
C  F   BARPRO   to get probability for Delta-Pi or Rho-N channels      *
C  F   BCHPRO   to get probability for 3-body decay out of collision   *
C  S   DELKST   to determine K* or Delta modes in K-N collision        *
C  F   ETAPRO   to get probability for eta production                  *
C  F   KNPRO    to get probability for K-N production                  *
C  F   KSTPRO   to get probability for K* production                   *
C  F   LAMPRO   to get probability for Lambda production               *
C  F   LPIPRO   to get probability for Lambda-Pi production in K-N     *
C  F   PKNPRO   to get probability for pi K-N production               *
C  F   RHOSC    to determine if F.F. for rho-N is called               *
C  F   SCPPRO   determines if F.F. is called for non-MM scattering     *
C  F   SCPRO    determines if elastic scattering for non-MM channels   *
C  F   DELPR    determine the probability for Delta N -> N N           *
C                                                                      *
C  F   SIGBR    get branching ratio for Sigma-Pi charge states         *
C  F   SIGLAM   get branching to K-Lambda from Pi-N                    *
C  F   SIGPRO   determine if Sigma is produced in low energy M B coll. *
C  S   SIGRHO   to determine Sigma* or Rho modes in K-N collision      *
C  F   SPIPRO   determine Sigma pi for low energy K-N collision        *
C  F   SRHPRO   determine Sigma pi pi for low energy K-N collision     *
C  F   SWAPRO   determine three modes from low energy K-N collision    *
C  S   RECPRO   to determine Delta or Rho modes in Pi-N collision      *
C  F   BCHKH    determines if three body in Pi-Hyperon interaction     *
C  F   LAMKH    determine if Lambda is produced in K-Hyperon scattering*
C  F   SIGKH    determine if Sigma is produced in K-Hyperon scattering *
C                                                                      *
C  S   MMSCAT   to collide 2 mesons and generate outgoing particles    *
C  S   SCAM     to do low-energy coll.'s M M scattering                *
C  F   SCPPI    determines if F.F. is called M M scattering            *
C  F   SCPI     determines if elastic scattering for M M               *
C  F   XEHC     determine charge exchange for K* system                *
C  F   CHEX     determine charge exchange low energy Pi Pi K Pi scatt. *
C  F   ETAETA   determine if eta eta is produced from M M scattering   *
C  F   KKBAR    determine if K Kbar is produced from M M scattering    *
C  F   PIPI     determine if pi pi is produced from M M scattering     *
C  F   ETPI     determine if eta pi is produced from M M scattering    *
C  F   PIK      determine if K pi is produced from M M scattering      *
C  F   BCHTB    determines if three body in M M interaction            *
C  S   RECP     determine delta or rho isobars in M N collision        *
C  S   ISOSCAT  collide isobar states N* Y* A1 ect. with isobar or reg.*
C                                                                      *
C  F   SCPA1    determines if F.F. is called A1 meson scattering       *
C  S   SCA1T    to do low-energy coll.'s - Pi-A1                       *
C  F   PIA1     determines if Pi-A1 is produced                        *
C  F   RA1PI    determines the charge mode for Pi-A1                   *
C  F   CHEXPI   determines the correct charge for the outgoing Pi Pi   *
C  F   KKB4PI   determines if K Kbar should be produced from Pi-A1     *
C  S   MAKISO   makes isobars out of 2 particles                       *
C  S   A1PRB    to determine probability for A1 formed from 2 particles*
C                                                                      *
C  S   PUTPNT   reuse array position that stores isobar decay products *
C  S   GETPNT   gets a reused position to store isobar decay products  *
C  F   CHARJ    returns charge of particle and works for isobars       *
C  F   A1MASS   returns mass of particle and works for isobars         *
C  F   PBARGT   determines energy dep. for diquarks at Ecm in F.F.     *
C  S   ANMESID  B Bar annihilation into mesons the meson id's are made.*
C  F   PANNIL   determines the B Bar annihilation probability.         *
C  S   BBARID   to decompose baryon id into flavor/spin of a baryon    *
C  F   YSTPRO   determines Y* production.                              *
C                                                                      *
C  S   PHIEXT   determines phi production in non M M scattering.       *
C  S   PHIEXTM  determines phi production in M M scattering.           *
C  F   PHIPI    determines if a phi pi state is produced in M M scatter*
C  F   PHIK     determines if a phi K state is produced in M M scatter *
C  F   PHIKST   determines if a phi K* state is produced in M M scatter*
C  F   PHIETA   determines if a phi eta state is produced in M M scatt.*
C  F   PHIPHI   determines if a phi phi state is produced in M M scatt.*
C  F   PHIPIA1  determines if a phi pi state decays from the rho prime.*
C  S   CHAGID   Puts in the correct OZI suppression for phi B scatters.*
C                                                                      *
C  F   IDISA    to convert particle id's from HEP to WIZ codes         *
C  F   IDHEP    to convert particle id's from WIZ to HEP codes         *
C  F   HEPMASS  to provide particle masses according to HEP standard   * 
C  F   RANF     to provide internal random number generator for WIZ    *
C                                                                      *
C  S   MBSET    Set parameters for generating minbias or beam jets.    *
C  S   MOVLEV   Moves one array into an other array.                   *
C  S   DECAY    The WIZ decay of particles not used outside of WIZ.    *
C  S   FLAVOR   The flavor decode of WIZ id's.                         *
C  F   LABEL    Returns character*8 string for WIZ id's.               *
C  F   AMASS    Mass of particles with WIZ id's.                       *
C  F   CHARGE   Charge of particles with WIZ id's                      *
C  F   DBLPCM   Calculate com momentum for A->B+C with double precision*
C  F   GETPT    generates pt for quark pairs for the F.F. chain.       *
C  S   IDGEN    Does nothing.                                          *
C  S   LOREN4   Does a general Lorentz transformation.                 *
C  S   JETGEN   Fragments jet into primary hadrons using Field Feynman.*
C  S   PRTEVT   Prints the event stored in the ISAJET part of WIZ.     *
C  S   QCDJET   Final state QCD jet evolution using Fox and Wolfram.   *
C  S   READIN   Setup parameters for the ISAJET part of WIZ.           *
C  S   RESCAL   Rescales momentum of particles to the total energy.    *
C  S   RESET    Resets all user defined variables for ISAJET part of WIZ
C  S   SETCON   Sets the constants for the ISAJET part of WIZ.         *
C  S   SETDKY   Sets the decay table that is used inside WIZ.          *
C  S   SETTYP   Sets jetype flags in ISAJET.                           *
C  S   HEAVYX   Generates X for heavy particle fragmentation (Peterson)*
C  S   QCDT     Auxiliary routine for QCDJET.                          *
C  S   QCDZ     Auxiliary routine for QCDJET.                          *
C                                                                      *
C                                                                      *
C          PART 8: Event study and data analysis routines              *
C                                                                      *
C  S   VNIANAL  to steer pre-programmed event analysis                 *
C  S   VNIANA1  statistics on global particle properties & observables *
C  S   VNIANA2  statistics on time evol., ln(1/x) spectra, BE effect   *
C  S   VNIANA3  statistics on pre-hadronic clusters                    *
C  S   VNIANA4  statistics on partons and produced hadrons             *
C  S   VNISPHE  to perform sphericity analysis                         *
C  S   VNITHRU  to perform thrust analysis                             *
C  S   VNICLUS  to perform three-dimensional jet cluster analysis      *
C  S   VNICELL  to perform jet cluster analysis in (eta, phi, E_T)     *
C  S   VNIJMAS  to give high and low jet mass of event                 *
C  S   VNIFOWO  to give Fox-Wolfram jet moments                        *
C                                                                      *
C                                                                      *
C          PART 9: Dummy routines and system dependent utilities       *
C                                                                      *
C  S   VNIPRIN  dummy routine: remove when linking 'vniprob.f'         *
C  S   VNIEVPR  dummy routine: remove when linking 'vniprob.f'         *
C  S   VNIPRND  dummy routine: remove when linking 'vniprob.f'         *
C  S   VNIPRIN  dummy routine: remove when linking 'vniprob.f'         *
C  S   PYINIT   dummy routine: remove when linking 'vnipythia.f'       *
C  S   PYEVNT   dummy routine: remove when linking 'vnipythia.f'       *
C  S   PDFSET   dummy routine: remove when linking PDFLIB              *
C  S   STRUCTM  dummy routine: remove when linking PDFLIB              *
C  S   VNICLOC  to measure and return the CPU time used  (sys. dep.)   *
C  S   VNIDATE  to give current date and time  (sys. dep.)             *
C                                                                      *
C                                                                      *
C***********************************************************************
\end{verbatim}
}
\end{center}

\bigskip
\bigskip

\newpage

\section{Bibiliography}

\end{document}